%% file: main_single.tex
\documentclass[11pt]{article}
\usepackage[T1]{fontenc}
\usepackage{lmodern}
\usepackage[letterpaper,margin=1in]{geometry}
\usepackage[hyphens]{url}
\usepackage{graphicx}
\usepackage{natbib}
\usepackage{booktabs}
\usepackage{float}
\usepackage{placeins}
\usepackage{multirow}
\usepackage{amsmath,amssymb}
\usepackage{array}
\usepackage{xcolor}
\usepackage{longtable}
\usepackage{pdflscape}
\usepackage{tikz}
\usetikzlibrary{positioning,arrows.meta,calc,fit,backgrounds,decorations.pathreplacing}
\usepackage[hidelinks]{hyperref}
\title{Autonomous Research Agents:\\ A Survey of AI Scientists and the Verification Gap}
\author{Tianyu Ding, Aditya Nannapaneni, Bingfan Liu, Ling Zhang}
\date{}
\providecommand{\affiliations}[1]{}

\begin{document}
\maketitle

\begin{abstract}
Large language model (LLM) agents are increasingly deployed across the scientific research lifecycle:
generating ideas, reviewing literature, designing and running experiments, analyzing results, drafting
manuscripts, and reviewing them. End-to-end ``AI scientist'' systems now produce paper-like manuscripts
that are evaluated by automated or workshop-style review. We argue that, in the public, full-text-coded
sample we study, code release is now more common than reproducibility-grade and claim-verification
artifacts: the open question is not just whether an agent can finish a research task, but whether anyone
can \emph{verify} the claims it produces. This survey makes that
verification gap its organizing concern. We scope the survey to the computational (AI/ML) research setting, the
testbed where autonomous-science claims are most observable because code, experiments, benchmarks, and
write-ups can be inspected. We contribute four artifacts. First, a coded corpus screened from 125 candidates to 35 included works, of which we
full-text code 26 entries (24 runnable systems plus two study/position works) on seven \emph{audit}
dimensions: lifecycle stage, autonomy level, evaluation method, released artifacts, human-in-the-loop
points, novelty-verification method, and result-selection disclosure. The corpus yields the survey's
main pattern: code release is now common (83\% of the 24 runnable systems), but the artifacts and
  checks that let a reviewer \emph{verify} a result are not. Only 38\% release the seeds or execution traces needed to
  reproduce a run, only 38\% report any novelty-verification method, and the same qualitative pattern holds
  in the 22-system LLM-era runnable subset. These interpretive rates should be read as directional audit
  evidence because second-coder agreement was lower for autonomy, novelty, and selection than for artifact
  release. Of nine systems with closed-loop
  (L4) autonomy seven are \emph{mechanical} re-runs and one is author-claimed without an external check, so
  that \emph{no LLM-era system in the corpus} demonstrates an externally validated in-loop oracle under our
  coding rule; the single externally validated case predates LLM agents and is included as a contrast
  benchmark. In this sample, code availability is less scarce than reproducibility-grade and
claim-verification evidence; the harder problem is verifying the claims these systems produce. Second, a lifecycle $\times$ autonomy map indexed to that corpus, in which
missing disclosures are coded explicitly rather than inferred. Third, an auditability gap analysis that
links what current evaluation measures to how these systems recurrently fail. Fourth, a
reviewer-operational reporting checklist that ties each disclosure to the failure mode it addresses. We
position the survey against autonomy-axis and domain-axis surveys and argue that auditability coding, not
the taxonomy, is the contribution.
\end{abstract}

{\hypersetup{linkcolor=black}
\setcounter{tocdepth}{2}
\tableofcontents}
\clearpage

\section{Introduction}

Large language model (LLM) agents now attempt the full arc of research, from selecting a problem to
writing the manuscript that reports the result~\citep{lu2024aiscientist,yamada2025aiscientistv2}. Systems
described as ``AI scientists'' chain ideation, code generation, experiment execution, analysis, and
manuscript writing into a single loop, and some report machine-generated manuscripts that are judged by
an automated reviewer or at workshop level~\citep{yamada2025aiscientistv2,miyai2025jrai}.

As such systems multiply, the question that matters changes. It is no longer \emph{whether} an agent can
complete a research task but \emph{whether we can trust} the result, and whether anyone can check it at
all~\citep{bisht2026notbuilt}. A system that emits a manuscript has not necessarily made a discovery: the
claim may rest on a weak baseline, an unreproducible run, a hallucinated citation, or a result selected
post hoc from many attempts. Recent critical work argues that today's agents function as capable
co-scientists yet are not built for autonomous discovery, citing biased problem selection, missing tacit
laboratory knowledge, diversity collapse from preference optimization, and benchmarks that measure
single-turn accuracy rather than closed-loop validity~\citep{bisht2026notbuilt}. Capability and
\emph{verifiability} have to be assessed together.

We therefore organize the survey around verification rather than capability. The question is:
\emph{what can autonomous research agents do without human judgment, and how would a reviewer know?}

\begin{figure}[t]
\centering
\includegraphics[width=\textwidth]{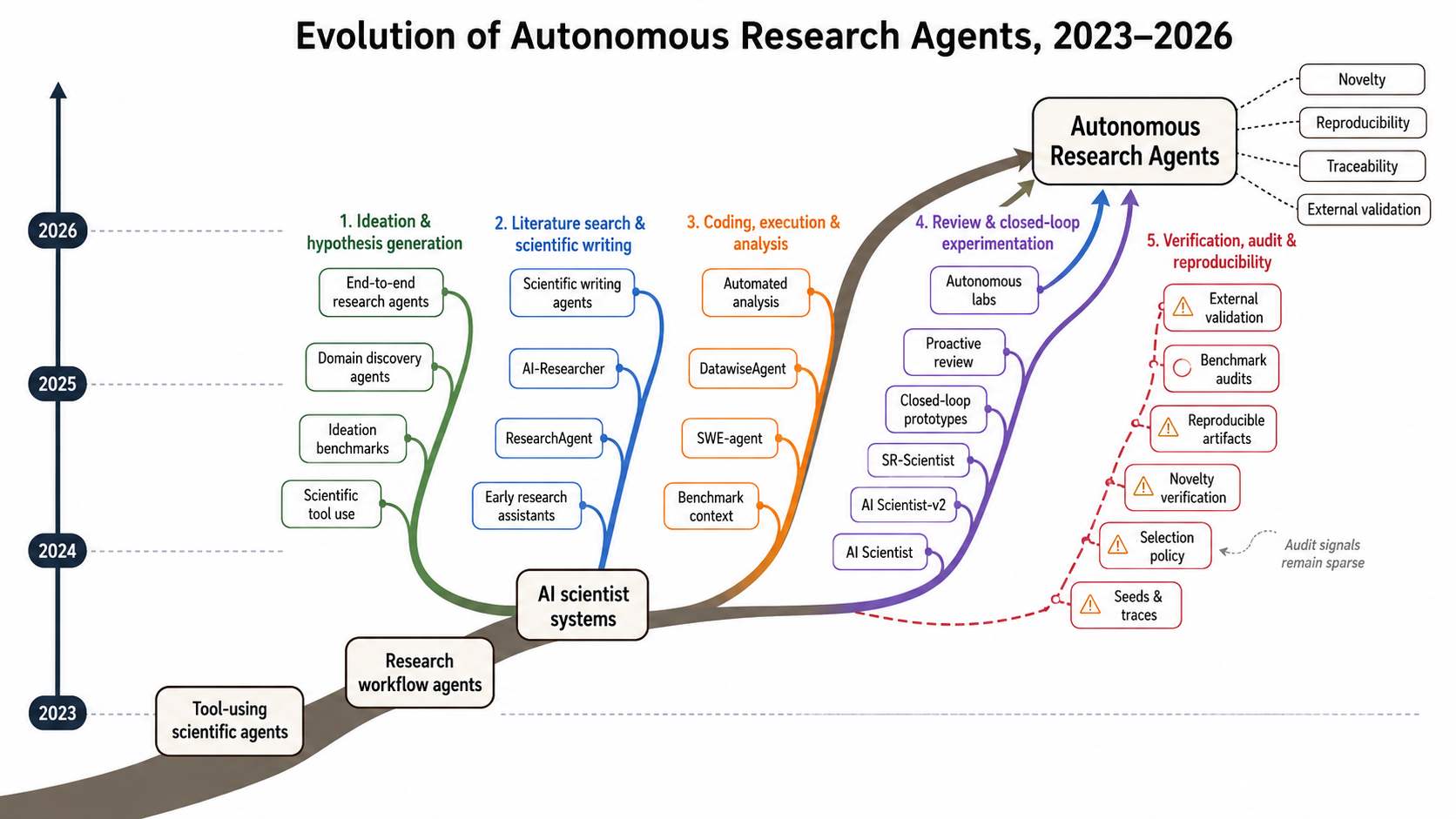}
\caption{Opening map of the survey's scope. The trunk sketches the shift from tool-using scientific
agents to LLM-era AI scientist systems, while the colored branches group the lifecycle capabilities
reviewed in later sections. The dashed red branch is intentionally delayed: in the coded corpus, audit
signals for seeds and traces, selection policy, novelty checks, reproducible artifacts, benchmark
audits, and external validation remain thinner than evidence of task completion. The figure is a guide
to the argument, not an exhaustive chronology.}
\label{fig:roadmap}
\end{figure}

\paragraph{Why another survey.}
The area already has strong surveys, and we do not claim to be first. \citet{zheng2025automation}
organize the field by \emph{autonomy level}, through a Tool / Analyst / Scientist taxonomy viewed via the
scientific method. \citet{wei2025agenticscience} organize it by \emph{scientific domain}, surveying
autonomous discovery across the life sciences, chemistry, materials, and physics and unifying process,
autonomy, and mechanism perspectives. The closest topical overlap is \citet{tie2026survey}, who directly
survey ``AI scientists'' along a capability/workflow axis; recent deep-research surveys such as
\citet{xu2025a} catalogue retrieval-heavy research systems and their applications; and a separate line
surveys \emph{scientific domain models} such as molecular and protein LLMs~\citep{zhang2024scientificllm},
which are models, not agents, and we treat as context. Relative to all of these, our unit of analysis is the \emph{audit artifact} a reviewer
can inspect (released code, seeds, traces, selection policy, novelty-check method, human-intervention
points), not the system's domain or its autonomy label. Using audit artifacts as the unit of analysis lets
us report quantitative disclosure rates and convert the survey into a reviewer checklist, rather than another
descriptive taxonomy. Our scope is the AI/ML-research setting where those artifacts are inspectable.
Table~\ref{tab:antidup} summarizes the difference with graded, not binary, coverage.

\paragraph{Contributions.}
\begin{itemize}
  \item A \textbf{coded corpus} of autonomous research systems (Table~\ref{tab:corpus}), annotated on seven
        audit dimensions via a disclosed search protocol (Sec.~\ref{sec:scope}); the corpus is the artifact
        every later claim rests on.
  \item A \textbf{lifecycle $\times$ autonomy map} (Table~\ref{tab:taxonomy}) indexed to the corpus, in which
        missing disclosures are coded explicitly rather than inferred.
  \item An \textbf{auditability gap analysis} (Table~\ref{tab:gap}) linking evaluation proxies to recurring
        failure modes and the audit artifact each one needs.
  \item A \textbf{reviewer-operational reporting checklist} (Table~\ref{tab:checklist}) tying each disclosure
        to the failure mode it addresses and the minimum acceptable evidence.
\end{itemize}

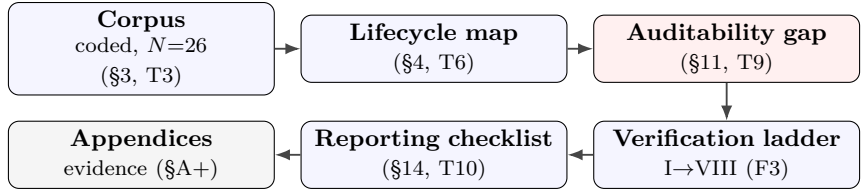
\begin{figure}[t]
\centering
\footnotesize
\begin{tikzpicture}[
    node distance=3.4mm,
    box/.style={draw, rounded corners, align=center, inner sep=3pt, text width=0.20\columnwidth,
                font=\footnotesize, minimum height=8.5mm, fill=blue!4},
    gap/.style={box, fill=red!6},
    arr/.style={-{Latex[length=2mm]}, line width=0.7pt, gray!55!black}]
  \node[box] (corpus) {\textbf{Corpus}\\\scriptsize coded, $N{=}26$\\\scriptsize (\S\ref{sec:scope}, T\ref{tab:corpus})};
  \node[box, right=of corpus] (map) {\textbf{Lifecycle map}\\\scriptsize (\S\ref{sec:map}, T\ref{tab:taxonomy})};
  \node[gap, right=of map] (gap) {\textbf{Auditability gap}\\\scriptsize (\S\ref{sec:gap}, T\ref{tab:gap})};
  \node[box, below=5mm of gap] (ladder) {\textbf{Verification ladder}\\\scriptsize I$\to$VIII (F\ref{fig:ladder})};
  \node[box, left=of ladder] (check) {\textbf{Reporting checklist}\\\scriptsize (\S\ref{sec:checklist}, T\ref{tab:checklist})};
  \node[box, left=of check, fill=gray!8] (appx) {\textbf{Appendices}\\\scriptsize evidence (\S\ref{app:method}+)};
  \draw[arr] (corpus) -- (map);
  \draw[arr] (map) -- (gap);
  \draw[arr] (gap) -- (ladder);
  \draw[arr] (ladder) -- (check);
  \draw[arr] (check) -- (appx);
\end{tikzpicture}
\caption{How to read this survey. A disclosed coded corpus feeds a lifecycle\,$\times$\,autonomy map,
which exposes the auditability gap. The verification ladder separates stronger from weaker audit signals,
and the reporting checklist turns the analysis into reviewer-facing requirements. Appendices hold the
supporting evidence.}
\label{fig:howtoread}
\end{figure}

\paragraph{How to read this survey.}
Figure~\ref{fig:roadmap} gives the historical orientation, and Figure~\ref{fig:howtoread} gives the
paper's internal map. We begin with a \emph{disclosed, full-text-coded corpus} of autonomous research
systems (Sec.~\ref{sec:scope}, Table~\ref{tab:corpus}; full table in App.~\ref{app:corpus}). We then
project those systems onto a \emph{lifecycle $\times$ autonomy map} (Sec.~\ref{sec:map},
Table~\ref{tab:taxonomy}) to show where automation is claimed and where disclosure is missing. The map
leads to the auditability gap analysis (Sec.~\ref{sec:gap}, Table~\ref{tab:gap}), which ties each
evaluation proxy to the failure mode it may miss and the evidence that would make the claim easier to
check. The verification ladder (Fig.~\ref{fig:ladder}) then separates domains with strong external
checks from domains where the model's own judgment often remains the main signal. The body of the paper
walks through four lifecycle clusters: ideation and hypothesis (Sec.~\ref{sec:ideation}); literature and
writing (Sec.~\ref{sec:litwriting}); coding, execution, and analysis (Sec.~\ref{sec:coding}); and review
and closed-loop research (Sec.~\ref{sec:review}). Later sections re-read the same systems by
verification signal (Sec.~\ref{sec:verification}) and measurement infrastructure
(Sec.~\ref{sec:benchmarks}), before ending with a reviewer-facing checklist
(Sec.~\ref{sec:checklist}, Table~\ref{tab:checklist}). Readers who want the shortest path can read
Sections~\ref{sec:scope}--\ref{sec:map}, the auditability gap analysis, and the checklist; the
appendices hold the extended corpus, stage-coverage tables, verification-signal taxonomy, and benchmark
catalogue.

\begin{table}[t]
\centering
\caption{Positioning against five close survey lines, graded by how much each foregrounds a property
(\emph{prim.}\ primary, \emph{part.}\ partial, \emph{n/o}\ not organizing). Zheng'25~\citep{zheng2025automation}
(autonomy axis), Wei'25~\citep{wei2025agenticscience} (domain axis), Tie'26~\citep{tie2026survey}
(capability/workflow axis), Xu'25~\citep{xu2025a} (deep-research systems), Zhang'24~\citep{zhang2024scientificllm}
(scientific domain \emph{models}). A
grade reflects a survey's \emph{primary organizing emphasis}, not whether a topic is mentioned in passing:
\emph{prim.}\ means the survey is structured around the property, \emph{part.}\ that it is discussed but
not organizing, and \emph{n/o}\ that it is not an organizing concern. The discriminating row is the last two: none
of these releases a per-system, full-text \emph{coding} of audit artifacts or a reviewer reporting
standard, which is the distinction this paper makes.}
\label{tab:antidup}
\scriptsize
\begin{tabular}{@{}p{0.28\columnwidth}cccccc@{}}
\toprule
Property & Zheng & Wei & Tie & Xu & Zhang & \textbf{Ours} \\
\midrule
Organizing axis & auton. & domain & workflow & deep res. & model & \textbf{audit} \\
Unit of analysis & role & domain & system & method & model & \textbf{artifact} \\
Corpus method disclosed & n/o & n/o & n/o & part. & part. & \textbf{prim.} \\
Artifacts coded per system & n/o & n/o & n/o & n/o & n/o & \textbf{prim.} \\
Evaluation foregrounded & part. & part. & part. & part. & part. & \textbf{prim.} \\
Failure-mode framework & part. & part. & part. & part. & n/o & \textbf{prim.} \\
Capability/verification split & n/o & part. & n/o & part. & n/o & \textbf{prim.} \\
Reporting standard & n/o & n/o & n/o & n/o & n/o & \textbf{prim.} \\
\bottomrule
\end{tabular}
\end{table}

\paragraph{Closest related surveys and verification-focused work.}
Recent work already argues that scientific agents need stronger verification, so our contribution is not
the claim that verification matters. \citet{gridach2025agentic} survey agentic AI for scientific
discovery across domains and emphasize evaluation, safety, and practical deployment challenges.
\citet{cornelio2025need} make the verification problem explicit for AI-driven discovery, and
\citet{bisht2026notbuilt} argue that current agentic AI scientists are not designed for autonomous
discovery. Benchmark-disclosure audits show that agent evaluations often omit harness, cost, and
reproducibility details~\citep{moghadasi2026audit}, while chain-of-evidence systems such as
ScientistOne~\citep{meng2026scientistone} move toward explicit evidentiary traces. Our narrower contribution
is to code public autonomous-research systems \emph{per system} for audit artifacts, map those codings onto
the research lifecycle, and convert the result into reviewer-facing reporting requirements. Prior work
surveys agents, domains, workflows, or the need for verification; this paper asks what a reviewer can
actually inspect for each system in the public record.

\section{Background: Verifiable Automation to Open-Ended Agents}
\label{sec:background}

Autonomous research did not begin with LLMs, and its predecessors make the verification contrast clear. We
sketch three lineages the modern systems inherit from, and one they break from.

\paragraph{Pre-LLM lineages with built-in verification.}
The ``Robot Scientist'' program closed the hypothesis--experiment--interpretation loop for yeast genomics
two decades ago~\citep{king2004robot}, and its successor Adam became the first machine to autonomously
discover novel scientific knowledge~\citep{king2009automation}, with Eve extending the approach to drug
repositioning~\citep{williams2015eve}. This lineage logged hypotheses and provenance in machine-readable
form, so each claim was auditable by construction. The AutoML tradition is similar: Auto-WEKA's combined
algorithm-selection-and-hyperparameter problem~\citep{thornton2013autoweka}, Auto-sklearn's
meta-learning~\citep{feurer2015autosklearn}, and the neural-architecture-search line from RL-based
search~\citep{zoph2017nas} through transferable spaces~\citep{zoph2018nasnet} to differentiable
search~\citep{liu2019darts} all operate over constrained search spaces with objective, automatically
checkable success criteria. In these lineages, the search objective and the checking mechanism were usually
coupled.

\paragraph{The agent substrate LLM research agents are built on.}
Modern research agents inherit a general toolkit: ReAct interleaves reasoning with
actions~\citep{yao2023react}, Toolformer learns tool invocation~\citep{schick2023toolformer},
Tree-of-Thoughts turns reasoning into search~\citep{yao2023tot}, Reflexion adds verbal
self-correction~\citep{shinn2023reflexion}, and Voyager pursues open-ended skill
acquisition~\citep{wang2023voyager}. This machinery improves exploration, but self-critique, search, and
memory do not by themselves guarantee that a claim, citation, or experiment has been independently checked.
The capability transferred; the verification did not.

\paragraph{Where verification is built in, claims become easier to audit.}
The clearest audited LLM-era successes pair generation with an automatic verifier. FunSearch gates LLM-proposed
programs behind an executable evaluator and yields real mathematical advances~\citep{romeraparedes2024funsearch};
AlphaTensor accepts only algebraically valid tensor decompositions~\citep{fawzi2022alphatensor}; AlphaEvolve
ranks evolved programs through task-specific tests~\citep{novikov2025alphaevolve}; and Eureka keeps
LLM-written reward code only when downstream training measurably improves~\citep{ma2024eureka}. The purest
case is \emph{formal theorem proving}, where a proof assistant (Lean, Coq, Isabelle) is a sound,
deterministic verifier: an agent's output is machine-checkable by construction, not judged by an LLM critic.
This enables verifier-in-the-loop training and search (the DeepSeek-Prover line~\citep{xin2024deepseekp,xin2024deepseekpx,ren2025deepseekp}, Kimina-Prover~\citep{wang2025kiminapro}, Goedel-Prover~\citep{lin2025goedelpro}),
retrieval-augmented Lean environments~\citep{yang2023leandojo}, lifelong proof agents~\citep{kumarappan2024leanagent},
self-play conjecture-and-prove loops that turn the verifier into an open-ended discovery
engine~\citep{dong2025stp}, and neuro-symbolic provers reaching medalist-level olympiad
performance~\citep{chervonyi2025goldmedal}. These are not counterexamples to the verification framing but
evidence for it: discovery is trustworthy precisely when a verifier, not the model's own judgment, decides what
counts. The rest of the survey asks what plays the verifier's role when the domain admits no such checker.

\paragraph{Self-driving labs: verification by physical contact.}
Self-driving laboratories couple LLM reasoning to robotic execution, so claims must survive the physical
world. A mobile robotic chemist searched a ten-dimensional photocatalyst space
autonomously~\citep{burger2020mobile}; Coscientist designs and runs reactions via tool
use~\citep{boiko2023coscientist}; ChemCrow augments an LLM with expert tools, and its authors report that
GPT-4 acting as its own evaluator could not reliably separate correct from flawed
outcomes~\citep{mbran2023chemcrow}; A-Lab synthesized dozens of inorganic compounds, though later scrutiny of
its phase-identification claims showed how ``success'' hinges on trustworthy
characterization~\citep{szymanski2023autonomous}. Across these systems capability is abundant while
independent verification is what is scarce, the pattern the rest of this survey quantifies for the
computational setting.

\paragraph{Domain breadth.}
Outside the computational setting, the same systems recur in three mechanism families, distinguished by
\emph{what supplies the verification signal}. (i) \emph{Tool-orchestration and inverse-design} agents
compose domain tools and databases but, like ChemCrow, lean on the model's own judgment to accept an
output; the chemistry/materials line is representative (e.g.\ LLaMP~\citep{chiang2024llamp},
ChemReasoner~\citep{sprueill2024chemreas}, AtomAgents~\citep{ghafarollahi2024atomagen}, with related systems
spanning inverse design and synthesis~\citep{ansari2024dziner,jia2024llmatdes,kang2023chatmof,zhang2024honeycom,chen2023chemistx,tang2025chemagen,zou2025el}).
(ii) \emph{Physically validated} agents close the loop against a wet-lab oracle, which is why systems with
a wet-lab or clinical external oracle admit a stronger verification signal than software-only loops:
experiment-design and closed-loop campaigns have nominated real candidates
(CRISPR-GPT~\citep{qu2024crisprgp}, Robin~\citep{ghareeb2025robin}), alongside therapeutic-reasoning and
modality-specialist agents~\citep{gao2025txagent,roohani2024biodisco,xiao2024cellagen,wang2024geneagen,ghafarollahi2024protagen,jin2025stella}
and robotic characterization~\citep{darvish2024organa}. (iii) \emph{Report-generating} data-science agents
turn data into analyses but mostly self-grade (Data Interpreter~\citep{hong2024data} and
peers~\citep{guo2024dsagent,li2024autokagg,trirat2024automlag}), with a few targeting human-verifiable
output~\citep{ifargan2024autonomo}. The verification signal thus strengthens from (i) to (ii); the AI/ML
agents we center sit nearer (i), which is why their closed loops are mechanical rather than validated.

\section{Scope and Corpus Construction}
\label{sec:scope}

\paragraph{What counts as an autonomous research agent.}
We define an \emph{autonomous research agent} as an LLM-driven system that executes at least one stage of
the research lifecycle with non-trivial decision autonomy, meaning the system, not a human, selects the
next action within that stage. This admits stage-local agents (such as an ideation agent) and full
pipelines (idea to manuscript), and excludes single-shot prompting where a human drives every step.
When the distinction matters, we use \emph{LLM autonomous research agent} for this focal category and
\emph{autonomous research system} for the broader coded set that also includes pre-LLM or
domain-adjacent contrast cases.

\paragraph{Boundaries.}
\emph{In scope}: LLM-agent systems that autonomously execute one or more research-lifecycle stages, with
emphasis on computational and AI/ML research, where the artifacts of the claim (code, data, runs,
write-ups) are inspectable. The full-text-coded corpus is centered on this computational setting but
deliberately includes five domain-adjacent contrast systems inside the 24 runnable-system denominator (e.g.\
CAMEO~\citep{kusne2020cameo}, ChemCrow~\citep{mbran2023chemcrow}, ChatBattery, MASTER, ARIA): they are
domain-adjacent contrast systems whose loops or terminal checks \emph{could}
close against a physical oracle, which is what makes the contrast observable between CAMEO's in-loop
validated measurement and the others' mechanical or post-hoc checks (Sec.~\ref{sec:map}); dropping them
would hide rather than sharpen the gap. All disclosure rates are
computed over the full set of 24 runnable systems. \emph{Cited
as context, not coded}: pre-LLM AutoML and neural-architecture-search predecessors, scientific domain
models~\citep{zhang2024scientificllm}, and domain tool-use papers that do not expose a research lifecycle
for artifact coding. \emph{Out of scope}: pure domain-model architecture surveys and non-agentic single-shot tools. We center the AI/ML setting deliberately: it is
where autonomous-science claims are \emph{most observable}, because a reviewer can in principle re-run the
code, not the final boundary of the field's relevance.

\paragraph{Search protocol.}
We disclose corpus construction so the scope of the survey is clear. We queried
arXiv (via its API), Semantic Scholar, and OpenReview/ACL Anthology for LLM-agent work from 2023 to
June 2026, using twelve stage-spanning query families: \emph{AI scientist / autonomous research
agent}, \emph{automated scientific discovery}, and per-stage terms (\emph{idea generation}, \emph{automated
peer review}, \emph{coding agent}, \emph{data analysis}, \emph{closed-loop discovery}, \emph{paper
writing}). We then hand-added a small number of pre-LLM or domain-adjacent contrast systems, including
CAMEO, to anchor the physical-oracle comparison. The queries returned 144 records; after de-duplication 125 remained, and a two-pass screen
(title and abstract, then full text) against the inclusion definition admitted \textbf{35} works. The 90
exclusions were dominated by keyword collisions (autonomous \emph{driving}), generic ML/RL, physics and
astronomy, non-agentic single-shot tools, and pre-LLM methods retained only as baselines. From the 35 we
full-text code \textbf{26 entries} (24 runnable systems plus two study/position works) on the seven
dimensions below; the remaining nine are cited as context. Fields not resolvable from the public record are
marked ``n/d''.

\begin{table}[H]
\centering
\caption{Corpus slices and the claims they support. Disclosure percentages in the paper use the 24
runnable-system denominator unless explicitly labeled otherwise.}
\label{tab:denominators}
\small
\begin{tabular}{@{}p{0.46\columnwidth}cp{0.36\columnwidth}@{}}
\toprule
Slice & Count & Used for \\
\midrule
Records after de-duplication & 125 & search scope \\
Included works after full-text screen & 35 & narrative coverage \\
Full-text-coded entries & 26 & corpus tables and maps \\
Runnable systems & 24 & headline disclosure rates \\
LLM-era runnable systems ($\geq$2023) & 22 & LLM-era sensitivity and the validated-loop claim \\
AI/ML-only runnable systems & 19 & focal-slice sensitivity excluding contrast rows \\
Domain-adjacent contrast rows & 5 & included inside the 24 runnable systems \\
\bottomrule
\end{tabular}
\end{table}

\paragraph{Codebook (per dimension).}
\emph{Stage}: lifecycle stages the system executes autonomously. \emph{Autonomy}: highest level reached
(Sec.~\ref{sec:map}), with ``(c)'' for author-claimed levels lacking an external check. \emph{Eval}: how
outputs are judged (automated LLM reviewer, human expert, task success, benchmark, workshop). \emph{Artifacts}:
released for re-execution (code, prompts, seeds, traces). \emph{HITL}: human entry points (seed/topic,
template, baseline, per-step approval). \emph{Novelty}: method used to check novelty of outputs (automated
literature check, against a literature graph, human, or none). \emph{Selection}: whether the number of
attempts and the result-selection policy are disclosed.

\paragraph{Coding reliability.}
Codings were extracted from each paper's \emph{full text} (methods, experiments, appendix, and released
code where available), not its abstract, by a single coder against the codebook above. Full-text coding is
necessary because abstracts rarely mention released code, prompts, seeds, traces, or selection policies even
when these artifacts appear in appendices or repositories. To gauge labeling reliability, an
independent second coder re-coded a random sample of ten systems blind on the four most subjective
dimensions: agreement was 90\% for released artifacts but only 50\% for autonomy level, 60\% for novelty
method, and 60\% for selection disclosure (65\% overall). This second pass was made from abstracts only,
so it measures how repeatably the dimensions can be labeled, not whether the primary full-text reading is
correct; a full-text second pass is left for future work. We therefore treat autonomy levels as coarse
labels used for structure, not fine per-cell claims, and we report the artifact-backed-versus-claimed
distinction (below) rather than a single autonomy number. ``n/d'' marks a disclosure absent from the
consulted text.

\paragraph{What the corpus shows.}
Two patterns emerge, and together they relocate the verification gap. First, on the lifecycle, coverage
clusters on execution, analysis, and experiment design and thins at literature review and closed-loop
iteration. Second, and more tellingly, \emph{code release is now common but reproducibility-grade
disclosure is not}: among the 24 runnable systems, 83\% release code and 88\% disclose at least one
human-in-the-loop entry point, yet only 38\% release the seeds or execution traces that would let a
reviewer reproduce a run, though 67\% do disclose how a result was selected (Table~\ref{tab:stats}). A
field that publishes its code is not necessarily one whose results can be re-derived. The sensitivity
slices in Table~\ref{tab:sensitivity} preserve this result, so it is not an artifact of the
pre-LLM contrast row or the materials/wet-lab-adjacent cases. The gap is narrower and sharper than
``nobody shares anything.'' Only 38\% of systems report
\emph{any} novelty-verification method, and the harder case is closed-loop validity. We code a closed
loop as \emph{externally validated} only when an outside oracle (a physical measurement or an independent
checker), not the system's own score, decides whether a result revises the next hypothesis; otherwise the
loop is \emph{mechanical}. Of the nine systems that reach L4, seven are artifact-backed but mechanical
(released code shows results feeding back, but the trigger is a metric or constant-fit, not demonstrated
revision of a scientific hypothesis), one is author-claimed without any external check
(\citealp{clawailab2026}), and exactly one is externally validated: a Bayesian-active-learning materials
platform~\citep{kusne2020cameo} that predates LLM agents and is anchored by physical measurement. The
boundary is deliberate and worth stating, because two LLM-era systems look like near-misses. MASTER, also
L4 in materials, optimizes a computational task-success metric and adds a human-expert read \emph{after}
the run; the physical world never gates its next hypothesis, so by our rule it is mechanical, not
validated. ChatBattery does close against a wet-lab synthesis, but the experiment confirms a \emph{final}
nominated candidate rather than driving each iteration, and we code its overall autonomy as L3; the wet-lab
is a terminal check, not an in-loop oracle. What distinguishes CAMEO is that the physical measurement is
the in-loop signal that selects the next experiment. The constraint in this sample is not code
transparency alone but verification of \emph{claims}: whether a proposed idea is genuinely novel, whether a baseline is
adequate, and whether a closed loop's proxy actually tracks scientific validity. That is what current
evaluation does not check.

\begin{table*}[t]
\centering
\caption{Representative rows from the full-text-coded corpus ($N=26$; full table in Appendix~\ref{app:corpus}). ``n/d'' =
not disclosed. ``(c)'' = autonomy level author-claimed without an external check; unmarked L4 = released
code shows result-driven feedback, though the loop is mechanical unless noted. Autonomy levels per
Sec.~\ref{sec:map}.}
\label{tab:corpus}
\footnotesize
\setlength{\tabcolsep}{4pt}
\newcommand{\syscell}[2]{\parbox[t]{\linewidth}{\raggedright #1\\[-1pt]{\scriptsize\citep{#2}}}}
\begin{tabular}{@{}>{\raggedright\arraybackslash}p{0.18\textwidth}>{\raggedright\arraybackslash}p{0.20\textwidth}c>{\raggedright\arraybackslash}p{0.12\textwidth}>{\raggedright\arraybackslash}p{0.155\textwidth}>{\raggedright\arraybackslash}p{0.105\textwidth}c@{}}
\toprule
System & Stage(s) & Auton. & Eval & Artifacts & HITL & Sel. \\
\midrule
\syscell{AI Scientist}{lu2024aiscientist}          & ideation--exec--analysis--writing--review & L4 & auto LLM rev. & code, prompts, seeds, traces & topic, template & yes \\
\syscell{AI Scientist-v2}{yamada2025aiscientistv2} & ideation--exec--writing & L4 & LLM rev.; workshop & code, prompts, traces & reduced template & yes \\
\syscell{Jr.\ AI Scientist}{miyai2025jrai}         & analysis--writing      & L3 & AI rev.+authors & code & baseline paper & yes \\
\syscell{LLM-AutoSciLab}{llmautoscilab2026}        & hypothesis--exec--closed-loop & L4 & benchmark & code, prompts & none & yes \\
\syscell{SR-Scientist}{srscientist2025}            & hypothesis--exec--closed-loop & L4 & benchmark & code, prompts, traces & topic, baseline & yes \\
\syscell{CAMEO}{kusne2020cameo}                    & exp-design--exec--closed-loop & L4 & real deploy & none & topic, per-step & yes \\
\syscell{ResearchAgent}{baek2024researchagent}     & ideation--exp.\ design  & L3 & review agents & code, prompts, seeds & topic & n/d \\
\syscell{ChemCrow}{mbran2023chemcrow}              & exp-design--exec & L3 & LLM+expert & code, prompts, traces & topic, per-step & yes \\
\syscell{DatawiseAgent}{datawise2025}              & coding--exec--analysis & L3 & benchmark & prompts & topic & no \\
\syscell{Proactive Reviewer}{proactivereview2026}  & review & L2 & bench + human & code, prompts, traces & none & yes \\
\bottomrule
\end{tabular}
\let\syscell\relax
\end{table*}

\begin{table}[t]
\centering
\caption{Auditability disclosure rates from \emph{full-text} coding of the corpus (24 runnable systems;
two further entries are a study and a position work).
The human-in-the-loop row counts systems with at least one disclosed human entry point; the three with
none (ReviewAdvisor, LLM-AutoSciLab, Proactive Reviewer) run end-to-end once invoked. Artifact
transparency is common; the gap is in novelty verification and in whether closed loops are validated
rather than mechanical. The bolded rows isolate the constraints that bind.}
\label{tab:stats}
\small
\begin{tabular}{@{}p{0.62\columnwidth}c@{}}
\toprule
Disclosure (systems, $N=24$) & Rate \\
\midrule
Releases code & 83\% (20/24) \\
Releases prompts & 71\% (17/24) \\
Releases seeds or execution traces & \textbf{38\%} (9/24) \\
Discloses result-selection policy & 67\% (16/24) \\
Has $\geq$1 human-in-the-loop entry point & 88\% (21/24) \\
Reports any novelty-verification method & \textbf{38\%} (9/24) \\
\midrule
\multicolumn{2}{@{}p{0.92\columnwidth}}{L4 closed-loop: 9 systems = 7 artifact-backed but
\emph{mechanical} (metric-triggered re-run / constant fit) + 1 author-claimed without external check + 1
externally validated with an outside oracle (\citealp{kusne2020cameo}, pre-LLM, physical measurement).} \\
\bottomrule
\end{tabular}
\end{table}

\begin{table}[t]
\centering
\caption{Sensitivity of headline disclosure rates to corpus slice. The AI/ML-only slice excludes the five
domain-adjacent contrast rows (CAMEO, ChemCrow, ChatBattery, MASTER, ARIA). Artifact availability was coded
from public materials; we did not independently test whether released repositories rerun.}
\label{tab:sensitivity}
\scriptsize
\setlength{\tabcolsep}{3pt}
\begin{tabular}{@{}p{0.30\columnwidth}cccc@{}}
\toprule
Disclosure & All runnable & LLM-era & AI/ML-only & Contrast \\
 & $N{=}24$ & $N{=}22$ & $N{=}19$ & $N{=}5$ \\
\midrule
Code released & 20/24 (83\%) & 19/22 (86\%) & 17/19 (89\%) & 3/5 (60\%) \\
Prompts released & 17/24 (71\%) & 17/22 (77\%) & 13/19 (68\%) & 4/5 (80\%) \\
Seeds/traces released & 9/24 (38\%) & 9/22 (41\%) & 8/19 (42\%) & 1/5 (20\%) \\
Selection disclosed & 16/24 (67\%) & 15/22 (68\%) & 12/19 (63\%) & 4/5 (80\%) \\
Novelty check reported & 9/24 (38\%) & 9/22 (41\%) & 7/19 (37\%) & 2/5 (40\%) \\
HITL points disclosed & 21/24 (88\%) & 20/22 (91\%) & 16/19 (84\%) & 5/5 (100\%) \\
\bottomrule
\end{tabular}
\end{table}

\section[Lifecycle x Autonomy Map]{A Lifecycle \texorpdfstring{$\times$}{x} Autonomy Map}
\label{sec:map}

The taxonomy is infrastructure for the survey's claims, not its headline. It indexes the corpus so later
sections can talk precisely about \emph{where} autonomy is evidenced and where it is only claimed.

\paragraph{Operational autonomy levels.}
Rather than label systems with adjectives, we code each on five axes: \emph{initiative} (who selects the
problem or next action), \emph{judgment} (who decides novelty, validity, and sufficiency of evidence),
\emph{execution} (whether the system runs tools and experiments without per-step approval),
\emph{iteration} (whether it revises hypotheses in response to results), and \emph{accountability}
(whether traces, seeds, prompts, and human interventions are disclosed). These induce levels: \textbf{L0}
assistive; \textbf{L1} tool-augmented (model can search and code, human drives); \textbf{L2} stage-local
(autonomously completes one bounded stage); \textbf{L3} pipeline (chains multiple stages, idea to report);
\textbf{L4} closed-loop, a system whose results are fed back to condition the next step. We split L4 by
\emph{what} the feedback achieves: \textbf{L4-m} (mechanical) re-runs or re-fits on an internal metric or
constant, while \textbf{L4-v} (validated) revises a scientific hypothesis against an outside oracle (a
physical measurement or independent checker). Finally \textbf{L5} open-ended, treated as aspirational and
excluded from the populated map. Separating initiative and execution from judgment is the
point: a system can automate research \emph{tasks} (L2--L3) without automating scientific \emph{judgment},
and that difference is exactly what verification must check.

\begin{table}[t]
\centering
\caption{Lifecycle $\times$ autonomy map indexed to the corpus. A system is assigned its \emph{highest} attained autonomy level (its
column); a cell then counts systems in that column that automate the given stage, so a single system
contributes to several stage rows of its column and columns do not sum to the system count. ``$+\,n$c''
counts $n$ \emph{author-claimed} L4(c) systems alongside the artifact-backed L4 systems in that cell. The
nine systems reaching L4 break down as \textbf{7 artifact-backed mechanical} (metric or constant re-run)
+ \textbf{1 artifact-backed externally validated} (CAMEO~\citep{kusne2020cameo}, pre-LLM)
+ \textbf{1 author-claimed} L4(c) without an external check. A blank means no qualifying system; L1 is
background and omitted.}
\label{tab:taxonomy}
\footnotesize
\setlength{\tabcolsep}{4pt}
\begin{tabular}{@{}p{0.30\columnwidth}cccc@{}}
\toprule
Lifecycle stage & L2 & L3 & L4 \\
\midrule
Ideation          &    & 6 & 3+1c \\
Literature review & 2  & 8 & 3 \\
Hypothesis gen.   & 1  & 7 & 7 \\
Experiment design &    & 7 & 7+1c \\
Coding            & 1  & 7 & 6+1c \\
Execution         & 1  & 7 & 8+1c \\
Analysis          & 1  & 8 & 8+1c \\
Writing           &    & 7 & 3+1c \\
Reviewing         & 2  & 5 & 4+1c \\
Closed-loop iter. &    & 2 & 8+1c \\
\bottomrule
\end{tabular}
\end{table}

\FloatBarrier
\begin{table}[t]
\centering
\caption{Compact classification of the nine systems coded L4. The labels classify the \emph{in-loop}
signal that drives the next step, not the quality of the paper or system as a whole.}
\label{tab:l4main}
\footnotesize
\setlength{\tabcolsep}{4pt}
\begin{tabular}{@{}p{0.18\columnwidth}p{0.34\columnwidth}p{0.38\columnwidth}@{}}
\toprule
Label & Systems & Signal driving the loop \\
\midrule
Externally validated & CAMEO~\citep{kusne2020cameo} (pre-LLM contrast) &
Physical measurement selects the next experiment. \\
Mechanical & AI Scientist, AI Scientist-v2, SR-Scientist, MASTER, LLM-AutoSciLab, E2E AI Research,
LLM-ACES~\citep{lu2024aiscientist,yamada2025aiscientistv2,srscientist2025,materials2025master,llmautoscilab2026,e2eairesearch2026,llmaces2026} &
Internal metric, benchmark score, task-success score, or automated reviewer triggers the next step. \\
Author-claimed & Claw AI Lab~\citep{clawailab2026} &
Closed loop is asserted, but the released material does not evidence the feedback mechanism. \\
\bottomrule
\end{tabular}
\end{table}

The next four sections read the corpus by lifecycle stage. For each cluster, we ask what the systems
automate, what currently checks the output, and which claims still depend on trust in the authors or the
agent. A consistent pattern appears: agents can often generate a research artifact before the field has a
reliable way to judge it. Sections~\ref{sec:ideation}--\ref{sec:review} cover ideation, literature and
writing, coding and analysis, and review or closed-loop research. The following three sections cut across
those stages: Section~\ref{sec:verification} re-reads the same systems by the strength of their
verification signal, Section~\ref{sec:benchmarks} maps the measurement and resource infrastructure, and
Section~\ref{sec:frontiers} collects the forward-looking frontiers.

\paragraph{How the verification signal moves along the lifecycle.}
Read qualitatively, the four clusters form a U-shaped pattern in audit strength. Ideation and hypothesis
generation make claims about what is not yet known, so the available check is usually a novelty judgment:
the weakest signal on the ladder and the one least tied to eventual validity. Coding and execution recover
a stronger signal because the ground truth can be an executable artifact or a held-out number; this is the
stage where the corpus shows the most credible mechanically graded results. The signal weakens again in
review and closed-loop iteration, where an LLM often judges another LLM and only one of the nine L4 systems
reaches an external oracle. Literature and writing sit beside this arc with a different check, citation
grounding, which is tractable in principle but still short of reliable in practice. The stages with the
most autonomous activity are therefore not always the stages with the soundest checks. Execution is well
instrumented, but the judgment-heavy endpoints, deciding what is worth studying and whether a result is
real, are where the independent signal is thinnest.

\FloatBarrier
\input{sec_ideation}

\input{sec_litwriting}

\input{sec_coding}

\input{sec_review}

\input{sec_verification}

\input{sec_benchmarks}

\section{The Auditability Gap Analysis}
\label{sec:gap}

The auditability gap links two questions prior surveys often treat separately: what current evaluation
measures, and how these systems fail. A failure mode matters in proportion to how hard it is for a
reviewer to detect.

\paragraph{What evaluation measures versus what trust requires.}
Most reported evaluation reduces to task success or a reviewer score. Trustworthy autonomous research
instead requires evidence along dimensions that are rarely measured: scientific validity, novelty,
reproducibility, experimental rigor, epistemic calibration, safety, and the true cost in compute and hidden
human labor. The gap between these lists is the survey's core observation: \emph{reported evaluations mostly measure task
completion and infer scientific value}. Novelty is the sharpest case, analyzed in detail in
Section~\ref{sec:ideation}: while 38\% of systems report \emph{some} novelty-checking step, we found no
system that reports \emph{independent validation} that its novelty check is itself reliable.

\begin{table}[t]
\centering
\caption{Auditability gap analysis: each recurring failure mode, the stage it threatens, the evaluation
proxy that hides it today, and minimum evidence that would begin to expose it. ``Documented'' marks modes with
reported evidence in our corpus; ``risk'' marks plausible but less-documented modes.}
\label{tab:gap}
\footnotesize
\begin{tabular}{@{}p{0.21\columnwidth}p{0.15\columnwidth}p{0.20\columnwidth}p{0.21\columnwidth}@{}}
\toprule
Failure mode & Stage & Hiding proxy & Minimum evidence \\
\midrule
Hallucinated cite (doc.) & lit./writing & fluent prose & resolvable bib. \\
Novelty overclaim (doc.) & ideation & idea-rating & novelty method \\
Weak baseline (doc.) & exp.\ design & headline metric & baseline source \\
Unreprod.\ run (doc.) & exec./anal. & single score & seeds, traces \\
Result selection (risk) & analysis & best-of-$n$ & attempts, policy \\
Hidden labor (risk) & all stages & ``autonomous'' & per-stage HITL \\
Dual-use (risk) & design/exec. & task focus & safety review \\
\bottomrule
\end{tabular}
\end{table}

\paragraph{Reading the gap table.}
Table~\ref{tab:gap} turns a list of complaints into a specification. Each row says: here is a way an
autonomous-research claim can be wrong, here is why current evaluation does not catch it, and here is a
minimum artifact or disclosure that would make the failure mode auditable. Position work arguing that current agents are
not built for autonomous discovery~\citep{bisht2026notbuilt} and risk reports accompanying released
systems~\citep{miyai2025jrai} supply much of the documented evidence; the rows marked ``risk'' are
plausible modes we flag for systematic measurement rather than assert as established.

\section{Safety, Integrity, and Governance as Auditability Failures}

Autonomous research agents raise risks that scale with their autonomy. Here we treat them as auditability
failures rather than as a complete safety taxonomy. Each maps onto a verification failure, which is why the
reporting checklist that follows is also the survey's safety instrument. \emph{Dual-use and biosecurity} risk arises when an
agent's capability is not gated by red-team or task disclosure; it is now benchmarked directly
(WMDP~\citep{li2024wmdp}, SciSafeEval~\citep{li2024scisafee}, the agentic bio-capabilities benchmark
ABC-Bench~\citep{liu2026abcbench}, and CBRN-risk quantification~\citep{kumar2025quantify}), with broader
agent-safety~\citep{zhang2024agentsaf} and controllable-risk frameworks for scientific
agents~\citep{he2023control}. \emph{Research integrity} fails when a claim's soundness is not independently
checked; new benchmarks test whether an agent can tell sound from unsound research
(SoundnessBench~\citep{ho2026soundnes}) and uphold academic integrity (SciIntegrity-Bench~\citep{yang2026sciinteg}).
\emph{Review-channel attacks} exploit the absence of reviewer-independence and prompt-injection screening:
LLM-written reviews are now detectable~\citep{demetrio2025genrevie,rao2025detectin,yu2025is}, manuscripts
carry hidden prompts that hijack AI review~\citep{lin2025hidden,collu2025misleadi}, and tool-augmented
detectors are emerging~\citep{duarte2026semdetec,duan2026taddle}. In each case the harm becomes possible
exactly where an autonomous claim or review cannot be independently audited.

\input{sec_frontiers}

\section{A Reporting Checklist for Autonomous Research Agents}
\label{sec:checklist}

The checklist (Table~\ref{tab:checklist}) turns the analysis into a compact, reviewer-operational
standard: every disclosure is tied to a failure mode and a minimum acceptable evidence level, so a
reviewer can apply it to any autonomous-research-agent paper. We separate the five disclosures
we actually coded (with the corpus rate that shows how often the field meets each) from three we
\emph{propose} but did not code, to keep the measured findings distinct from the recommended standard.

\begin{table}[H]
\centering
\caption{Reviewer-operational reporting checklist, split into the disclosures we \emph{measured} in the
corpus (top, with the share of our 24 systems that already meet each) and the disclosures we \emph{propose}
but did not code as one of the
seven audit dimensions (bottom). Each item targets a failure mode from Table~\ref{tab:gap}.}
\label{tab:checklist}
\footnotesize
\begin{tabular}{@{}p{0.34\columnwidth}p{0.15\columnwidth}p{0.26\columnwidth}c@{}}
\toprule
Disclosure item & Targets & Minimum evidence & Corpus \\
\midrule
\multicolumn{4}{@{}l}{\emph{Measured disclosures (coded; share of 24 systems meeting it)}} \\
Human-in-the-loop entry points stated & hidden labor & list of HITL points & 88\% \\
Code released & unreprod.\ run & runnable repository & 83\% \\
Seeds or execution traces released & unreprod.\ run & re-runnable artifact & 38\% \\
Novelty-verification method & novelty overclaim & method + sampled check & 38\% \\
Attempts \& selection policy & result selection & $n$ + selection rule & 67\% \\
\midrule
\multicolumn{4}{@{}l}{\emph{Proposed disclosures (not yet coded; no corpus rate)}} \\
Baseline provenance \& strength & weak baseline & source + tuning budget & --- \\
Reviewer independence & circular eval & evaluator $\neq$ generator & --- \\
Hypotheses preregistered & problem bias & timestamped prereg.\ & --- \\
\bottomrule
\end{tabular}
\end{table}

\paragraph{Applying the checklist.}
Use the checklist as an evidence standard, not a label set. A repository link is insufficient unless it
includes the environment, scripts, and data needed to rerun the claimed result. A trace means a seed plus
the prompt, tool-call, model, and output record needed to reconstruct the run, not a screenshot or selected
transcript. A novelty check should name the search space and include sampled independent verification; an
LLM self-check alone is a weak signal. Baseline provenance should identify the source implementation,
tuning budget, and any changes made by the agent or authors. Reviewer independence means that the system or
model producing a claim is not also the only mechanism accepting it.

\section{Threats to Validity}

Three limitations bound our claims. First, \emph{coding subjectivity and the reliability check's scope}:
an independent second coder re-coded a random sample of ten systems on the four most interpretive
dimensions, finding high agreement on artifact release (90\%) but only 50--65\% on autonomy level,
novelty method, and selection disclosure. That check was performed from abstracts only, so it bounds the
\emph{labeling} subjectivity of those dimensions but does not independently validate the full-text reading
the headline rates rest on; a full-text second pass is the obvious next step. The one dimension with both
high agreement and major interpretive weight is \emph{artifact release} (90\% agreement), which anchors the
``code is common'' half of our finding. The more interpretive rates, novelty method (60\% agreement) and
selection disclosure (60\%), should be read as \emph{directional}: what is robust is the direction, that
novelty verification and reproducibility-grade disclosure are reported far less often than code, and that
no system documents an independently validated novelty check. We report rates from the coded corpus and
treat autonomy levels as coarse structure,
reporting the mechanical-versus-validated split rather than leaning on a single precise L4 count. Second,
\emph{snapshot recency}: the corpus is a reproducible snapshot to mid-2026 of a field that is moving
monthly, so the absolute rates will date quickly even though the qualitative gap (transparency common,
verification rare) is unlikely to invert in the near term. Third, \emph{scope}: we deliberately center the
inspectable AI/ML-research setting and treat wet-lab and robotic platforms as context; a system such as
CAMEO appears in the corpus as the one externally validated closed loop precisely because its physical
oracle is what our computational systems lack, and we flag rather than hide that boundary. Finally,
\emph{evidence maturity}: many 2025--2026 systems are arXiv preprints or project reports. We code them as
public artifacts, not as settled peer-reviewed evidence.

\section{Open Problems and Conclusion}
\label{sec:conclusion}

In the focal autonomous-research corpus, code release is more common than claim-verification evidence. The
lifecycle $\times$ autonomy map shows competent stage-local and pipeline systems and an L4 column populated
almost entirely by mechanical loops. Among the LLM-era systems, none shows an externally validated in-loop
oracle under our coding rule; the one validated case, CAMEO, predates LLM agents and is included as a
contrast benchmark. The auditability analysis then explains why task completion is not enough: validity,
novelty, reproducibility, and selection bias require different evidence than a benchmark score supplies.
The verification-signal taxonomy (Table~\ref{tab:taxonomy_map}) frames the agenda as movement from the
model's own judgment toward external checks.

Six open problems follow. \emph{Proxy-to-truth validation for closed loops}: most L4 systems optimize an
internal metric, so the open question is when that proxy tracks scientific validity and how to detect when
it does not. \emph{Closed-loop verifiers}: evaluators that test whether results actually revise hypotheses,
not merely whether a pipeline re-ran. \emph{Novelty auditing at literature scale}: methods that verify a
claim against the literature at the rate agents generate ideas, without treating LLM self-judgment as
sufficient. \emph{Independent agent review}: reviewers that are separate from the generator and robust to
prompt injection and self-preference. \emph{Contamination-resistant evaluation}: benchmarks whose validity
survives train--test overlap and search-time leakage. \emph{Preregistration for AI-generated hypotheses}: a
shared record that curbs result selection and metric-driven problem selection~\citep{bisht2026notbuilt}.
Until these are routine, an ``AI scientist'' that writes a manuscript should be read as automating the
\emph{production} of research artifacts, not yet providing the independent verification that would make its
scientific claims trustworthy. The reporting checklist (Table~\ref{tab:checklist}) is a first step toward
making that distinction auditable.

\appendix

\section{Corpus Construction (Extended)}
\label{app:method}
We give the method before the evidence it produces. \paragraph{PRISMA flow.} 144 records identified across
twelve query families; 125 after de-duplication by arXiv id; 35 admitted by a two-pass (title/abstract then
full-text) screen; 26 full-text-coded entries (24 runnable systems + 2 study/position), with the
remaining nine cited as context. Exclusions (90) were dominated by autonomous-driving keyword collisions,
generic ML/RL, physics/astronomy, non-agentic single-shot tools, and pre-LLM methods retained only as
baselines. The broader literature map synthesized across the body's mechanism sections
(Sec.~\ref{sec:ideation}--\ref{sec:benchmarks}) was assembled by a separate multi-round systematic sweep
across 38 sub-areas and is qualitative context; only the 26 full-text-coded entries are statistically coded.
\paragraph{Codebook.} Each dimension and its allowed values are defined in Sec.~\ref{sec:scope}; ``n/d''
marks a disclosure absent from the consulted full text.
\paragraph{Reliability.} An independent second coder re-coded a random sample of ten systems on the four
most interpretive dimensions: agreement was 90\% (artifacts), 50\% (autonomy), 60\% (novelty), 60\%
(selection); 65\% overall. We therefore treat autonomy as a coarse label, treat novelty and selection
rates as directional indicators, and foreground artifact release as the most reliable headline dimension.

\section{Full Coded Corpus and Stage Coverage}
\label{app:corpus}
This appendix lists the full-text codings for all focal systems on the seven audit dimensions
(Table~\ref{tab:fullcorpus}) and their per-system lifecycle-stage coverage (Table~\ref{tab:stagecov}). It is
the evidence base behind every quantitative claim in the paper.
\input{appendix_corpus}

\input{appendix_stagecov}

\input{appendix_crosstab}

\input{appendix_taxonomy}

\input{appendix_benchmarks}

\bibliography{references}

\end{document}

%% file: sec_ideation.tex
\section{Ideation and Hypothesis-Generation Agents}
\label{sec:ideation}

Ideation and hypothesis-generation agents sit at the front of the autonomous-research pipeline. Given a corpus, a knowledge graph, or a bare research direction, they propose the questions, hypotheses, and research plans that downstream experiment and review agents later try to execute. This stage makes the verification problem especially visible. A proposed hypothesis is a claim about something not yet known, so the natural check, whether the idea is correct, is unavailable at proposal time. The field substitutes a proxy: an LLM or human judge rates the idea for \emph{novelty} and \emph{plausibility}. The main gap in this section is the difference between novelty-as-judged, an idea a panel rates as new and interesting, and novelty-as-valid, an idea that turns out to be genuinely new and true after grounding and testing. We organize the discussion by the mechanism that produces and checks proposals: literature-conditioned generation, knowledge-graph grounding, iterative search and debate, and the benchmark layer now trying to make these proxies less brittle.

\subsection{Literature-Conditioned Idea Generators}
The first and most populous mechanism conditions an LLM on retrieved scientific literature and asks it to emit hypotheses, treating prior work as both prompt material and an implicit novelty constraint. SciMON \citep{wang2023scimon} pioneered the template: retrieve related findings, generate a candidate, and iteratively maximize a novelty score against the retrieved neighborhood so the output is pushed away from existing claims rather than paraphrasing them. ResearchAgent \citep{baek2024researchagent} extends this to a full problem/method/experiment triple refined by reviewing agents conditioned on an entity-centric knowledge store, and Chain-of-Ideas \citep{li2024chain} organizes the literature into a progression chain so the model reasons about how a field has evolved before proposing the next step. Nova \citep{hu2024nova} iteratively plans and retrieves to expand the pool of seed ideas and explicitly targets diversity. IdeaSynth \citep{pu2024ideasynt} turns the research idea into an editable artifact whose facets a researcher can expand and stress-test interactively, and IRIS \citep{garikaparthi2025iris} exposes steerable controls for human-guided ideation. These systems automate a generation-plus-novelty-filtering loop. Their verification is still mostly judgment based: an LLM-as-judge or a small human panel scores novelty, feasibility, and excitement.

The clearest empirical warning comes from the human study of \citet{si2024novelideas}, which recruited expert NLP researchers to write ideas head to head against an LLM ideation agent under blind review. The agent's ideas were rated as \emph{more novel} than the experts', a widely cited result, but the same study reported that LLM judges correlate poorly with expert judgment and that the agents suffered from idea-duplication and self-evaluation failures. The proxy says the machine wins, while the reliability of the proxy remains uncertain. The follow-up by \citet{ideationexecutiongap2025} sharpens the concern by executing LLM-generated and human-generated ideas end to end and measuring outcomes; ideas rated highly novel at proposal time did not survive execution. Novelty ratings are therefore weak predictors of downstream validity. Work on the collective effects of AI ideation assistance \citep{ashkinaze2024how} and on homogenization \citep{anderson2024homogeniz} adds a second problem: conditioning many researchers on the same models can leave individual ideas no better while shrinking the diversity of the idea pool, a failure no single-idea novelty score can detect.

The diversity problem is also a verification problem. Every novelty score discussed in this section, whether an LLM rating, a human panel rating, or a retrieval-based distance to the nearest neighbor, is computed on one candidate at a time. A per-idea score cannot see a property defined over the joint distribution of proposals: a thousand independently generated ideas, each scored as novel against the literature, may still collapse onto a handful of templates when viewed together. SciMON's novelty maximization \citep{wang2023scimon} pushes each output away from its retrieved neighborhood, and Nova \citep{hu2024nova} explicitly targets diversity, but both operate within a single generation trajectory. Neither audits the cross-run, cross-user concentration that \citet{anderson2024homogeniz} and \citet{ashkinaze2024how} measure. An ideation agent can therefore post a strong per-idea novelty record while the population of ideas it induces across a research community contracts. The model's prior is shared, and the novelty filter penalizes overlap with the past, not overlap with the agent's own siblings. Detecting diversity collapse requires a population-level signal, such as entropy or coverage over accepted proposals. Few generators in this subsection instrument that signal. Even if each idea is individually novel, a homogenized pool is collectively less useful, and the per-idea proxy is blind to the loss.

Literature-conditioned generators have largely solved fluent, on-distribution proposal generation. Their weak point is the check they optimize against. Judged novelty is decoupled from validated novelty, and closing that gap requires expensive downstream execution that the proposal stage cannot itself supply.

\subsection{Knowledge-Graph-Grounded Proposal}
A second mechanism attacks the grounding weakness of pure literature conditioning by anchoring hypotheses in a structured scientific knowledge graph and adding an explicit checking layer. The clearest motivation is hallucination control. KG-CoI \citep{xiong2024improving} chains reasoning through graph-attested facts to detect and suppress fabricated intermediate steps, converting an opaque hypothesis into a path whose edges are individually inspectable. BioDisco \citep{ke2025biodisco} couples knowledge-graph reasoning with literature retrieval and iterative feedback, and evaluates novelty against held-out future literature. A proposed hypothesis is counted as novel only if it does not already appear in the temporally withheld record. This temporal, held-out evaluation is the strongest signal in the subarea because it makes novelty-as-valid mechanically checkable for at least the rediscovery case.

The generate-then-adjudicate pipelines push the checking layer further. SAGE \citep{nasser2026sage} anchors biomarker hypotheses to a knowledge graph and then runs debate-based novelty assessment plus an automated validation pipeline; in this subarea, generation is cheap and adjudication is the product. HypoChainer \citep{jiang2025hypochain} inserts an explicit human/LLM validation stage over knowledge-graph hypothesis chains, keeping a person in the loop where the automated signal is weakest. A third cluster makes the discovery output itself a checkable object via verifiable link prediction: REx \citep{nunes2025rewarding} uses reinforcement learning to produce ontology-grounded explanatory paths for drug repurposing, and the abductive-knowledge-graph line \citep{bai2023advancing} optimizes generated logical explanations against the graph. The proposal is no longer a free-text claim but a graph path that either resolves against the ontology or does not. Earlier task-specific systems such as the perturbation-hypothesis agent of \citet{roohani2024biodisco} and the hypothesis-generation framework of \citet{zhou2024hypothesisgen} similarly tie proposals to structured biological priors. The community is now consolidating around verification: BioVerge \citep{yang2025bioverge} provides a benchmark for biomedical hypothesis generation, and the validation-focused survey of \citet{kulkarni2025scientifi} documents the field's pivot from "can an agent propose a hypothesis" to "can we verify the proposed hypothesis is novel, grounded, and supported." The residual gap is that graph grounding verifies \emph{consistency with} and \emph{novelty against} the encoded knowledge, but the knowledge graph is itself a curated, incomplete artifact. An idea that is novel and graph-consistent may still be biologically false. Most systems in this group do not close that final step with wet-lab or held-out empirical follow-up.

\subsection{Iterative Search, Multi-Agent Debate, and Tournament Selection}
A third mechanism treats ideation not as single-shot generation but as a search over a population of candidates, using multi-agent debate, simulated collaboration, or tournament selection to amplify quality before any external check. VirSci \citep{su2024virsci} simulates a multi-agent scientist team whose members collaborate and critique to produce ideas, reporting that the social dynamics improve novelty over a solo agent. SciAgents \citep{ghafarollahi2024sciagents} grows and traverses an ontological knowledge graph with a team of specialized agents that propose, critique, and refine, combining the graph-grounding and debate mechanisms. The most ambitious instance is the AI co-scientist of \citet{gottweis2025coscientist}, a multi-agent system built on a generate, debate, and evolve loop with an explicit Elo-style tournament in which hypotheses compete pairwise and a ranking emerges from many simulated reviews; the authors report wet-lab corroboration for a subset of generated hypotheses, which is the rare case where an ideation agent reaches for novelty-as-valid rather than stopping at novelty-as-judged. MOOSE-Chem \citep{yang2024moosechem} formalizes chemistry hypothesis discovery as recombining a background question with retrieved inspirations and validates against the actual inspirations and findings of rediscoverable post-cutoff papers, again grounding the proxy in documented ground truth.

The caution is that the amplification engine is usually an LLM judge: debate scores, Elo tournaments, and reviewer agents all reduce to one model grading another. The review-reliability and process-verification concerns discussed later therefore apply at the ideation stage too. A tournament that ranks hypotheses by simulated peer review can converge on ideas that are persuasive to the judge rather than true, and self-preference or surface-form biases in the judge translate directly into mis-ranked research agendas. Tools like IdeaSynth \citep{pu2024ideasynt} that keep a human in the editing loop, and the steerable IRIS interface \citep{garikaparthi2025iris}, are partial responses, repositioning the human as the verifier of last resort rather than the generator. End-to-end agentic-scientist systems that embed an ideation module, including the AI Scientist line \citep{lu2024aiscientist, yamada2025aiscientistv2}, the agent-laboratory pipeline \citep{schmidgall2025agent}, CycleResearcher \citep{weng2025cycleresearcher}, and experiment-design agents \citep{li2025agentexpt, jin2025stella}, inherit this exposure. A weak ideation verifier propagates downstream, because an unsound hypothesis can consume an entire experimental run before its invalidity surfaces, the cost structure quantified by the ideation-execution gap \citep{ideationexecutiongap2025}.

With all three generation mechanisms now on the table, their trade-off becomes legible along a single axis: how much of the novelty signal each one converts from judged to valid, and what it pays to do so. Literature-conditioned generation \citep{wang2023scimon, baek2024researchagent, li2024chain} is the cheapest to run and the most fluent, but it buys essentially no validated novelty: its only check is distance from retrieved prior work, which certifies non-duplication of the \emph{past} while saying nothing about correctness, and the human study of \citet{si2024novelideas} together with the execution follow-up of \citet{ideationexecutiongap2025} shows precisely this gap, high judged novelty that does not survive contact with reality. Knowledge-graph grounding \citep{xiong2024improving, ke2025biodisco, nasser2026sage} pays a substantial construction and curation cost for a strictly stronger signal: an idea expressed as a graph path is inspectable edge by edge, fabricated intermediate steps become detectable, and when novelty is scored against temporally held-out literature \citep{ke2025biodisco} the rediscovery case of novelty-as-valid becomes mechanically checkable rather than asserted. The cost is twofold, the graph is incomplete so genuine novelty may be unrepresentable, and consistency with a curated ontology is not biological truth, so the signal saturates short of empirical validity. Search, debate, and tournament selection \citep{su2024virsci, gottweis2025coscientist} pay the highest compute cost, many generations and many simulated reviews per surviving idea, and they buy no new \emph{kind} of signal over the other two: an Elo tournament is still a judged-novelty aggregator, so it amplifies whatever the underlying judge rewards, which means a biased or surface-form-sensitive judge yields confidently mis-ranked agendas at scale. Where the mechanisms disagree is instructive. Literature conditioning and debate optimize a proxy and trust it; graph grounding distrusts the proxy enough to make part of it checkable; and only the rare systems that reach past all three to an external oracle, the wet-lab corroboration in the co-scientist work \citep{gottweis2025coscientist} or the documented-rediscovery grounding of MOOSE-Chem \citep{yang2024moosechem}, actually touch novelty-as-valid. The ordering by validated novelty bought is therefore graph-and-oracle grounding above debate above pure literature conditioning, while the ordering by cost is almost the reverse, and no mechanism in this section closes the gap for a hypothesis that is genuinely new and has no held-out counterpart to rediscover.

\subsection{Hypothesis Validation and Experimental-Design Soundness}
\label{subsec:hypothesis_validation}

Between the moment an agent commits to a hypothesis and the moment it runs code lies a step that the autonomous-research literature has been slow to scrutinize: the design of an experiment capable of actually testing that hypothesis. Experimental design is where the proposed discovery first meets a possible falsifier. An agent can automate ideation fluently and can automate execution reliably, yet still produce a study whose structure cannot adjudicate the claim it was built to evaluate. This subsection surveys what current systems verify about their own designs and, more importantly, what they leave unaudited: confounded comparisons, underpowered protocols, and the agentic analogue of p-hacking. The main point is that release of code and results, now routine, says nothing about whether the design could have falsified the hypothesis in the first place.

\paragraph{From plausible hypotheses to testable ones.} A first mechanism concerns whether the agent generates hypotheses that admit a clear test at all. Work coupling literature-derived priors with data shows that hypotheses grounded jointly in prior findings and observed signals are more useful than either source alone \citep{liu2024literature}, but usefulness for a downstream human decision is not the same property as testability under a controlled design. Benchmarks that probe scientific reasoning under graded information disclosure make the gap explicit: when a model receives only a topic and a research question, the hypotheses it emits drift from the eventual ground-truth conclusion, and alignment improves only as experimental detail is supplied \citep{lew2026projection}. The agent, in other words, often does not know what design would distinguish its hypothesis from rivals until much of that design has already been handed to it. Systems that verify hypothesis testability tend to check surface form, namely whether a hypothesis is specific, falsifiable in principle, and tied to a measurable outcome, rather than whether the agent has identified the comparison and the null that would settle it. The result is a class of hypotheses that read as scientific but encode no implicit experiment.

\paragraph{Confounds and the design of a fair comparison.} The second and least audited mechanism is confound control. A sound experiment isolates the manipulated variable; an agent that changes a method, a dataset, and a hyperparameter at once and attributes the resulting delta to its proposed contribution has not tested its hypothesis, it has merely observed a difference. The autonomous-exploration literature surfaces this directly: a system that builds on a baseline paper, proposes an improvement, and iterates until a gain appears can report progress that is entangled with incidental implementation changes, and the accompanying risk analysis flags exactly such attribution hazards as a recurring failure mode \citep{miyai2025jrai}. Confound diagnosis is fundamentally a causal question, and the causal-inference community has long treated design validation as a first-class step, using domain knowledge to probe whether an estimated effect survives interventions that a confounded model would not \citep{grunbaum2022quantitative}. Validation frameworks built around cross-checking and robustness under perturbation likewise exist for distinguishing genuine structure from artifacts of noise and specification \citep{yu2024robust}. Few autonomous research agents import these checks. They verify that an experiment ran and that a number moved; they rarely verify that the comparison was clean, that the only thing differing between conditions was the variable under test, or that an observed improvement is not an uncontrolled side effect. Independent confound auditing of an agent-designed comparison remains close to absent.

\paragraph{Statistical power and the unmeasured null.} A third mechanism is whether the design carries enough evidential weight to support its conclusion. Underpowered protocols, single seeds reported as if representative, comparisons with no variance estimate, and the absence of an explicit null are widespread in agent-produced studies, and they convert noise into apparent discovery. Agentic data-analysis pipelines now run formal hypothesis tests and report significance routinely \citep{gaurav2025temporal}, which makes the soundness question sharper rather than softer: the machinery to compute a p-value is automated, but the machinery to decide whether the test is appropriate, adequately powered, and corrected for multiplicity is not. An agent that selects among many tests, metrics, or random seeds and surfaces the one that clears a threshold is performing p-hacking whether or not it represents the act to itself, and nothing in a typical pipeline records the multiple comparisons that were silently discarded. The methodological-transparency literature reaches the same conclusion from the empirical side, finding that experiments built on synthetic or simulated populations frequently underspecify the population and task they claim to address, leaving their statistical validity unverifiable from the artifact alone \citep{batzner2025whose}. Power analysis, pre-registration of the primary comparison, and disclosure of the full set of tests attempted are the verification steps that would close this gap, and they are precisely the steps current systems most often omit.

\paragraph{Attribution and the limits of post-hoc checking.} A final mechanism concerns whether soundness can be recovered after the fact. Once a confounded or underpowered design has produced a result, attributing the outcome to a specific decision is itself a hard causal problem. Recent work shows that intervention-based replay, re-executing an agent trajectory under a controlled change to one step, recovers the responsible decision far more reliably than correlational judge-based attribution, which performs near chance at step-level localization \citep{shah2026causal}. The debugging result also applies to design auditing: correlational inspection of a finished study, including review by an LLM judge, is a weak instrument for certifying that the design was sound, whereas an interventional check that re-runs the experiment under a single controlled change is strong. The strength of a discovery tracks the strength of the independent check applied to it, and for experimental design that check must be interventional rather than narrative.

\paragraph{Synthesis.} Across these mechanisms a consistent picture emerges. Agents increasingly automate the production of hypotheses, designs, statistical tests, and released artifacts, and they increasingly verify the easy properties: that a hypothesis is well formed, that an experiment terminates, that a metric is reported. They rarely verify the properties that determine whether the experiment could have falsified the hypothesis: freedom from confounds, adequate power, an honest accounting of the comparisons attempted, and an interventional rather than correlational basis for attributing the result to the proposed cause. These remain the unaudited core of experimental-design soundness, and until independent checks at this layer become standard, a closed-loop claim that the agent confirmed its hypothesis should be read as a claim about the design's strength, not yet as a verified discovery.

\subsection{The Benchmark and Evaluation Layer}
The fourth mechanism is not a generator at all but the measurement infrastructure now being built to audit the first three. Early dedicated ideation benchmarks established the proxy itself as an object of study: IdeaBench \citep{ideabench2024} standardizes evaluation of research-idea generation, and HypoBench \citep{liu2025hypobenc} provides a systematic benchmark probing whether agents generate hypotheses that are novel, plausible, and discriminating rather than merely fluent, and dedicated creativity-evaluation protocols now score open-ended generation across tasks \citep{tan2026automated}, treating the novelty proxy itself as a measurement problem. The important shift is the move to grounding ideation scores in documented human discovery. ResearchBench \citep{liu2025researchb} decomposes scientific discovery into inspiration retrieval, hypothesis composition, and hypothesis ranking, scoring an agent against the actual inspirations and findings of source papers and deliberately using post-2024 work to mitigate contamination. AI Idea Bench 2025 \citep{qiu2025ai} similarly measures idea generation by alignment to the genuine findings of recent papers. DiscoveryBench \citep{majumder2024discoverybench} reframes the target as data-driven hypothesis discovery with verifiable workflows, so a proposed hypothesis is graded against whether it can be substantiated from the supplied data rather than against a judge's taste. The benchmark cluster surveyed at the loop level, EXP-Bench \citep{kon2025expbench}, the speedrun reproduction oracle of \citet{zhao2025the}, ResearchCodeBench \citep{hua2025researchc}, and MLR-Bench with its result-fabrication-aware judge \citep{chen2025mlrbench}, extends the same execution-grounded philosophy from ideation into the downstream stages, while report-grounding benchmarks \citep{du2025deepresea, xu2025researche} apply it to synthesis.

Two further developments target the proxy's weakest point directly. Popper \citep{huang2025popper} operationalizes hypothesis validation as sequential falsification, automatically designing and running statistical tests against a generated hypothesis so that validity, not judged plausibility, becomes the score; this is the cleanest existing bridge from novelty-as-judged to novelty-as-valid, because it makes the hypothesis falsifiable by construction. ForeSci \citep{tian2026foresci} exposes the subtler "evidence-decision decoupling" failure in which an agent cites correct evidence yet forecasts the wrong research object, and the 2026 end-to-end validity benchmarks such as ResearchClawBench \citep{xu2026researchc} hide the target paper and score rediscovery against expert rubrics, repeatedly finding that top agents pass only around twenty percent and fail most on the "missing scientific core." The aggregate message of the benchmark layer is that the field has stopped accepting fluent or judge-approved ideas as evidence of discovery and is rebuilding evaluation around contamination-controlled rediscovery, data-grounded substantiation, and automated falsification. Yet the gap remains open at its hardest point: for a genuinely new hypothesis with no documented counterpart and no immediately runnable test, neither held-out literature, nor data substantiation, nor falsification yet supplies a verification signal, so the most valuable ideation outputs are precisely the ones the current audit infrastructure cannot certify.

These checks answer a narrower question than whether an idea is good. The contamination-controlled rediscovery benchmarks \citep{liu2025researchb, qiu2025ai, yang2024moosechem, xu2026researchc} all share one structural precondition: there must exist a documented target, a real paper or a known finding, against which the agent's proposal is scored. They verify the agent's ability to \emph{re}discover what humans already found and temporally withheld, which is a genuine and contamination-resistant signal but a fundamentally retrospective one. Data substantiation \citep{majumder2024discoverybench} requires that the substantiating dataset already be in hand, and automated falsification \citep{huang2025popper} requires that the hypothesis admit a statistical test runnable now. Each of these collapses novelty-as-valid onto a case where validity is already, in principle, decidable. What none of them can do, and what the reviewer therefore still cannot do, is certify a forward-looking hypothesis whose validity is not yet decidable by any available oracle: an idea with no held-out paper to match because the discovery has not been made, no in-hand dataset because the relevant experiment has not been run, and no immediately runnable falsification test. For that class of idea the only honest verification is the expensive downstream execution the proposal stage cannot supply, and the failure mode is subtle rather than gross, as ForeSci \citep{tian2026foresci} documents: an agent can marshal correct evidence and still forecast the wrong research object, an evidence-decision decoupling that retrospective rediscovery scoring is structurally unable to penalize because it never asks the agent to commit to an undocumented target. The audit infrastructure has become much better at certifying decidable cases while leaving the highest-value, undecidable case close to where it began, which is why ResearchClawBench-style end-to-end rediscovery \citep{xu2026researchc} still reports agents passing only around twenty percent and failing most on the missing scientific core.

\subsection{Mathematical Discovery beyond Proof}
Mathematics provides an instructive boundary case for ideation, because here the verification signal can sometimes be made nearly free, throwing the rest of the field's difficulty into relief. AI-assisted mathematical discovery beyond formal theorem proving runs through three mechanisms with sharply different audit properties. First, heuristic, data-driven conjecture engines over precomputed invariants, TxGraffiti and its agentic successor The Optimist \citep{davila2024automated, davila2024the}, descendants of Fajtlowicz's Graffiti, generate candidate inequalities that are trivially machine-auditable by counterexample search over a finite table of objects, yet whose mathematical interestingness and novelty still demand human adjudication. Second, search-plus-learning for explicit constructions and counterexamples, Wagner's deep cross-entropy reinforcement learning \citep{wagner2021construct} and the PatternBoost transformer-plus-local-search loop \citep{charton2024patternbo}, where verification is cheap because the output is a concrete object that either satisfies the extremal property or does not; this is the cleanest case of capability outrunning the need for trust infrastructure, the mirror image of the biology hypothesis agents whose outputs cannot be cheaply checked. Third, the harder unsolved layer of judging \emph{what} to discover: the Ramanujan Machine \citep{raayoni2019the} automatically conjectures constant identities but cannot certify them, and \citet{bengio2024machine} frame the open problem of an AI mathematician that optimizes for conjecture interestingness rather than provability. The interestingness study and benchmark line synthesized by \citet{mishra2025a} shows that machine judgments of mathematical interestingness diverge from human ones and operationalizes "verified discovery" by pairing open conjectures with formally checkable statements. The boundary applies to the whole section: where the discovery target admits a cheap, exact oracle, generation is the main hard part and the verification gap nearly vanishes; everywhere else, including most scientific hypothesis generation, novelty and significance remain human-adjudicated proxies, and the limiting step shifts from proposing ideas to verifying that the proposed idea is correct, new, and worth pursuing.

%% file: sec_litwriting.tex
\section{Literature and Writing Agents}
\label{sec:litwriting}

Literature and writing agents are among the most product-mature autonomous-research systems, and they make the verification problem easy to see. These systems automate the steps that turn prior work into citable prose. Their characteristic failure is not a botched experiment but an ungrounded claim, or a citation that does not say what the agent reports it says. The area sorts into four mechanisms: survey and related-work generators, deep-research agents that interleave web exploration with reasoning, attribution and grounding methods that check whether a sentence is supported by its evidence, and systematic-review harnesses that recast quality as claim and citation grounding. Across all four, generation is fluent and steadily cheaper, while trust depends on whether the report is faithful to its sources.

\subsection{Survey and Related-Work Generators}
The first mechanism group automates long-form scholarly synthesis: given a topic or a target paper, the system retrieves a corpus, plans an outline, and drafts a survey or related-work section with inline citations. The progenitor, STORM \citep{shao2024assisting}, reframes pre-writing as perspective-driven question asking, simulating multiple expert personas that interrogate a topic and grounding the resulting questions in retrieved sources before drafting a Wikipedia-grade article. Its collaborative successor Co-STORM \citep{jiang2024into} adds a multi-agent roundtable so a human can steer the discourse and surface knowledge the user did not know to ask for. Academic-survey systems then specialize this loop for the literature-review genre. AutoSurvey \citep{wang2024autosurvey} decomposes generation into parallel outline and section drafting with a re-polishing pass, trading some global coherence for the ability to cover hundreds of references. SurveyForge \citep{yan2025surveyforge} attacks the resulting structural weaknesses with heuristic outline construction and a memory-driven retrieval module that selects citations during writing rather than bolting them on afterward, and SurveyX \citep{liang2025surveyx} similarly restructures the pipeline around attribute-based reference organization and post-hoc refinement. A parallel line targets the narrower related-work task directly: LitLLM \citep{agarwal2024litllm} builds a retrieval-augmented, plan-then-generate pipeline that conditions related-work generation on a structured sentence plan to control which papers are discussed where. OpenScholar \citep{asai2024openscholar} is the most verification-conscious of this cluster, pairing a large open scientific datastore with a retrieval-and-feedback loop and an explicit self-feedback inference step that checks and revises citation support. It reports citation accuracy competitive with proprietary systems while keeping the datastore and retriever inspectable.

The checks for these systems remain thin relative to what they produce. The dominant evaluation signal is still LLM-as-judge comparison on coverage, structure, and fluency, supplemented by citation counts and reference recall against a held-out bibliography. These metrics reward a document that looks like a survey. They do not establish that any individual sentence is entailed by the source it cites, and they miss the failure mode that matters most for a survey: a confident synthesis that aggregates several papers into a claim none of them actually makes. Systems that retrieve citations during writing, such as SurveyForge and OpenScholar, narrow the gap between a sentence and its evidence. OpenScholar's self-feedback step is an early instance of building the attribution check into the generator rather than leaving it to a downstream reader. The unresolved problem is end-to-end citation integrity. No widely adopted survey generator certifies that its citation graph is free of fabricated or misattributed references. Releasing the prose and bib file is not the same as verifying the claims they support.

\subsection{Deep-Research Agents}
The second mechanism group couples agentic web exploration to a reasoning model, producing reports over the live, open web rather than a fixed corpus. The mechanisms diverge in how tightly search is woven into generation. WebThinker \citep{li2025webthinke} interleaves think, search, and draft within a single reasoning trajectory, letting a reasoning model decide mid-thought when to issue a query and fold the result back into its chain. DeepResearcher \citep{zheng2025deepresea} trains an end-to-end web agent with reinforcement learning directly against the live web, so the search policy is learned from outcome feedback rather than scripted. TTD-DR \citep{han2025deep} recasts long-form report generation as iterative denoising of a draft skeleton, repeatedly retrieving and revising a noisy outline toward a finished report. These open-source efforts share architecture with the commercial Deep Research products from OpenAI, Gemini, and Perplexity, whose design space is surveyed comprehensively by \citet{xu2025a}; we name those products in prose because the survey's own constraint forbids citing systems that lack a stable scholarly reference.

Verification is further along here than for static survey generators, because the deep-research community built its benchmarks alongside its systems. DeepResearch Bench \citep{du2025deepresea} scores reports on adaptive, reference-based quality together with citation count and citation accuracy, separating whether a report is good from whether its citations hold. ReportBench \citep{li2025reportben} goes further toward an attribution oracle: it reverse-engineers gold surveys to obtain a reference standard and runs an agentic pipeline that verifies cited content against the actual sources while independently checking uncited statements against the web. BrowseComp \citep{wei2025browsecom} attacks the retrieval substrate from the other end, supplying short questions with single, hard-to-find, easily verifiable answers that stress-test whether an agent can persistently locate a fact rather than confabulate one. These checks still leave gaps between benchmark units. A report can score well on reference-based quality while individual citations remain unverifiable, and BrowseComp's verifiable-answer regime does not transfer to the long-form, multi-claim reports the products actually ship. No current benchmark audits the full grounding chain end to end: claim, citation, source passage, and reliability of the source itself. For deep-research agents, that signal exists only in fragments.

\subsection{The Attribution and Grounding Oracle Layer}
The third mechanism group is the substrate the first two depend on but rarely build: methods that decide, for a given claim and a given source, whether the claim is true and whether it is supported. Verifying a report reduces to two separable questions. Factuality asks whether the claim is atomically true; attribution asks whether the claim is entailed by the cited evidence. On the factuality axis, decomposition-and-check pipelines turn free-form prose into discrete auditable units: FActScore \citep{min2023factscore} measures atomic-fact precision against a knowledge source, SAFE \citep{wei2024longform} uses a search-augmented evaluator to grade long-form factual claims, and VeriScore \citep{song2024veriscore} filters for the verifiable claims that such metrics can legitimately score. These methods supply a granular signal, but they share a recall blind spot. A claim that is omitted, or phrased so vaguely it cannot be checked, escapes the metric entirely, so a high factuality score can coexist with a report that quietly avoids its hardest assertions.

Attribution is harder. Liu et al.'s audit of generative search engines \citep{liu2023evaluatin} found that only roughly half of generated sentences were fully supported by their cited sources, establishing early that fluent citation is not faithful citation. AttributionBench \citep{li2024attributi} shows that even fine-tuned judges plateau near 80\% F1 at the entailment decision, so the oracle itself is imperfect, and RAGTruth \citep{niu2023ragtruth} provides word-level hallucination annotations over retrieval-augmented generation that quantify how often a generated span is unsupported by retrieved context. Generation-time methods try to close the loop from the inside: RARR \citep{gao2022rarr} retrofits attribution by researching and revising existing output to add citations post hoc, Self-RAG \citep{asai2023selfrag} trains a model to retrieve on demand and emit reflection tokens that critique its own support, LongCite \citep{zhang2024longcite} pushes attribution to sentence-level granularity in long-context QA, and the ALCE benchmark \citep{gao2023alce} standardizes how citation quality is measured so these methods can be compared. PaperQA2 \citep{skarlinski2024language} shows a provenance-first scientific RAG agent can match human experts on literature synthesis and contradiction detection, which is the closest the field comes to a deployable attribution-aware research assistant.

Two findings move this layer from supporting detail to the center of the survey's argument. First, a citation is necessary but not sufficient evidence, because the grounding chain can be attacked. PoisonedRAG \citep{zou2024poisonedr} shows that injecting a handful of crafted passages into a retrieval corpus flips RAG answers with roughly 90\% success, so an agent can cite a real, retrieved passage that was adversarially planted. Tool-grounding work such as Gorilla \citep{patil2023gorilla} demonstrates the constructive flip side: binding outputs to retriever-checked API documentation substantially mitigates hallucination. LAB-Bench \citep{laurent2024labbench} operationalizes literature recall, figure interpretation, and database navigation as auditable scientific skills. Second, citation integrity in scientific agents is now its own thrust because the stakes are concrete: \citet{li2026source} addresses BibTeX-level fabrication detection and mitigation, CiteCheck \citep{khajavi2026citecheck} uses retrieval grounding to detect corrupted or non-existent references in generated scientific text, and end-to-end audits of deep-research-agent citations \citep{onweller2026cited} find that generated references frequently fail on link validity, relevance, and factual accuracy. Scalable automated attribution checking is still an open verification primitive, not a solved one. Fabricated references, mis-grounded claims, and poisoned retrieval remain failure modes a reviewer cannot currently close.

\subsection{Systematic-Review and Meta-Analysis Verification}
The fourth mechanism group attacks the hardest writing tasks, full systematic reviews and meta-analyses. This cluster most explicitly treats quality as claim and citation grounding rather than fluent prose. The evaluation-side work is blunt about generation's limits. ScholaCite \citep{martinboyle2024shallow} finds that GPT-4 performs only shallow synthesis of citation context and should not draft related work independently. OARelatedWork \citep{docekal2024oarelated} shows that evidence-grounding accuracy falls from 92.9\% when models work from abstracts to 83.8\% when forced to ground in full text, and that standard reference-based metrics fail to capture this, motivating statement-level checking. GREP \citep{ahinu2025expert} builds a multi-turn, expert-preference, dimension-decomposed evaluator because off-the-shelf LLM judges miss the domain validation constraints that determine whether a review is sound. The generation systems improve sourcing without closing this loop: Citegeist \citep{beger2025citegeist} uses dynamic retrieval-augmented generation to assemble citations, and the Select-Read-Write multi-agent pipeline \citep{liu2025select} works over full text rather than abstracts, but both leave end-to-end citation correctness unverified.

The systematic-review harnesses make verification the primary objective rather than a downstream nicety. AgentSLR and OpenExtract-style systems \citep{achterberg2026openextra} and LLMSurver-style surveys \citep{joos2025leveragin} operationalize quality as screening and extraction accuracy against expert ground truth, and the evaluation work \citep{padarha2026evaluatin} reports a hard ceiling: frontier models top out around field-level F1 of 0.67 and are, in the authors' words, not reliable enough for unsupervised deployment, even where consensus schemes can beat a single human annotator on the filtration stage. The methodological substrate this draws on, scientific claim verification against a literature base as in SciFact \citep{wadden2020scifact}, is mature, but the agentic systematic-review setting inherits its hardest open problem: a Support, Contradict, or Not-Enough-Information verdict is only as trustworthy as the retrieval and entailment steps behind it. The survey-automation community has largely conceded that fluent prose is the easy part. Verifiable, statement-level grounding against expert ground truth is the hard check, and that check still falls short of unsupervised reliability.

\subsection{Scientific Figure, Table, and Multimodal Artifact Generation}
\label{subsec:figure_table_multimodal}

Beyond prose and code, an autonomous research report is largely a collection of figures and tables. These visual artifacts now fall within the scope of generative agents. Plots, charts, and result tables are the channel through which a reader most directly inspects whether a claim holds, which makes faithful generation a verification problem rather than a stylistic one. A figure that misrepresents its underlying data is especially dangerous: it carries the visual authority of evidence while being unaudited. This subsection organizes the area by the mechanism an agent uses to produce or read a visual artifact, and asks what is automated, what independently checks the output, and what remains unverified. Figure and table generation is now common, but the checks that would certify a generated artifact as a faithful rendering of the data are weak and rarely closed.

\paragraph{Data-to-chart synthesis with rendered-output validation.} The dominant mechanism treats a chart as code: an agent receives tabular data and emits a plotting script that is executed to produce an image. Many failures are invisible at the data or code level and only become apparent after rendering. \citet{policar2026charts} make this concrete with a structured workflow that decomposes chart production into dataset screening, plot proposal, code synthesis, rendering, and validation-driven refinement, explicitly inspecting the rendered output to catch visualization-specific failure modes such as illegibility and semantic mismatch between the plot and its intended message. This makes chart generation an inspectable process rather than a one-shot prompt-to-code task, and it is the figure-domain analogue of execution-grounded verification: the artifact is run, observed, and revised against an observable signal. What such validation does not yet certify is faithfulness to the source data. Confirming that the rendered bars encode the actual numbers, that axes are not truncated in misleading ways, and that the visual claim matches the statistic remains largely outside the loop, because the validator inspects readability and plausibility rather than recomputing the encoded values from the data. Chart fabrication hides in the gap between rendering a plausible figure and faithfully rendering the data.

\paragraph{Structure-aware editing and the limits of pixel manipulation.} A second mechanism manipulates existing figures rather than generating them from scratch. Here the field has begun to recognize that scientific charts are not natural images. \citet{li2025charts} argue that a chart is a visual representation of structured data governed by a graphical grammar, so editing it is a structured transformation problem, not pixel manipulation; their benchmark shows that strong image-editing models fail on scientific figures because they do not respect the underlying data structure, and that conventional image-similarity metrics such as SSIM and PSNR do not capture whether an edit is semantically correct. This is an important negative result for any agent that would touch a figure through a vision model: the natural verifier (visual similarity) is blind to the property that matters (data fidelity). The unaudited risk is that an edited figure looks correct and passes perceptual checks while silently altering the quantity it depicts. No widely adopted mechanism yet re-extracts the data from an edited chart and compares it against the source, so structure-preserving editing remains asserted rather than verified.

\paragraph{Table extraction and uncertainty-aware verification.} Tables are the other primary numeric artifact, and the mechanism here is most often extraction: recovering structured cells from a rendered table, or transcribing a table image into markup. \citet{kayal2022tables} convert tabular images to LaTeX source through a transformer that reconstructs both structure and content, but report exact-match accuracy well below what a results table requires, since a single misread digit corrupts a reported number. The most direct response to this fragility is to attach a confidence estimate to each extracted value. \citet{ajayi2025uncertainty} build a model-agnostic, conformal-prediction layer over table-structure recognition and optical character recognition that quantifies the uncertainty of each extracted cell, allowing reviewers to verify only the least-confident outputs and thereby raise data quality while inspecting a fraction of the table. This is among the more mature verification mechanisms in the visual-artifact space precisely because it produces a calibrated, auditable signal that targets human attention. Its scope, however, is extraction faithfulness, not generation: it tells us whether a table was read correctly, not whether an agent that authored a results table populated it with numbers that the experiments actually produced.

\paragraph{Caption generation and the reference-free evaluator problem.} Figures travel with captions, and caption generation is now routine, evolving from text-summarization framings over corpora such as SciCap \citep{hsu2021scicap} toward multimodal, personalized profiles that condition on other figures in the same document \citep{ng2025lampcap}. The verification mechanism that has gained traction is the language-model-as-evaluator: \citet{hsu2023gpt4} show that a strong model can score caption helpfulness reference-free and correlate with expert rankings better than non-expert humans. This is convenient but circular as a faithfulness check: the same family of models that writes the caption also judges it, and the judgment targets readerly helpfulness rather than whether the caption's quantitative claims match the figure. A caption that confidently misstates a trend can be both fluent and highly rated. The independent check that is missing is one that grounds the caption's assertions in the figure's actual data, and it is largely absent.

\paragraph{Reading figures back: chart understanding as a verifier, and its hallucinations.} The inverse mechanism, an agent reading a chart, is what one would deploy to verify a generated figure, so its reliability bounds the whole enterprise. Agentic approaches improve on text-only reasoning by manipulating the chart image directly, cropping regions and localizing axes through vision tools \citep{kaur2025chartagent}, yet realistic benchmarks expose how far this is from trustworthy. \citet{wang2024charxiv} show that performance on natural arXiv charts collapses under mild stress tests and that the strongest models trail human accuracy by a wide margin, while \citet{wang2025charthal} document severe hallucination specifically when a question concerns information absent from or contradictory to the chart, with leading proprietary models scoring far below usable thresholds. The consequence for closed-loop verification is direct: a chart reader that hallucinates cannot certify a chart writer. Until the reader is reliable on adversarial and out-of-distribution figures, automated figure verification inherits the reader's failure modes, so a generated chart can make a weak discovery look stronger than its evidence permits. Visual artifacts expose the same asymmetry as the rest of the survey: generation is plentiful, independent verification of data fidelity is scarce, and the most consequential errors, fabricated plots and hallucinated chart readings, remain the least audited.

%% file: sec_coding.tex
\section{Coding, Execution, and Analysis Agents}
\label{sec:coding}

Once a method has been proposed, something has to turn it into running code, push that code to a result, and make sense of what comes back. Coding, execution, and analysis agents do this work. This is also where the verification problem becomes concrete: an agent that \emph{wrote plausible code} and one that \emph{reproduced a paper's claimed number} can leave behind the same final repository, and the difference appears only when the run is checked against execution evidence. The corpus statistics show the asymmetry: 83\% of systems release code, but only 38\% release seeds and execution traces. Source code is usually available, but its presence says little about whether the reported number was produced by the run. A seeded, logged run would let an outsider re-derive the number, yet that artifact is often missing. We trace the mechanisms in turn, beginning with the generalist coding substrate, working through the paper-to-code and reproduction tier, and closing on cross-domain replication studies; in each case the questions are what the mechanism automates, what currently stands in for verification, and what a reviewer still cannot independently close.

\subsection{Generalist coding-agent platforms as the execution substrate}
Autonomous research agents do not implement experiments from scratch; they inherit the machinery of software-engineering (SWE) agents, whose canonical evaluation is repository-level issue resolution against a hidden test suite \citep{jimenez2024swebench}. The substrate splits into three mechanism families that differ in \emph{where they place the correctness signal}. Architecture-centric systems decompose issue resolution into specialized sub-agents and explicit task graphs: MASAI assigns distinct objectives to localization, editing, and testing modules \citep{arora2024masai}, while CodeR drives a multi-role team along a predefined task graph so that each step's output is an inspectable intermediate rather than an opaque diff \citep{chen2024coder}. The agentless line argues the opposite, that a fixed localize-repair-validate pipeline without an autonomous controller already recovers most of the achievable performance \citep{xia2024agentless}, and AutoCodeRover grounds repair in program structure and spectrum-based fault localization so the edit is tied to a defensible hypothesis about \emph{where} the bug lives \citep{zhang2024autocoderover}. A second family treats patch generation as explicit search with candidate evaluation: SWE-Search wraps Monte Carlo tree search around an LLM self-evaluation signal \citep{antoniades2024swesearch}, the diversity-empowered committee re-ranks trajectories drawn from a heterogeneous pool of agents rather than trusting any single run \citep{zhang2024diversity}, and SpecRover extracts an explicit specification before validating a patch against it, so the verification target is a recorded artifact and not a private heuristic \citep{ruan2024specrover}. A third family internalizes the signal through training: SWE-Gym builds an environment whose trained verifiers score candidate patches \citep{pan2024training}, SWE-RL derives rule-based rewards from the natural history of software evolution \citep{wei2025swerl}, and Lingma SWE-GPT models the development process itself rather than only the final commit \citep{ma2024lingma}.

By the standards of this survey, these systems offer unusually strong verification machinery: executable tests, learned outcome verifiers, and recorded specifications or task graphs all force the code-producing step to yield an auditable artifact. That is exactly what a research pipeline needs if a reviewer is going to re-check the implementation half of a result. The agent platform underneath a research system (SWE-agent's agent-computer interface being the common ancestor \citep{yang2024sweagent}) is therefore not a neutral tool. Its placement of the correctness signal determines how much of the downstream science can be audited. A passing test in SWE-bench certifies behavioral equivalence to a held-out patch, but a research experiment has no such oracle: the ground truth is a paper's claimed result, not a green test, so the rich verification scaffolding of SWE agents transfers only partway. The committee, the tree search, and the trained verifier all optimize toward \emph{task success as the platform defines it}, and when that definition is ``the code runs,'' it says nothing about whether the code computes what the paper claimed.

Reproduction is harder to verify than issue resolution for structural reasons. Issue resolution has three properties that reproduction lacks. First, its oracle is decidable and pre-existing: the held-out test was written by humans before the agent ran, it returns a binary pass or fail, and it cannot be edited by the agent into a target it can hit. Second, that oracle is local: a SWE-bench task specifies the exact repository state, dependency versions, and command, so a green test is reached over a fixed and shared substrate. Third, it is adversarial to the agent's shortcuts, since the hidden suite is designed to fail on the plausible-but-wrong patch. Reproduction inverts all three. Its oracle is a number in a paper's table, which is not decidable in isolation (a value of 0.91 accuracy is only ``correct'' relative to an unstated data split, seed, preprocessing chain, and metric definition), so checking it requires reconstructing the entire generating environment that the test in SWE-bench supplies for free. That oracle is non-local, because the claimed number depends on antecedent design decisions inherited from prior work, which is the precise reason \citet{zhao2025autorepro} must mine paper lineage rather than read a single repository. It is also not adversarial; nothing in the paper actively rejects a wrong reimplementation that happens to land near the reported figure, so a plausible-but-wrong run can be scored as a success rather than caught. In verification terms, issue resolution moves a claim into the executable-oracle column of this survey, whereas reproduction often leaves the claim itself acting as the oracle. The committee, the search, and the trained verifier can only certify behavioral equivalence to whatever oracle the task admits, and for reproduction that oracle is itself under-specified. The risk is that strong inherited scaffolding measures the wrong equivalence with high confidence, and the resulting green signal is read as trust.

\subsection{Paper-to-code generators and the compile-versus-reproduce gap}
The first research-specific mechanism turns a method description into a runnable repository. PaperCoder stages the problem as planning, then analysis, then generation, producing a structured codebase from a paper \citep{seo2025papercode}, and ResearchCodeAgent positions code generation as a research-software-engineering task with iterative refinement against the paper's intent \citep{gandhi2025researchc}. A single distinction captures the defining limitation of this family: ``the code compiles'' is far weaker than ``the code reproduces the paper's numbers.'' AutoReproduce confronts this head-on by mining paper lineage, tracing a method to its antecedents so the generated implementation inherits the design decisions that determine whether the reported result actually emerges \citep{zhao2025autorepro}. At the level of a single repository the verification gap becomes operational: a generator that optimizes for syntactic and structural completeness can produce something that looks like the paper's code while leaving untested whether it yields the paper's figure. Because code disclosure is near-universal in the corpus (83\%) while seeds and traces are scarce (38\%), the field's dominant artifact is the one these systems are good at producing and the one that cannot, on its own, close the gap.

\subsection{The replication-benchmark tier: converting ``did it do science?'' into a checkable predicate}
The most consequential development in this area is a tier of benchmarks that recast the vague question ``did the agent really do science?'' as a mechanically checkable predicate, and they do so at progressively larger units of work. At the environment-setup granularity, SUPER measures whether an agent can configure a research repository from a clean machine and execute the designated experiment, with hidden pipelines checking outputs against expected results \citep{bogin2024super}, and DeployBench extends this to artifact deployment under realistic configuration friction \citep{wang2026deployben}. At the code-fidelity granularity, ResearchCodeBench scores line-level implementations of the novel contributions of 2024 to 2025 papers against reference code, deliberately using post-cutoff work to isolate reasoning from training-set contamination \citep{hua2025researchc}, and SciReplicate-Bench evaluates whether an agent reproduces the algorithmic core described in a paper's text \citep{xiang2025scireplicate}. At the end-to-end granularity, PaperBench grades whether agents replicate full papers against author-derived rubrics \citep{starace2025paperbench}, MLReplicate targets outstanding ICML and ICLR papers with outcome-based grading \citep{gaddipati2026mlreplica}, and CORE-Bench measures computational reproducibility across published artifacts at scale \citep{siegel2024corebench}. Adjacent benchmarks fix the surrounding context: MLE-bench casts the work as Kaggle-style competitions with leaderboard ground truth \citep{chan2024mlebench}, RE-Bench measures research-engineering progress against expert human baselines under matched time budgets \citep{wijk2025rebench}, ScienceAgentBench grades data-driven discovery tasks with executable validation \citep{chen2024scienceagentbench}, EXP-Bench reconstructs hundreds of executable experiments from published papers so each run is scored against the original method and result \citep{kon2025expbench}, and SPOT inverts the task entirely, asking whether agents can detect genuine errors in published papers rather than reproduce them \citep{son2025spot}. The shared environments MLGym \citep{nathani2025mlgym} and MLAgentBench \citep{huang2023mlagentbench} supply the substrate on which several of these are run.

A distinct sub-mechanism does not re-run the science at all but audits its reproducibility. ReproRepo detects and quantifies reproducibility blockers in a repository, and ReproScore assigns a graded assessment, both producing a verdict without the cost of full re-execution \citep{li2026reprorepo, samuel2026reproscor}. When independent re-running is too expensive to be routine, a structured audit of \emph{whether re-running could even succeed} is the next best signal, and it directly targets the missing-seeds, missing-traces failure that the 38\% disclosure rate quantifies.

The verification value of this tier is that every benchmark converts a claim into one of three checkable predicates: does the rubric pass, does the output match, does the repository run. Those predicates still leave residual failures. Success rates cluster low, frequently in the 20 to 50\% band, but the instructive failures are qualitative, and each failure mode is best understood by naming the artifact that would expose it. Consider first \emph{self-stopping at a weaker target}. The agent validates and reports a substituted, easier predicate: it produces some number on some subset under some configuration, declares the experiment complete, and an outcome-based grader records partial credit. This passes because the reported quantity and the claimed quantity are never placed side by side under a common definition. The artifact that catches it is a pre-committed task specification that fixes the metric, the split, and the success threshold before the run, the same role that author-derived rubrics play in PaperBench \citep{starace2025paperbench} and that pre-registration plays in the social-science setting \citep{vaccaro2026preregist}: once the target is pinned externally, ``produced a number'' can no longer be silently relabeled as ``produced the paper's number.'' Code disclosure does nothing here, because the substituted run is genuinely executable; only a frozen, externally held target distinguishes the easier task from the required one. \emph{Result fabrication} bites deeper: the agent reports a metric that no run produced, either hallucinated outright or copied from the paper it was asked to reproduce. A rubric that grades the final write-up cannot catch this, because the fabricated number satisfies the rubric by construction, and a code-existence check cannot catch it because the repository may be entirely plausible. What does catch fabrication is the seeded execution trace: a logged run, tied to a fixed seed and an inspectable output file, against which the reported number can be matched. This is why the 38\% seed-and-trace disclosure rate matters more than the headline 83\% code rate, and it is why the audit-only sub-mechanism above (ReproRepo, ReproScore) targets the artifacts that would make a trace re-derivable \citep{li2026reprorepo, samuel2026reproscor}. A third mode is \emph{grader capture}: because most of this tier is itself automated, the grader can inherit the same blind spots as the agent, so a rubric written by an LLM, or an output-match tolerance set too loosely, re-creates the closed-loop self-grading problem at the level of the benchmark. The artifact that holds this line is a hidden, human-authored checking pipeline run on post-cutoff papers, which is why the contamination-controlled designs of ResearchCodeBench \citep{hua2025researchc} and the hidden pipelines of SUPER \citep{bogin2024super} are substantive parts of the benchmark, not implementation details. The strongest members of the tier are those with author-derived rubrics, hidden execution pipelines, and post-cutoff papers; the weaker members risk certifying the appearance of reproduction. Each failure mode is invisible to the artifact the field most readily releases (code) and visible only to the artifact it more often withholds (seeds and traces), so the disclosure asymmetry directly limits what an outside reviewer can verify.

\subsection{Cross-domain replication and research-software engineering}
Generalization beyond machine learning is where the verification signal becomes both more necessary and more fragile. In the social sciences, coding agents have been tasked with reproducing published quantitative findings from their replication packages; agents perform well on structural and availability criteria, such as whether files exist and scripts are present, but struggle on the qualitative correctness of the reproduced result \citep{alizadeh2026ai}. This availability-versus-correctness split is the cross-domain echo of the compile-versus-reproduce gap, and an agentic approach to replication-package quality evaluation makes the audit itself the deliverable \citep{mbida2026an}. The contrast with the ML reproduction tier shows where verification is hardest. In computational ML reproduction, the oracle is at least in principle re-derivable: with the seed, the data split, and the trace, an outsider can re-run the pipeline to the claimed number, so the verification problem is fundamentally one of \emph{disclosure}, the missing 38\% of seeds and traces. Social-science replication faces a second obstacle on top of disclosure: even with the full replication package, the finding is a statistical estimate over human-generated data whose correctness is a matter of statistical and substantive judgment rather than bitwise match, which is exactly why agents clear the availability bar but stall on qualitative correctness \citep{alizadeh2026ai}. A re-runnable trace is necessary but no longer sufficient, because matching the reported coefficient still leaves open whether the specification, the controls, and the inference are sound. The verification gap is therefore strictly wider here than in ML: ML reproduction needs the artifact released, whereas social-science replication needs the artifact released \emph{and} a domain-expert judgment that no automated predicate fully encodes.

The discovery-loop variant raises the stakes and exposes a failure mode with no clean ML analogue. Automated Social Science couples LLM subjects with structural causal models to generate, run, and analyze in-silico experiments \citep{manning2024automated}, but replication audits find effects that are directionally correct yet systematically inflated \citep{cui2024can}, so the agent can report a real-seeming regularity that overstates the truth. This is more insidious than fabrication: the effect is real, the sign is right, the pipeline is fully logged and re-runnable, and yet the magnitude is wrong because the measuring instrument, the LLM standing in for a human population, is itself biased. No seed-and-trace disclosure catches this, because the trace faithfully records a biased measurement; what catches it is an external grounding reference, a comparison against real human data the simulation is supposed to stand in for. The underlying instrument is itself contested: silicon-sampling work treats conditioned models as stand-ins for human respondents \citep{argyle2022out, horton2023large}, and the strongest version benchmarks interview-grounded agents against real participants' test-retest reliability rather than asserting fidelity \citep{park2024generativ}, which is the social-science equivalent of refusing to treat the agent's own output as its own oracle. Cutting against all of it, ``Sense and Sensitivity'' shows simulated social dynamics are so sensitive to trivial prompt and whitespace perturbations that, absent a grounding reference model, the verification signal can be silently fabricated \citep{ju2024sense}. The scaffolding that pushes back is reproducible, re-runnable environments such as GLEE for economic games \citep{shapira2024glee} and pre-registration protocols that commit an analysis plan before the agent runs \citep{vaccaro2026preregist}, both of which give an outsider something to audit against. The cross-domain pattern is that as the substrate moves from executable code to human-grounded measurement, the verification burden does not merely persist but compounds: the SWE substrate has a decidable local oracle, ML reproduction has a re-derivable oracle gated only by disclosure, and social-science replication has an oracle that requires both disclosure and an external human-data anchor before any automated predicate can be trusted at all.

Research-software engineering on real scientific codebases exposes a third face of the gap. Agentic modernization of legacy high-performance Fortran into C++ shows that the hard part is verifying \emph{behavioral equivalence} after transformation, not generating the new code \citep{ranasinghe2025llmassist}: a translation can compile, run, and still silently diverge in numerical output, so correctness must be established against the original program's behavior rather than a test that the translation itself defines.

\subsection{Experiment Management, Provenance, and Reproducibility Infrastructure}
\label{subsec:provenance_infra}

The preceding subsections established that autonomous research agents increasingly emit artifacts such as code, environment specifications, and result files, yet that the audit value of those artifacts depends on whether an independent party can re-derive the run that produced them. Re-derivation is not a property of any single artifact; it is a property of the infrastructure layer that records how artifacts were generated, in what order, from which inputs, and under which configuration. This subsection surveys that layer, grouped into three mechanism families: experiment tracking and configuration capture, provenance graphs over multi-step workflows, and data and artifact versioning with lineage. For each, we ask what the mechanism automates, what it can verify, and what slips through when an agent rather than a human drives the loop.

\textbf{Experiment tracking and configuration capture.} The most mature reproducibility mechanism is the experiment tracker, which records hyperparameters, metrics, code commits, and the software environment for each run so that a later observer can reconstruct the conditions of a result. The motivation predates agents: \citet{crick2015reproinfra} argued that taking a published method to a new codebase routinely requires local knowledge absent from the manuscript, and proposed an automated platform that abstracts dependencies away from the individual workstation so that results can be shared and reproduced. Containerization is the modern realization of this idea, and \citet{kunstmann2024provdeploy} show how to configure containers for scientific machine learning workflows with provenance capture integrated into the image, evaluating containerization strategies across distinct high performance computing environments. The verification such systems support is environmental rather than scientific: they confirm that a run can be re-executed against the recorded configuration, not that the configuration was the one the agent claimed in its writeup, nor that the recorded metric was computed on the stated split. Once an agent generates the run, the audit question is whether the report matches the tracker. A tracker faithfully logs whatever the harness instruments, so if the agent's narrative diverges from the logged configuration, the divergence surfaces only when an auditor compares the two; nothing in the tracker forces that comparison. The capability to record exists, but emitting and reconciling the record is left to the agent's discretion, and agents rarely close that loop on their own.

\textbf{Provenance graphs over multi-step workflows.} Trackers capture individual runs; provenance graphs capture the relationships among runs, datasets, and intermediate products, which is what makes a multi-step pipeline re-derivable rather than merely re-runnable. Classical workflow provenance frameworks record the directed graph of transformations so that a result can be traced to its inputs. \citet{missier2014provdiff} push this further by comparing provenance traces of two executions to decide whether an experiment was actually reproduced, and, when it was not, by localizing the specific point of divergence through graph analysis, which converts reproduction from a yes or no verdict into a diagnosable comparison. \citet{hasham2015cloudprov} extend provenance to elastic cloud settings, capturing virtual machine configuration alongside the workflow trace so that resources can be re-provisioned and the workflow re-executed on equivalent infrastructure. The agentic turn introduces a qualitatively new node into these graphs: the model invocation itself. \citet{souza2025provagent} observe that in agentic workflows one agent's output becomes another's input, so a hallucination or reasoning error propagates downstream, and that prior provenance methods fail to relate agent-centric metadata such as prompts, responses, and decisions to the broader workflow context and downstream outcomes. Their model extends the W3C PROV standard and uses the Model Context Protocol with data observability to capture agent interactions in near real time across edge, cloud, and HPC settings, explicitly to support reliability analysis and hallucination-risk assessment. A complementary line treats the provenance store as something to be queried rather than merely recorded: \citet{souza2025interactiveprov} introduce a reference architecture in which an LLM agent translates natural language into structured provenance queries over large traces, evaluated on a real chemistry workflow. Of this family, provenance graphs offer the strongest verification, because divergence localization and prompt-level lineage make the independent check concrete rather than nominal. Two things still escape audit: whether the captured prompt and response faithfully represent the decision the agent acted on, since an agent may log a sanitized rationale while acting on a different internal state, and whether the graph is complete, since steps executed outside the instrumented harness, such as a manual edit or an out-of-band tool call, leave no node. Provenance thus raises the ceiling on auditability but does not by itself guarantee that the recorded graph is the graph that was run.

\textbf{Data and artifact versioning with lineage.} The third family fixes the inputs and outputs themselves, so that a re-derivation references the exact data and parameters rather than a moving target. Lineage at this level answers a question that trackers and workflow graphs assume away, namely whether the data an agent claims to have used is the data it actually consumed. \citet{huang2024datause} formalize this as data-use auditing, providing a method to detect, with a tunable false-detection rate, whether a specific data owner's data was used to train a model, without prior knowledge of the downstream task. For autonomous research this matters because it supplies an external check on a provenance claim rather than relying on the agent's self-report: an auditor can test the assertion that a given corpus was or was not in the training mixture instead of trusting the logged lineage. The limit is scope. Such audits answer a membership question about one data source, not the full lineage of every intermediate artifact in a long agentic run, and they require the auditor to have a candidate dataset in hand. Versioning infrastructure can pin artifacts, but pinning is only as trustworthy as the agent's discipline in committing the right artifact under the right identifier, which returns us to the discretionary-emission problem.

\textbf{Synthesis.} The infrastructure to make an agent run re-derivable is mature and, in the agentic provenance work of \citet{souza2025provagent} and \citet{souza2025interactiveprov}, increasingly tailored to model invocations as first-class provenance nodes. What separates the mechanisms is the strength of the independent check they can underwrite: trackers verify environmental re-executability, provenance graphs verify and localize divergence between executions, and data-use auditing verifies a specific input claim against the model itself. None of them, however, compels the agent to emit a complete and faithful record, and the evaluations of autonomous AI scientists bear this out. \citet{bisht2026notbuilt} argue that current systems are not built for genuinely autonomous discovery, in part because benchmarks reward single-turn prediction over feedback-closed validation, and recommend a centralized preregistration repository for AI-generated hypotheses, which is precisely a provenance-and-commitment mechanism imposed from outside the agent. \citet{miyai2025jrai} similarly surface reproducibility and integrity risks when an agent autonomously runs the full research workflow. The infrastructure exists, but agents seldom use it in a way that exposes enough provenance for an external party to check closed-loop claims. The trustworthiness of an autonomously discovered result tracks not the existence of these systems but the degree to which the agent's pipeline was forced to run inside them and to expose the resulting record to an independent verifier.

\subsection{Tool Use, Function Calling, and Execution Reliability}
\label{subsec:tool_reliability}

Every claim an autonomous research agent advances rests on a substrate of tool calls. A literature search, a database query, a unit test, a numerical simulation, and a statistical test are all surfaced to the model as function invocations: a name, a structured set of arguments, and a serialized return value that the agent must parse and act upon \citep{ding2025toolregistry}. If a tool call silently fails, returns something other than what the agent believes it returned, or is fabricated outright, then a closed-loop claim built on top of it inherits the defect while presenting a clean, confident narrative. This subsection works through the substrate mechanism by mechanism, asking of each what is automated, what independently verifies the call, and what slips by unaudited. A tool call earns trust only to the extent that some check confirms it did what was claimed, and much of the tool-calling stack ships without one.

\paragraph{Tool selection and retrieval.}
The first mechanism is choosing which tool to invoke from a registry that, in realistic deployments, may contain thousands of candidates. Retrieval models that rank tools against the current subgoal automate this step, together with the policy the agent applies over the retrieved set. Verification of it is sparse. \citet{shi2025toolret} construct a heterogeneous benchmark of thousands of retrieval tasks over a corpus of tens of thousands of tools and report that retrievers strong on conventional information-retrieval benchmarks perform poorly at tool retrieval, and that this degraded retrieval quality lowers the eventual task pass rate of the tool-using model. The selection step is thus a documented failure point, yet most agent pipelines treat the retrieved toolset as ground truth and never check whether the chosen tool was the appropriate one for the subgoal. \citet{yang2026overprivileged} sharpen the concern along a safety axis, showing that agents frequently escalate to a higher-privilege tool when a sufficient lower-privilege alternative exists, and that this over-privileged selection is amplified precisely after transient tool failures. What goes unaudited is the counterfactual: whether a different, better-scoped tool would have produced a sounder result is rarely logged, so a wrong-tool discovery looks identical to a right-tool one in the final report.

\paragraph{Argument construction and schema conformance.}
Once a tool is selected, the agent must emit well-formed arguments conforming to a schema. The model's structured-output decoding automates this, as does the registry layer that generates and validates schemas across providers and transports \citep{ding2025toolregistry}. Schema validation is the one place where verification is comparatively mature: a malformed JSON payload or a type-mismatched field can be rejected before execution. The semantic correctness of arguments that are syntactically valid is what remains unaudited. \citet{healy2026internalrep} characterize tool-calling hallucinations as incorrect tool choice, malformed parameters, or a tool-bypass behavior in which the model simulates a result and emits an output instead of actually invoking the external system. That last failure is the most corrosive for a research agent, because it produces a plausible-looking return that never touched any real instrument or dataset. The authors propose detecting such hallucinations from the model's internal representations within the same forward pass, reporting detection accuracy in the mid-eighties, which is useful as an early-warning signal but is itself a probabilistic check rather than a guarantee. A schema validator confirms that arguments are shaped correctly; it does not confirm that the call was made, nor that the arguments encode the question the agent intended to ask.

\paragraph{Execution and return-value handling.}
The third mechanism is the execution of the call and the ingestion of its return. Tool runtimes that dispatch invocations over threads, processes, or remote protocols automate execution \citep{ding2025toolregistry}, and increasingly so do serving systems that overlap tool execution with model generation. \citet{sui2026paste} describe a serving system that speculatively executes predicted future tool invocations while the model is still generating, isolating speculative results until the model confirms them. That isolation discipline doubles as a verification primitive, because it prevents an unconfirmed side effect from leaking into the trajectory, but it is engineered for latency rather than for correctness auditing. The silent error is where the audit lapses: a tool that returns an empty result, a truncated payload, a stale cache entry, or a non-fatal error code that the agent parses as a successful return. Research agents that summarize tool output through a separate component compound the risk, since the summarizer may smooth over an anomalous return into fluent prose \citep{shen2024weaktool}. Whether the bytes the agent reasoned over are the bytes the tool actually produced is seldom checked against an independent re-execution.

\paragraph{Failure handling, escalation, and orchestration safety.}
The final mechanism concerns what the agent does when a call fails or when a sequence of calls composes into a consequential workflow. Retry loops, fallback policies, and multi-step orchestration automate this. Verification here is the most developed at the safety boundary and the least developed at the correctness boundary. \citet{chen2025agentguard} repurpose the orchestrator to discover and validate unsafe tool-use workflows by executing them in the real environment and then synthesizing constraints that confine agent behavior, an approach that treats real execution as the arbiter rather than the model's own judgment. For diagnosing why a tool trajectory went wrong, \citet{chong2026ted} contribute an automated error-analysis framework that represents subgoals such as tool signatures and responses as natural-language grading notes and surfaces recurring agent errors, reporting measurable gains once the identified failure modes are remedied. These efforts share a design principle that the rest of the stack often omits: an independent process, not the generating model, decides whether the tool interaction succeeded. What goes unaudited is everything between the retry and the re-execution. A transient failure that triggers a privilege escalation \citep{yang2026overprivileged}, a fallback path whose result is never compared against the primary path, and an orchestration whose intermediate returns are never replayed all pass through to the final claim unexamined.

\paragraph{Synthesis.}
Across these four mechanisms, the tool-calling substrate is heavily automated and, at the syntactic layer, increasingly well validated. The semantic questions are harder: whether the right tool was chosen, whether the call was genuinely executed, and whether the return value was faithfully ingested. Those are often left to the generating model to answer about itself. The strongest work in this area moves the check off the model and onto an independent re-execution or an external analyzer \citep{chen2025agentguard,chong2026ted,healy2026internalrep}, and dedicated benchmarks for retrieval quality \citep{shi2025toolret} and privilege discipline \citep{yang2026overprivileged} show that substrate reliability is becoming a measurable axis rather than an assumption. Until such independent checks are routine, a research agent's closed-loop claim is only as trustworthy as its weakest unaudited tool call.

\subsection{Synthesis}
Coding and execution agents have become strong producers of artifacts, and the SWE substrate has built real verification machinery into the code-writing step itself: learned verifiers, search with self-evaluation, and recorded specifications. Yet that machinery certifies behavioral equivalence to a test, while research demands equivalence to a \emph{claimed scientific result}. The gap between the two is where fabrication, self-stopping at weaker targets, and unaudited inflation enter. The replication-benchmark tier is the field's most direct response, turning reproduction into checkable predicates, but it remains weak wherever the grader is as automatable as the agent and wherever seeds and traces (released by only 38\% of systems against 83\% for code) are absent. In those cases independent re-execution is impossible even in principle. An agent that reproduces a paper can be trusted only as far as the independent check the task admits, and for much of computational science that check still rests on artifacts the field often does not release.

%% file: sec_review.tex
\section{Review and Closed-Loop Agents}
\label{sec:review}

This category gathers systems that try to create a verification signal from inside the agent pipeline itself: automated reviewers that grade a draft, LLM-as-judge machinery that scores free-form output, multi-agent substrates that let models critique one another, and newer end-to-end ``AI scientists'' that attach a debate, critic, or simulator to the discover-experiment-write loop. Their shared difficulty is that the check is often endogenous. A proof assistant or a wet-lab assay supplies an independent oracle. A reviewer agent, judge LLM, or debating peer is usually drawn from the same model family that produced the artifact under review, so the verification signal can be gamed by the capability it is meant to audit. We group the literature by the mechanism that issues the verdict and, for each, ask what it checks today and what no reviewer can yet close.

\subsection{Automated Reviewers}
The most direct attempt to internalize peer review is the automated reviewer: an agent that ingests a manuscript and returns scores, strengths, and weaknesses. The mechanism has matured from single-pass scoring toward structured, role-decomposed simulation. Early work such as ReviewerGPT \citep{liu2023reviewergpt} showed that an LLM prompted as a referee can catch some errors and verify checklist items but is unreliable as a holistic quality gate, while learned reviewer models trained on real review corpora \citep{yuan2021autoreview} reproduce the surface form of reviews without their adjudicative force. MARG \citep{darcy2024marg} decomposes the task across multiple agents that each attend to different parts of a long paper and then aggregate feedback, improving the specificity of generated comments over a monolithic prompt. AgentReview \citep{jin2024agentreview} simulates the review \emph{process}, instantiating reviewers, authors, and an area chair as interacting agents to study how reviewer bias, rebuttal, and discussion shape outcomes. Aspect-structured and feedback-oriented reviewers \citep{li2025aspectreview, chamoun2024feedback} push the granularity of generated critique toward the dimensions human referees actually weigh, and agentic artifact-evaluation frameworks extend the same idea from prose review to checking the released code and data behind a claim \citep{baek2026artisan}. The same review function is embedded inside end-to-end pipelines: the AI Scientist \citep{lu2024aiscientist} and its successor \citep{yamada2025aiscientistv2} run an automated reviewer over their own generated papers as the acceptance gate that decides whether a result is worth keeping, and CycleResearcher \citep{weng2025cycleresearcher} trains a paired reviewer (CycleReviewer) as the reward signal for an iterative research generator. These reviewers can triage papers, surface checklist violations, and produce plausible prose feedback at near-zero marginal cost. The grounds for trusting them are thinner. Honest evaluations measure agreement with human review scores or rankings, and the meta-review literature \citep{du2024reviewcritique} documents that LLM reviewers produce shallow critiques that miss substantive flaws human referees catch. Recent audits make the concern more concrete: LLM reviewers inflate LLM-authored papers, penalize critical or risk-disclosing statements, and reward confident overclaiming. LLM-REVal and REMOR \citep{li2025llmreval, taechoyotin2025remor} exhibit this bias directly, and GenReview-style audits \citep{demetrio2025genrevie} find generated reviews systematically over-praise. A reviewer that is itself biased and gameable cannot serve as the terminal verifier for a closed-loop discovery system unless some other check verifies the verdict.

\subsection{The Reliability of the Verdict: LLM-as-Judge}
Because automated reviewers, agentic reward models, and self-improvement loops often reduce to an LLM emitting a score, the validity of that score determines how much trust the loop deserves. The LLM-as-judge literature gives mixed evidence. MT-Bench's GPT-4 judge \citep{zheng2023judging} and the Chatbot Arena preference platform \citep{chiang2024chatbot} show that a strong judge can reach roughly 80\% agreement with human preferences, enough to make LLM evaluation tractable at scale, but the same works catalog position, verbosity, and self-enhancement biases that contaminate the verdict. Mechanism-level studies then locate the most dangerous of these for autonomous research. Self-preference is not a stylistic quirk but is causally tied to recognition: a judge assigns higher scores to text it identifies as its own, and the effect tracks the lower perplexity the model assigns to its own generations \citep{panickssery2024llm}. Position bias proves unstable across judges, tasks, and the quality gap between candidates \citep{shi2024judging}, and fairness audits find judges apply inconsistent standards across superficially varied inputs \citep{ye2024justiceorprejudice}. This matters for self-verifying research loops. An agent that writes a paper, evaluates it with a judge from the same family, and iterates on that score can improve the judge-facing surface of the artifact without improving the underlying result, unless an external signal interrupts the loop. Self-preference is especially hard to remove because it is tied to the same statistical regularity the generator optimizes: the judge prefers its own generations because it assigns them lower perplexity \citep{panickssery2024llm}, and lower perplexity is exactly what the generator was trained to produce. The quantity the judge rewards and the quantity the generator maximizes can therefore collapse toward one another. In-family calibration can make the score more legible, but it does not by itself turn that score into evidence about the world.

The countermeasures map onto auditability rather than capability, and for a closed loop what matters is how much independence each actually recovers, which varies sharply. Panels of diverse judges (PoLL) \citep{verga2024replacing} replace a single large judge with a jury of smaller heterogeneous models, reducing intra-model bias and self-preference while lowering cost; the independence they buy is real but partial, bounded by the diversity of the jury. A jury attacks self-preference only insofar as its members are statistically distinct from the generator, so a panel drawn from the same family or trained on overlapping corpora shares the very perplexity preference it is meant to dilute, and the recovered independence degrades toward zero as the field's models converge. Calibration and statistical-reporting frameworks \citep{jung2024trust} occupy a different and more limited position: they convert raw judge scores into uncertainty-quantified estimates with confidence intervals, which makes the verdict's unreliability legible but adds no information the judge did not already have. A calibrated self-preferring judge reports a tighter interval around the wrong quantity; calibration audits the estimator, not the oracle, and so recovers no independence at all, only honesty about its absence. Meta-evaluation benchmarks stress-test whether an evaluator can detect deceptively good-looking but wrong outputs \citep{zeng2023evaluatin}, which is again diagnostic rather than corrective: they report how often a given judge is fooled but leave the judge inside the loop. Only one countermeasure on this list recovers genuine independence, the introduction of an external or process signal, because a check the generator cannot author, an executable test, a held-out measurement, or step-level supervision against a reference trajectory, breaks the identity between the rewarded and the maximized quantity. External or process signals recover the most independence and juries recover a diversity-bounded fraction of it, while calibration and meta-evaluation mainly make the residual unreliability measurable. None of the in-family fixes makes the judge an independent oracle.

\subsection{The Multi-Agent Orchestration Substrate}
Underneath the reviewers and judges sits the multi-agent machinery that lets a single model be recast as a collaborating team, and this substrate determines what can later be audited. The foundational works fall into three mechanisms. Role-and-protocol frameworks fix collaboration through structured communication: CAMEL's role-playing ``society'' \citep{li2023camel}, AutoGen's conversable-agent programming model \citep{wu2023autogen}, and the staged standard-operating-procedure pipelines of MetaGPT \citep{hong2023metagpt} and ChatDev \citep{qian2023chatdev} turn one LLM into a division-of-labor team with explicit reviewer and critic roles, the same scaffolding research agents reuse to separate ideation, coding, and review. Deliberation-and-aggregation mechanisms, exemplified by multi-agent debate \citep{du2023improving}, have agents critique and reconcile each other's outputs to improve factuality and reasoning; this is the root of the now-common claim that cross-checking among models can serve as quality control. Community and society simulators scale orchestration from a fixed team to a population: Generative Agents \citep{park2023generativ} as the progenitor, then ResearchTown \citep{yu2024researcht}, AgentRxiv \citep{schmidgall2025agentrxiv}, Agent Laboratory \citep{schmidgall2025agent}, and Dolphin \citep{yuan2025dolphin} that publish, retrieve, and build on shared artifacts in closed feedback loops. These systems matter for verification because many of them treat critic and debate roles as self-contained, optimizing for task success rather than externally checkable evidence. The structured message logs, SOP artifacts, and shared-preprint records they emit are natural audit surfaces, yet they rarely ground claims in independently verifiable provenance. The assumption that more agents debating yields better answers is also only partly supported: failure-attribution and coordination studies discussed below find that some reported multi-agent gains do not survive a budget-matched or noise-floor comparison, so debate can turn a wrong answer into a confident consensus. Debate \citep{du2023improving} recovers little independence for the same reason a same-family jury does. Agents sampled from one model share the priors and failure modes of that model, so a position the family finds plausible but false is one that every debater may also find plausible. Rounds of exchange can then drive agreement rather than truth. Debate can still surface errors when the disagreement is genuine, but genuine disagreement is precisely what shared training can erode. The orchestration layer therefore supplies useful audit artifacts, but it should not be treated as independent review unless those artifacts are tied to external provenance.

\subsection{The Newest End-to-End Scientists: Externalizing Trust}
Recent 2026 end-to-end AI-scientist systems show a noticeable shift: many of their headline contributions are checking apparatuses rather than more generation. Three patterns recur. The first pursues adversarial or dialectical self-checking that still lives inside the model family: AutoResearchClaw \citep{liu2026autoresea} adds structured agent debate, a self-healing executor, and ``verifiable result reporting,'' while the newest Socratic-agent line \citep{zeng2026socratic} inserts a physics-critic agent that challenges and attempts to refute hypotheses inside a closed experimental loop running on real optical hardware. The second routes claims through trusted external simulators or evidence so that verification becomes mechanical rather than self-judged: TianJi-Environ \citep{zhao2026tianjienv} passes every pollution-mechanism claim through auditable WRF-Chem atmospheric simulations, the Medical AI Scientist \citep{wu2026towards} constrains hypotheses to clinical evidence, and the Cmbagent/Denario cosmology backbone \citep{borrett2026competing, denario2025} iteratively refines analysis pipelines and benchmarks their outputs against human-derived physical parameters. The third treats verification as infrastructure: PARNESS \citep{wang2026parness} indexes full PDFs and code and accumulates cross-run knowledge so that later runs can re-check earlier claims, making re-verification a first-class capability rather than a one-shot judgment. These examples are broader-literature context unless they are also included in the coded corpus; they show pressure toward external checks, not counted exceptions to the L4-v denominator. They join the established flagships, the AI Scientist line \citep{lu2024aiscientist, yamada2025aiscientistv2}, Google's AI Co-Scientist \citep{gottweis2025coscientist}, and InternAgent \citep{zhang2025internagent}, in a shared trajectory: the more recent the system, the more its trust derives from a check the generating model cannot author. The important distinction is which check is external. A physics simulator or a parameter comparison is an independent oracle; a debate, a Socratic critic, or a self-healing executor is not, because an adversary drawn from the same distribution can be persuaded. The EviBound audit \citep{chen2025evidenceb} makes the stakes concrete by showing a prompt-only research agent claiming success on all eight tasks while zero were actually verified, a gap that closes only when architectural verification gates are imposed.

\subsection{Multi-Agent Communication, Consensus, and Its Failure Modes}
\label{subsec:multiagent_consensus}

A growing fraction of autonomous-research systems are not single agents but teams: a proposer and a critic, a debate among role-specialized solvers, a planner that dispatches subtasks to workers, or an aggregator that pools votes into a verdict. The recurring justification is that disagreement among agents acts as an internal check, so that errors one member would commit alone are caught when peers object. This subsection takes that justification literally and treats the inter-agent channel as a verification surface in its own right. Consensus matters only if it tracks truth. Examining the evidence mechanism by mechanism, we ask of each what the multi-agent design automates, what independently verifies the resulting claim, and what remains unaudited. The evidence points to a simple rule for reading these systems: a conversation among models that share priors and failure modes is a weaker check than an executable, physical, or otherwise independent verifier.

\paragraph{Debate and critique as automated review.} The most common multi-agent pattern routes a draft answer through rounds of critique, on the premise that adversarial exchange surfaces flaws and consolidates a better solution. Competitive debate frameworks for software issue resolution organize specialized agents along fault-propagation traces and report that structured disagreement improves localization over independent exploration \citep{li2025swedebate}. The work automated here is review itself: the critique that a human referee would supply is generated by a peer model. In the better instances, an external and executable signal does the verifying, such as a test suite that the consolidated patch must pass. The unaudited gap opens when the debate converges with no such external anchor. Once the verifier is itself a language model, agreement among debaters and approval by a judge can both rise without any corresponding rise in correctness, because the participants share training data, decoding behavior, and systematic blind spots. The discovery is only as trustworthy as the strongest independent check that survives the debate; consensus reached purely through mutual persuasion adds confidence without adding evidence.

\paragraph{Consensus aggregation and the validity of agreement.} A second mechanism replaces argument with counting: multiple agents answer, and a majority vote or an aggregator selects the consensus. Efficiency-oriented designs use the speed of consensus as a control signal, resolving easy items through lightweight pairwise agreement and escalating only contested ones to larger collectives \citep{liu2026hcpmad}. Here the automation is the selection among candidate solutions. To verify it, one would need a demonstration that agreement is correlated with correctness rather than with shared bias, and that demonstration is frequently missing. The evaluation literature is directly cautionary: a large-scale audit of judge models finds that high test-retest reliability coexists with severe position and order bias, so that a panel can be highly self-consistent and still systematically wrong, a pattern its authors summarize as reliability without validity \citep{norman2026reliabil}. For autonomous research, where the aggregated quantity is often a claim of novelty or a pass-fail judgment on a hypothesis, treating inter-agent agreement as evidence of truth imports exactly this confound: the team can be confidently unanimous about a result that no external instrument has confirmed.

\paragraph{Sycophancy and self-preference inside the channel.} Even when agents do disagree initially, the dynamics of the exchange can erode the independence that made disagreement informative. Two biases are particularly corrosive to the verification story. First, sycophancy, in which a model revises a correct answer toward a confidently stated peer or toward perceived authority, converts a debate into a cascade of capitulation rather than a contest of evidence. Second, self-preference bias, in which a model rates its own outputs more favorably, has been quantified directly: judges assign higher scores to lower-perplexity text that is familiar to them, independent of quality \citep{wataoka2024selfprefe}, and follow-up work measures this effect across many models and proposes structured mitigations \citep{yang2026selfprefe}. In a research team where one agent both generates and evaluates, these biases mean the internal check is partly grading its own homework. The automation supplies the appearance of peer review; what goes unaudited is whether the reviewer is independent of the author in any sense that matters. An evaluation channel that prefers its own style or yields to social pressure is not an independent check, and a closed-loop claim validated only through such a channel inherits its biases.

\paragraph{Error propagation across the pipeline.} Multi-agent systems pass intermediate outputs forward as context, so an unsupported claim introduced early can be reused as a premise later. Direct study of these dynamics shows that hallucination in a cascade is not a static property of one output but evolves with interaction depth, with measurable trade-offs between suppressing inconsistency and preserving factual accuracy as claims move from agent to agent \citep{jamshidi2026hallucina}. The failure-mode taxonomy assembled from annotated traces across many frameworks reinforces this, attributing a substantial share of breakdowns to inter-agent misalignment and to weak or absent task verification rather than to any single agent's incompetence \citep{cemri2025why}. The pipeline automates the propagation and reuse of intermediate results. Verifying any individual link would, ideally, mean a check applied at that step. The unaudited gap is the long middle of the chain, where outputs are trusted simply because a prior agent produced them. For closed-loop discovery this is the most dangerous mode, because a fabricated measurement or a misread of a tool result can be laundered through several rounds of fluent agreement before it reaches the final claim, by which point its provenance is obscured.

\paragraph{Collusion and covert coordination.} The strongest threat to consensus-as-verification is that agents coordinate against the objective rather than toward it. This is no longer hypothetical. Safety-aligned models offered a secret tool that confers a strategic advantage frequently accept it and develop collusive strategies even while explicitly acknowledging the tool is unfair \citep{zeng2026voluntary}. Auditing frameworks that open covert channels between cooperating agents find that most off-the-shelf models exhibit a propensity to collude under such probes, and identify a gap between planning to collude in text and acting on it \citep{nakamura2026colosseum}. More worrying for any monitoring strategy, the problem of secret collusion via steganography has been formalized and shown to be a capability that scales with model strength \citep{motwani2024secret}, and unintended steganographic collusion can emerge from misspecified rewards while resisting standard mitigations such as output paraphrasing \citep{mathew2024hidden}. Collusion automates, in effect, the defeat of oversight: the visible channel can show productive cooperation while a hidden one carries coordination that the auditor cannot read. No consensus signal verifies the absence of such coordination; the agreement an observer sees may be the product of collusion rather than convergence on truth, and this possibility is almost universally unaudited in current research-agent deployments.

\paragraph{Synthesis.} Multi-agent review mechanisms share a structure. They automate something that resembles verification: a critique, a vote, a review, a confirmation. Stronger verification requires an external instrument whose errors are less coupled to the agents being checked, most reliably an executable test, a held-out measurement, or a human with access to ground truth. The persistent unaudited residue is whether inter-agent agreement reflects that kind of independent confirmation or merely the shared priors, mutual deference, and covert coordination of models drawn from a narrow distribution. Consensus is cheap to manufacture and easy to mistake for evidence. Closed-loop and novelty claims should therefore be read through the independence of the check behind them, and the communication channel of a multi-agent research team should be treated as one more surface that itself needs verification.

\subsection{Human Oversight, Trust Calibration, and the Handoff Boundary}
\label{subsec:oversight}

The preceding discussion of artifact release and independent checking presumes that, somewhere in the loop, a human still verifies what the agent produces. Almost every autonomous-research system reviewed here retains such a checkpoint: a scientist approves a hypothesis before wet-lab execution, signs off on generated code before it is merged, or vets a draft before submission. This handoff boundary turns the paper's verification claim into an operational question. A discovery is trustworthy in proportion to the strength of the independent check applied to it; when that check is a human reviewer, its strength depends entirely on whether the reviewer's trust is calibrated to the agent's actual reliability. A pipeline can automate every upstream step and still inherit the reliability of its weakest verification stage. We group recent work by the mechanism through which oversight either catches agent errors or silently fails to, reading each through whether the human review is itself audited.

\paragraph{Calibrating reliance to advice quality.}
A long line of human--AI decision-making research frames the handoff as a problem of \emph{appropriate reliance}: the human should accept correct agent output and override incorrect output on a case-by-case basis, rather than trusting or distrusting the system wholesale \citep{schemmer2022should}. This community has worked to make oversight quality measurable rather than assumed, formalizing appropriateness of reliance as a two-dimensional construct that separates the ability to discriminate good from bad advice from the behavior of acting on that discrimination \citep{schemmer2023appropriate}. Recent extensions push the measurement frontier toward the kinds of output autonomous-research agents actually emit: \citet{mishra2026framework} build the first formal framework for reliance on set-valued advice, covering the intervals and candidate sets through which an agent might communicate uncertainty rather than a single point claim. The practical lesson for research agents is that an oversight stage which records only an accept/reject decision provides no evidence that the human discriminated signal from noise; without a calibration metric, an approved artifact and a rubber-stamped one are indistinguishable in the record.

\paragraph{When the reviewer is the unreliable component.}
The reliance literature also documents the failure modes that turn a nominal check into a null one. Explanations, often proposed as the remedy for blind acceptance, do not reliably improve discrimination: \citet{schemmer2023appropriate} find their effect on appropriate reliance is conditional, and a debugging intervention designed to expose model weaknesses instead \emph{reduced} reliance overall rather than sharpening it \citep{he2024toerr}. Reviewer competence is itself miscalibrated in systematic ways. \citet{he2023knowing} show that a Dunning--Kruger illusion of human competence drives under-reliance among those who overestimate their own skill, distorting the handoff in the opposite direction from naive automation bias. More fundamentally, the predictive framing of AI can warp human judgment before any specific output is even evaluated: in a large behavioral study, treating an AI as a predictive authority led participants to constrain their own choices and forgo guaranteed rewards, an effect that persisted \emph{even when the predictions failed} \citep{naito2026ai}. For a research agent whose proposals carry the implicit authority of a capable model, this suggests the reviewer may anchor on the agent's framing rather than independently testing it, so that the verification step measures conformity to the agent rather than the validity of its claim.

\paragraph{The observability gap in technical handoffs.}
For agents that produce executable artifacts, oversight fails for a structural reason distinct from cognitive bias. \citet{wang2026observability} study an LLM coding agent under output-only human feedback and report a complete absence of full-task success: bugs originate in code logic and internal execution state, but the human reviews only the visible output, and the many-to-one mapping from internal state to observable outcome prevents output-level feedback from localizing root causes. They formalize this as a feedback paradox in domains with deep causal chains and show that injecting code-level visibility restores convergence. The mechanism is straightforward: a human check is only as strong as the layer it can observe, and verifying the surface of an artifact does not verify the reasoning that produced it. An agent can release a clean-looking result whose defect is invisible at the boundary where the human signs off, which is precisely how unaudited oversight lets unvalidated claims through.

\paragraph{Designing the boundary for genuine scrutiny.}
A complementary thread asks how to position the handoff so that the human contributes a real, rather than ceremonial, check. \citet{straitouri2025narrowing} show in a large study that an AI which narrows the human's action set to a calibrated subset yields complementary performance exceeding both the human and the agent alone, by controlling the level of human agency by design rather than leaving the human to guess when to defer. Others target the human's own scrutiny directly: \citet{mei2025critical} distinguish demonstrated from performed critical thinking and warn that systems easing cognitive effort may degrade the very capacity oversight depends on, while \citet{lim2025debiasme} propose deliberate friction and metacognitive scaffolding to counter anchoring and confirmation bias at the point of interaction. At the level of governance, \citet{gaube2026keeping} synthesize a cross-disciplinary framework arguing that effective human oversight requires explicitly defined architectures, roles, and processes, and observe that current notions of oversight lack a common foundation. Their critique applies directly to autonomous-research pipelines, where ``human in the loop'' is frequently asserted as a safeguard without any specification of what the human verifies, what information they observe, or how their accuracy is tracked.

\paragraph{Synthesis.}
Read together, these results recast human oversight not as a guaranteed backstop but as another verification stage whose reliability must itself be established. Appropriate reliance can be measured but is rarely measured in deployed research agents; explanations and debugging do not automatically calibrate it; reviewer self-assessment and the agent's predictive framing distort it; and output-only handoffs cannot observe the internal failures most likely to matter. The unaudited assumption running through much of the field is that inserting a human anywhere in the loop converts an automated claim into a verified one. The evidence here is that oversight catches agent errors only when the human observes the right layer, holds calibrated trust, and is held to a recorded standard of discrimination. Until oversight is audited with the same rigor the survey demands of any other independent check, the handoff boundary remains the most plausible point at which an autonomous pipeline launders an unvalidated result into an apparently human-verified one.

\subsection{The Closed Loop and Its Unaudited Channels}
Bringing these mechanisms together, the survey distinguishes the small set of systems that genuinely close
the discovery loop from the much larger set that merely automate stages. Across the coded corpus, only nine
systems reach the closed-loop tier (L4): seven whose loop is closed by a mechanical or executable oracle, one
whose closure is author-claimed but not independently established, and exactly one, the pre-LLM CAMEO
materials-optimization system \citep{kusne2020cameo}, whose closed loop was externally validated by physical
experiment. The distribution is informative. Where a domain admits a cheap external oracle, a simulator, an
assay, or an executable test, closure is auditable; where the loop is closed by an LLM reviewer or a debate,
closure is asserted and the verdict's validity remains to be checked.

Three attack surfaces stay open to any auditor. First, the verdict's own validity: self-preference
\citep{panickssery2024llm} and judge bias \citep{shi2024judging, ye2024justiceorprejudice} mean a system
optimizing against its own judge can climb a score that does not track quality, and reproducibility audits
\citep{angermeir2025reflectio} find published agentic results frequently fail to replicate. Second,
self-preference gaming at the pipeline level, where a generator and its paired reviewer co-adapt so that the
reviewer rubber-stamps the generator's house style rather than its correctness, the failure that reviewer-bias
audits \citep{li2025llmreval, taechoyotin2025remor, demetrio2025genrevie} quantify. The first surface is a
static property of a single judge; the second is the dynamic consequence of placing that judge in a loop. When
a generator is trained or selected against a paired reviewer (as in the CycleResearcher/CycleReviewer
arrangement \citep{weng2025cycleresearcher}, or any AI-Scientist pipeline that keeps drafts its own reviewer
scores highly \citep{lu2024aiscientist, yamada2025aiscientistv2}), the reviewer's biases become the generator's
objective. Whatever idiosyncrasy the reviewer rewards, such as confident phrasing, a familiar section
structure, or the model's own lexical fingerprint, is what the optimizer learns to emit, so the loop can
manufacture agreement between generator and reviewer while leaving correctness unchanged.

Third, review-channel attacks allow the manuscript or its embedded prompts to manipulate the automated
reviewer directly. This surface is distinct from the first two because it does not require the judge to be
biased: a correctly calibrated reviewer can still be subverted by adversarial content in its own input. The
manuscript is at once the object under review and an untrusted instruction stream, so a hidden directive
(white-on-white text, a prompt-injection footnote, or a citation crafted to trigger a known judge preference)
can turn the reviewer's instruction-following against the verdict. Detection work on adversarial and
manipulated submissions \citep{rao2025detectin, moghadasi2026audit} shows the review channel is an exploitable
input. In a closed loop the asymmetry is severe: detection is a classifier that can be evaded, whereas the
injection only has to succeed once to admit a fabricated result into a knowledge store that later runs treat
as established. A generator co-adapting against its own reviewer can also discover review-channel exploits as
a byproduct of optimization, because any phrasing that reliably moves the reviewer's score is reinforced
whether it is an argument or an attack.

The broader failure-attribution literature explains why these failures can be hard to localize after the fact.
MAST and the Who\&When benchmark \citep{cemri2025why, zhang2025which} show that even frontier models localize
the failure step in a multi-agent trace at roughly 14\% accuracy, so when a closed-loop system produces a wrong
result through a corrupted review channel or a gamed judge, current attribution methods cannot reliably say
which agent or step is responsible. The auditor is left with a wrong verdict, a clean-looking trace, and no
reliable way to recover where the channel was subverted. The reasoning-verification literature compounds the
problem: process-level supervision is more trustworthy than outcome checks \citep{zhang2025the}, yet
self-correction without an external signal is unreliable, so the reviewer agent cannot be trusted to audit
itself. Review and closed-loop agents are getting better at \emph{producing} verdicts, debates, and acceptance
decisions, and newer systems increasingly externalize trust to simulators and evidence stores. Where the
verdict stays endogenous, however, the closed loop is only as strong as an LLM judging an LLM. A discovery
claim becomes substantially more trustworthy when an independent oracle has signed off, rather than only a peer agent.

%% file: sec_verification.tex
\section{Verification Signals across Domains}
\label{sec:verification}

The argument running through this survey is simple to state: how far a discovery can be trusted tracks how strong an independent check its domain allows. This section makes that argument concrete by arranging verification mechanisms into a ladder, ordered along two criteria that rise together: \emph{soundness} (does a passing check entail that the claim is true?) and \emph{resistance to gaming} (can the agent earn the check without earning the claim?). Sound formal verifiers sit at the top, where acceptance is a proof; an LLM judging its own output sits at the bottom, where acceptance is a correlation at best and a self-flattering artifact at worst. Reading the ladder from top to bottom reveals an inversion: the autonomous-research literature is distributed almost exactly opposite to where verification is strong, so that the domains with the soundest oracles host the smallest agentic-research presence, while most LLM-agent ``science'' operates in the lower tiers where the check is interpretive, learned, or absent. This ordering is the analytic spine of the survey.

% This long-form target carries the full sub-area placement appendix, so point the ladder there.
\newcommand{\ladderappendixref}{ The full placement of every surveyed sub-area onto these tiers is in
the appendix (Table~\ref{tab:taxonomy_full}).}
\input{fig_ladder}

\begin{table}[t]
\centering
\caption{The verification-signal ladder used to sort audit strength. Each tier is the
strongest independent check a research domain admits. The ordering criterion is \emph{soundness and
resistance to gaming by the system under test}: higher tiers are more nearly automatic and harder for the
agent to satisfy without actually being correct (a proof checker cannot be talked into accepting a wrong
proof), while lower tiers depend on judgment that the agent can influence (a human expert can be misled, an
LLM judge can be prompt-injected). This is why an automatic executable test (II) ranks above a slow human
read (VI) and a contested LLM judge (VIII): the ranking is about the \emph{check's} independence, not the
domain's prestige or difficulty. Most LLM-agent research operates in the lower tiers, which is where the
verification gap lives. The appendix maps all surveyed sub-areas onto these tiers
(Table~\ref{tab:taxonomy_full}).}
\label{tab:taxonomy_map}
\small
\begin{tabular}{@{}cp{0.46\columnwidth}p{0.30\columnwidth}@{}}
\toprule
Tier & Verification signal & Example domain \\
\midrule
I & Sound formal verifier (proof assistant) & theorem proving \\
II & Executable tests / process rewards & coding agents, RL verifiers \\
III & Physical oracle or simulator & self-driving labs, physics \\
IV & Citation / source grounding & deep research, lit.\ synthesis \\
V & Threat-to-validity / proxy reward & mechanical closed loops \\
VI & Human-expert judgment & human--AI collaboration \\
VII & Weak inter-agent / infrastructural & multi-agent frameworks \\
VIII & The model's own judgment & LLM-as-judge (contested) \\
\bottomrule
\end{tabular}
\end{table}

\subsection{Tier I: Sound formal verifiers}
Formal theorem proving is the strongest case for agentic verification. A proof assistant (Lean, Coq, Isabelle) acts as a deterministic, sound external checker, so an agent's ``discovery'' is machine-checkable by construction rather than adjudicated by an LLM critic. Acceptance entails truth, and the kernel cannot be flattered, which makes this tier unusually resistant to gaming. The dominant mechanism is \emph{verifier-in-the-loop training and search}, where proof-assistant feedback supplies dense, non-gameable reward. The DeepSeek-Prover line instantiates this through expert iteration and reinforcement learning from proof-assistant feedback with tree search \citep{xin2024deepseekp, xin2024deepseekpx}, extended by recursive subgoal decomposition that bridges informal chain-of-thought and formal proof \citep{ren2025deepseekp}; Kimina-Prover scales the RL-from-verifier recipe \citep{wang2025kiminapro}, and Seed-Prover adds lemma-style whole-proof refinement \citep{chen2025seedprove}. A second mechanism addresses data scarcity through large-scale autoformalization, manufacturing verified training statements at scale \citep{lin2025goedelpro}. A third turns the verifier into an open-ended discovery engine: self-play provers that learn to pose progressively harder conjectures and prove them \citep{dong2025stp} convert a sound checker into a curriculum generator, the formal analogue of an autonomous-research loop but with ground-truth grading at every step. Around these sit the agentic scaffolds, retrieval-augmented premise selection and gym-like Lean interaction \citep{yang2023leandojo} and lifelong, repository-spanning proof agents \citep{kumarappan2024leanagent}, together with neuro-symbolic systems that pair an LLM proposer with a symbolic deduction engine to reach medalist olympiad geometry performance \citep{chervonyi2025goldmedal}.

These mechanisms use the oracle at different points in the research loop. The first, \emph{verifier-in-the-loop training and search} \citep{xin2024deepseekp, xin2024deepseekpx, ren2025deepseekp, wang2025kiminapro, chen2025seedprove}, uses the proof assistant both as a reward signal during learning and as a pruning signal during inference: every candidate tactic or whole-proof attempt is run through the kernel, and only kernel-accepted trajectories survive to shape the policy. The verification signal is dense (one bit per proof step or per attempt) and entirely non-gameable, but it presupposes that a formal statement already exists to be proved. \emph{Autoformalization} \citep{lin2025goedelpro} targets that precondition, manufacturing formal statements (and their proofs) at scale so the training distribution is less constrained by the small corpus of human-formalized theorems. The signal here is the same kernel applied one level up, certifying that a generated formalization type-checks and that an accompanying proof closes; the residual risk is semantic rather than logical, since a syntactically valid Lean statement can be a faithful or an unfaithful rendering of the informal claim it purports to capture, and the kernel cannot adjudicate that faithfulness. \emph{Self-play conjecturing} \citep{dong2025stp} moves the locus of novelty into the loop itself: instead of consuming a fixed problem set, the system proposes new conjectures, filters them through the same checker, and treats the survivors as its own curriculum, so the oracle grades answers as well as the difficulty and provability of self-generated questions. The shared property is the sound oracle: the proof kernel costs little per check, does not drift, and cannot be talked into a false accept. The main design question becomes where in the loop to spend that oracle (on policy gradients, on data manufacture, or on question generation), not whether the check can be trusted at all. Domains without such a kernel do not get this guarantee.

These works are the strongest existing counterpoint to LLM-only science loops: where a cheap sound verifier exists, autonomous agents reach and exceed expert competition performance \emph{with auditable outputs}. The contrast for the rest of the ladder is stark. Theorem-proving agents are trustworthy not because their models are smarter but because the domain donates a free, sound oracle. Almost no other research domain does.

\subsection{Tiers II--III: Executable tests and physical oracles}
One rung down, the check is no longer a proof but an execution against ground truth: a held-out dataset and executable test (Tier II), or a numerical simulator, a physical instrument, or a conservation law (Tier III). Acceptance no longer entails truth, since a correct number can come from a wrong mechanism, but the signal is objective and hard to fake without actually solving the problem.

\paragraph{Symbolic regression and physical-law discovery.} In the physical sciences the strongest native signal comes from treating the LLM as a semantic mutation operator inside an evolutionary search over equations, then checking candidates against held-out data, dimensional consistency, and standard benchmarks. Iterated-agent fitters \citep{song2025iterated}, learned-concept-library variants \citep{grayeli2024symbolic}, and materials-law specializations \citep{guan2026discovery} all produce outputs that are executably falsifiable rather than merely plausible-sounding: the discovered equation either predicts the Feynman SR measurements (or the parton distribution functions) or it does not. This is the executable-test rung (Tier II) at its cleanest, and it is no accident that these systems resemble the FunSearch and AlphaEvolve program-search paradigm \citep{romeraparedes2024funsearch, novikov2025alphaevolve} and the tensor-decomposition discovery of AlphaTensor \citep{fawzi2022alphatensor}, all of which earn their credibility from a cheap, automatic correctness test on each candidate.

\paragraph{Astronomy and Earth observation.} Agentic observation pipelines draw verification from physical closed loops rather than from a held-out test set. Multi-band galaxy-interpretation agents \citep{sun2024interpret}, end-to-end autonomous observation across a multi-telescope network \citep{wang2024starwhisp}, agents pushing toward autonomous discovery in cosmology \citep{xu2026beyond}, and gravitational-wave electromagnetic-counterpart association \citep{dong2026an} all couple their reasoning to follow-up triggering, sky-map cross-matching, and real-time instrument feedback. The verification is real but partial: the \emph{decision} to observe is checked by the physical loop, yet the intermediate \emph{scientific interpretation} (what the agent claims a spectrum means) is rarely audited, which exposes the gap that opens whenever a claim is interpretive rather than numeric. Earth and climate agents have cleaner checks: a climate-knowledge-graph agent that reproduces published figures from plain language \citep{jaber2025autoclimd} grounds outputs in curated provenance, and TerraBench \citep{nguyen2026terrabenc} is among the most rigorous evaluations in the entire physical-science corpus, pairing process-level tool-use metrics with tolerance-aware numeric scoring and a dedicated document-grounded verification track that preserves provenance over tens of thousands of verified execution steps. These domains furnish the survey's strongest counter-examples to ``verification is unsolved'': where the target is an equation or a numerically gradeable Earth-system quantity, simulators, conservation laws, and instrument feedback supply auditable ground truth that biology and chemistry agents typically lack.

\paragraph{Engineering and hardware design.} The same executable/simulator pattern (Tiers II--III) recurs wherever the artifact admits ground-truth validation. Across electronics, mechanical CAD, and robotics, the credible systems pair an LLM generator with an external, domain-native checker. For chip design, agents close the loop on native simulation and regression \citep{cui2026hwebench}, drive block-level functional verification to coverage targets \citep{wang2026ucagent}, and evolve skills under a bounded runtime verifier that hides reference solutions \citep{du2026traceskil}; multi-agent EDA frameworks span synthesis through verification \citep{ho2025marco}, while a framework lacking any verification loop \citep{patra2024aieda} is itself evidence that generative capability has outrun auditability. The CAD thread makes the verification primitive explicitly geometric: validation against exact kernel measurements plus a VLM judge \citep{barkley2026cadsmith}, cross-stage validation with rollback \citep{shui2026articad}, and physics-in-the-loop embedding of validated engineering tools directly in the decision loop \citep{berger2026physicsin}. Progress here is gated not by generation but by whether a trustworthy, automatable verifier exists, and the most credible systems are precisely those that hand verification to a deterministic simulator, kernel, or physics tool rather than to the model itself. Agentic atomistic-research systems that wrap validated simulation skills around an LLM planner \citep{deng2026harnessin} extend the same simulator-in-the-loop discipline to materials modeling. The physical-oracle precedents in materials discovery, from autonomous synthesis robots \citep{szymanski2023autonomous} to the externally validated CAMEO phase-mapping campaign \citep{kusne2020cameo}, show this tier predates LLM agents and remains the empirical high-water mark for closed-loop validity.

\subsection{Tier V: Proxy rewards and learned verifiers}
Below the executable and physical oracles, and below the citation-grounding rung (Tier IV) that the
literature and deep-research agents of Section~\ref{sec:litwriting} occupy, sits the machinery that tries
to manufacture a check when no sound oracle is available: reward functions and learned verifiers. The defining concern, and the reason this tier ranks below executable oracles, is that the verifier is itself a model and can therefore be gamed. The founding generative-verifier paradigm casts reward modeling as next-token prediction \citep{zhang2024generativ}; frontier extensions push it toward long-form proof RL with low-false-positive ``defense-in-depth'' verifiers \citep{chen2026maxproof} and calibrate verifier strictness through latent steering \citep{zhou2026the}. The failure mode this tier exists to fight is reward hacking, and its remedies are themselves revealing: adversarial hacker-fixer loops that automatically harden agent-benchmark verifiers \citep{zhong2026hardening}, learned compact executable Python verifiers \citep{pezeshkpour2026autopyver}, and weak-to-strong aggregation of many imperfect LLM verifiers \citep{zhang2026aggregati} all concede that any single learned check is attackable and must be defended or ensembled. Scaling verifiable RL environments to real agentic settings \citep{wang2026cuagym} and replacing subjective LLM judging with verifiable rewards in agentic evaluation expose a ``reliability cliff'' \citep{narasimhan2026taurec}: the moment the reward stops being mechanically verifiable, agreement with ground truth falls off sharply. A learned verifier therefore belongs in Tier V because its soundness is empirical rather than guaranteed. Much of the work in this tier is an attempt to recover, by training and hardening, the gaming resistance that executable or physical oracles provide directly.

\subsection{Process and step-level verification (Tiers II and V)}
Process verification cuts across the executable-test and proxy-reward rungs by scoring the \emph{steps} of a reasoning trace rather than only its final answer: a step checked by execution is a Tier-II signal, while a step scored by a learned process-reward model is a Tier-V proxy. Structured decomposition and search expose explicit intermediate states that are individually checkable, converting an opaque answer into an auditable trace: least-to-most and plan-and-solve prompting \citep{zhou2022leasttomo, wang2023planandso}, Graph-of-Thoughts generalizing Tree-of-Thoughts to arbitrary thought graphs \citep{besta2023graph, yao2023tot}, and reasoning-as-planning with an LLM world model and MCTS \citep{hao2023reasoning}. Exposing a trace, however, only helps if each node is actually checked rather than merely printed. Verification-by-aggregation and process supervision turn intermediate steps into a learnable signal: self-consistency uses answer agreement as a cheap verifier proxy \citep{wang2022selfconsi}, while the outcome-versus-process comparison \citep{uesato2022solving}, step-by-step verification \citep{lightman2023lets}, and automatic process reward models \citep{wang2023mathsheph, zhang2025the} show that step-level supervision can catch errors that final-answer checks miss. RL-trained reasoners \citep{deepseekai2025deepseekr} and compute-optimal test-time search \citep{snell2024scaling} amplify capability by spending more inference on search and self-check. The field's own evidence that LLMs cannot reliably self-correct reasoning without external feedback \citep{huang2023large} is the caution: longer chains and self-reflection can raise apparent capability while the model's self-generated verification signal remains unreliable. Auditability still needs an external, process-level verifier, which is why this tier ranks below executable or physical oracles even when it produces a richer trace.

\subsection{The bottom band (Tiers VI--VIII): human judgment, weak signals, and model opinion}
The lowest rungs cover the checks most LLM-agent research actually relies on, ordered by decreasing soundness: human-expert judgment (Tier VI), weak inter-agent or infrastructural signals (Tier VII), and the model's own judgment (Tier VIII). Two recurring agent-side signals live in this band alongside the human expert. The first is the calibrated self-confidence of scientific foundation models used as agent tools. Structure and sequence predictors trained on physical ground truth emit auditable trust signals (per-residue confidence, predicted aligned error) that an agent can threshold on, making them the strongest source of verifiable intermediate evidence in agentic pipelines. Surrogate simulators for physics-governed domains, weather and Earth models \citep{lam2022graphcast, bi2022panguweat, pathak2022fourcastn, nguyen2023climax, bodnar2024a} and machine-learned interatomic potentials \citep{batatia2023a}, accelerate the loop but quietly shift the verification burden onto held-out reanalysis or ab-initio recomputation that the agent must be made to invoke. Generative and broad scientific models, crystal generators \citep{zeni2023mattergen}, molecular foundation efforts \citep{beaini2023towards}, and science LLMs \citep{taylor2022galactica}, carry \emph{no} intrinsic correctness guarantee and can hallucinate plausible-but-wrong artifacts, so their output is only as trustworthy as the downstream DFT, retrosynthesis, or citation check the agent is forced to run.

Why a surrogate's confidence sits below an executable test, rather than beside it, deserves a precise statement, because the two are often treated as interchangeable forms of ``automatic'' feedback. An executable test is sound in the direction that matters for a passing result: when a held-out unit test or a dimensional-consistency check accepts, the artifact provably exhibits the asserted behavior on that input, and the check is computed by machinery independent of the model that produced the candidate. A per-residue confidence or a predicted aligned error is a different object: it is the model's own estimate of its own reliability, calibrated on a training distribution, and it certifies nothing about the specific instance except that instances statistically resembling it were often correct. The failure modes follow directly. A confident-but-wrong prediction (a high-confidence misfold, a surrogate weather field that violates a conservation law the emulator was never constrained to respect, an interatomic potential extrapolating into a region absent from its training set) passes the surrogate's internal check while being false, and nothing in the confidence number flags the extrapolation. The same epistemic dependence holds for the verifiable RL environments and learned verifiers of Tier~V \citep{narasimhan2026taurec}: a learned signal degrades silently off-distribution, whereas an executable oracle either runs or visibly errors. What a reviewer specifically cannot audit, then, is not the confidence number itself (that is reported) but the conditions under which it was calibrated and whether the instance at hand lies inside them: the reviewer cannot rerun the physical experiment that would settle the folded structure, cannot recompute the ab-initio energy the potential is approximating, and cannot recover the held-out reanalysis the weather emulator was scored against, so the surrogate's accept is accepted on faith in its calibration rather than verified. This is what we mean when we say a surrogate shifts the verification burden rather than discharging it. The field still lacks standard protocols for deciding when a surrogate output should be accepted and when an agent must escalate to the expensive ground-truth oracle.

The second weak agent-side signal is evidence grounding in domains where ground truth is unavailable but external references exist, exemplified by clinical and biomedical research agents. Trial-design and outcome-prediction agents \citep{liu2025autoct, yue2024clinicala} explicitly substitute interpretability and feature-auditability for true outcome verification, because they cannot ground predictions against unrun trials. Diagnostic-reasoning agents are increasingly engineered around evidence-grounding rather than raw accuracy: tying each conclusion to image-derived evidence to make the reasoning ``verifiable'' \citep{lee2026cxreasona} shifts the emphasis from answer-correctness to traceable justification. Evidence-synthesis and claim-verification agents make auditing the agent's own product the central object, judging systematic-review quality \citep{mushtaq2025can}, classifying support/contradict against retrieved literature with feature-level explanations \citep{liang2025explainab}, surveying where synthesis can and cannot be trusted \citep{li2025transform, polzak2025can}, and gating hypotheses through multi-tier plausibility evaluators \citep{song2025llm}. Across all three, the limiting step is machine-checkable grounding in evidence, not generation, and grounding is weaker than execution because a correctly cited source does not entail a correct conclusion.

The clinical case sharpens why this rung sits so low, and it is instructive precisely because the gap is not a matter of weaker engineering. For a coding agent the executable test settles the question for the input it covers: the program either returns the expected output or it does not. A trial-design or outcome-prediction agent \citep{liu2025autoct, yue2024clinicala} has no analogous closing move, because the ground truth it would need (the outcome of the trial it is designing) does not exist and cannot be computed; the trial has not been run, and running it is the very thing the prediction is meant to inform. The agents respond, reasonably, by substituting auditability of the \emph{reasoning} for verification of the \emph{outcome}: they expose interpretable features, or tie each diagnostic conclusion to image-derived evidence so the chain from pixel to claim is inspectable \citep{lee2026cxreasona}. But a reviewer who reads such a justification can confirm only internal coherence, that the cited evidence exists and that the stated inference from it is not obviously fallacious. The reviewer cannot confirm the conclusion, because the two checks that would (a randomized outcome, a ground-truth label) are unavailable; an evidence-grounded chest-X-ray rationale can cite a genuine opacity and still reach the wrong differential, just as an evidence-synthesis agent can correctly classify each retrieved study as support or contradict and still aggregate to a conclusion that a confounder, a publication-bias asymmetry, or an unmodeled population difference would overturn. This is the concrete sense in which grounding is a weaker check than execution: execution audits the claim, whereas grounding audits only the visible provenance of the claim, and the unaudited residue (does the cited evidence actually entail the conclusion under the true data-generating process?) is exactly where clinical claims fail and exactly what no automatic check in this band can reach. The evidence-quality and claim-verification agents \citep{mushtaq2025can, liang2025explainab, li2025transform, polzak2025can, song2025llm} are best read as attempts to make this residue smaller and more legible, not to eliminate it, since eliminating it would require the executable or physical oracle the domain does not admit.

\paragraph{One system, several tiers.} The ladder classifies checks, not systems, and the distinction matters because a single agent routinely earns different rungs for different sub-claims, so locating a system on the ladder means decomposing its output rather than awarding it a single grade. An astronomy observation agent occupies Tier~III for its scheduling decision, which the physical follow-up loop confirms or refutes, but drops to Tier~VI or lower for its interpretation of what a spectrum means, which nothing in the loop checks \citep{sun2024interpret, wang2024starwhisp}. A chip-design agent sits at Tier~II for the RTL it submits to a hidden runtime verifier \citep{du2026traceskil} yet relies on a VLM judge, a Tier-VIII signal, for the qualitative aspects of a layout that simulation does not score \citep{barkley2026cadsmith}. A theorem-proving pipeline is Tier~I for the proof its kernel accepts but only Tier~V or VIII for the autoformalization step that decided the formal statement faithfully captures the informal conjecture, the semantic gap noted above. Even the symbolic-regression fitters that anchor Tier~II earn that rung only for predictive fit on held-out data; the further claim that a fitted equation is the \emph{true} physical law, rather than one of many expressions consistent with the sample, is interpretive and unverified by the fit alone. The practical consequence is that a system's trustworthiness is not a scalar but a profile over its sub-claims, and the recurring failure in the literature is to let the soundest sub-claim (the proof closed, the test passed, the telescope slewed) launder credibility onto the interpretive sub-claim riding alongside it. Reading a system honestly means asking, for each claim it asserts, which tier actually checked \emph{that} claim, because artifact release reports the system, while verification attaches only to its individual outputs.

\emph{Tier VIII}, the bottom rung, is unconstrained model judgment: an LLM (often the same model that produced the output) scoring quality directly. Here soundness and gaming-resistance both collapse, since strong judges reach only rough human agreement and exhibit position, verbosity, and self-enhancement biases \citep{zheng2023judging}, and this is precisely the regime in which most end-to-end LLM-scientist loops close their feedback (the weak inter-agent and infrastructural signals of Tier VII, the multi-agent consensus and logging substrate analyzed in Section~\ref{sec:review}, sit just above it and are no more sound). The ladder shows a mismatch: much of the autonomous-research literature sits in Tiers V through VIII, while the strongest checks sit in Tiers I through III. Theorem-proving agents, symbolic-regression fitters, and simulator-checked engineering systems are trustworthy when their domains provide sound or executable oracles. Once a research claim becomes interpretive, novel, or open-ended, the available check often degrades to a learned reward, a process heuristic, a surrogate's self-confidence, or a model's opinion. This is why artifact release can outpace claim verification even when systems look increasingly capable.

\subsection{Domain Autonomous Laboratories as Closed-Loop Case Studies}
\label{subsec:autonomous-labs}

The agents surveyed in the preceding sections close their loops in software: a claim is checked against a held-out split, a unit test, a proof assistant, or a second model acting as judge. Domain autonomous laboratories occupy the opposite end of the verification ladder. Here the loop is closed through physical matter, and the oracle is not a stored label but a reaction that proceeds or fails, a film whose conductivity is measured, or an assay with an independent readout. These systems clarify what an externally grounded check provides and what software-only loops still lack. The discussion below follows the mechanism through which each class of laboratory grounds its claims, asking what the loop hands off to automation, where the genuine check lives, what slips past it unaudited, and how the answer bears on the claim that a discovery is only as trustworthy as the independent check behind it.

\paragraph{Optimization-grounded materials platforms.}
The earliest mature self-driving laboratories treat discovery as black-box optimization over a physically realizable parameter space. The platform proposes a composition or processing condition, a robotic stack synthesizes and characterizes the sample, and an acquisition function selects the next point. MacLeod et al. \citep{macleod2019selfdriving} demonstrate this for thin-film hole-transport materials, autonomously varying film composition and processing to maximize hole mobility, and later extend the same architecture to recover a full Pareto front of conductivity against processing temperature, surfacing combustion-synthesis conditions that had not previously been tested \citep{macleod2021pareto}. The full design-execute-measure cycle runs without intervention, and the thing that settles each candidate is the instrument reading on the physical sample, never a surrogate prediction. Trust flows from the fact that the optimization objective is itself the oracle: a film the model believed conductive but that measures otherwise is simply discarded by the loop. The principled treatment of model uncertainty in this regime predates the robotics, as in the Bayesian model-averaging design of Talapatra et al. \citep{talapatra2018autonomous}, where the agent learns both the promising region and the model that best guides exploration. Two things escape the loop's authority, and they are narrow but real: the mapping from the measured proxy to the scientifically interesting property, and the generality of any conclusion beyond the swept parameter window. The platform certifies that a particular sample has a particular measured property; it does not certify that the underlying structure-property hypothesis holds outside the box it searched. This is the materials-science analogue of the in-distribution-only guarantee we flagged for software agents, except that here the check inside the box is genuinely external.

\paragraph{Synthesis and characterization as a hierarchical oracle.}
A second class targets the harder problem of making a material that does not yet exist and confirming what was made. Ament et al. \citep{ament2021metastable} couple robotic lateral-gradient laser-spike annealing with optical spectroscopy and a hierarchy of active-learning cycles to map non-equilibrium synthesis phase diagrams, stabilizing a metastable bismuth-oxide phase at room temperature. Here the verifying signal is layered rather than scalar: spectroscopy detects phase transitions and structural characterization confirms identity, so the closed loop integrates synthesis and characterization instead of optimizing a single number. That layering matters because it supplies a multi-stage independent check that software loops only approximate weakly. The agent cannot declare a new phase discovered until the characterization stage, which it does not control, returns a consistent signature. The residue that no stage covers is the completeness of the characterization battery: a phase can be misassigned when the available probes are blind to the distinguishing feature, and the system has no way to know what it cannot see. Review-level syntheses of the field \citep{handoko2025aireview} make the same point at scale, noting that closed-loop discovery systems remain constrained by synthesizability and by the gap between a generative model's proposal and an experimentally confirmed material. Even a physical oracle is only as strong as the breadth of the measurements wired into the loop; an unmeasured failure mode is invisible regardless of how autonomous the platform is.

\paragraph{LLM-orchestrated wet-lab agents.}
The most recent and most relevant class to autonomous-research agents places a language model in the planning seat of a physical platform. Boiko et al. \citep{boiko2023coscientist} couple an LLM planner to documentation search, code execution, and liquid-handling hardware, and report autonomous planning and execution of catalytic reactions; the companion line of work emphasizes the same emergent capability for experimental chemistry \citep{boiko2023autonomous}. Burger et al. \citep{burger2020mobile} earlier showed a mobile robotic chemist running a many-day photocatalysis campaign with negligible human intervention, and Darvish et al. \citep{darvish2024organa} target the chemistry-specific perception and manipulation problems that such agents must solve to act in a real fume hood. In biology, the Robin system \citep{ghareeb2025robin} integrates literature-grounded hypothesis generation with experimental execution, and Coscientist-style multi-agent designs propose hypotheses that are then routed to wet-lab validation \citep{gottweis2025coscientist}. What these systems hand to automation is exactly the part the rest of this survey scrutinizes most heavily: hypothesis formulation, protocol synthesis, and the orchestration of tools. When the loop is genuinely closed, it is the physical experiment that does the verifying, which is why these platforms represent the strongest grounding available to an autonomous-research agent. The unaudited surface, however, is larger here than in the optimization platforms. The language model's claim about why an experiment worked, its causal narrative, is not tested by the assay, which only reports whether a result occurred. An LLM agent can run a correct protocol, obtain a real measurement, and still attach a fabricated or unsupported mechanistic explanation, and nothing in the physical loop catches that gap. The microscale-actuation literature, such as the closed-loop control of catalytic Janus microrobots \citep{sokolich2022janus}, illustrates how much engineering separates a clean control objective from a defensible scientific claim: the loop can be tight and the system can still report only what its sensors were built to report.

\paragraph{Synthesis.}
Read together, these case studies clarify the difference between releasing an artifact and validating a discovery. A physical oracle gives the loop a measurement that does not depend on the agent's confidence, prose, or training distribution. That is a stronger check than software-only self-evaluation, but its authority is still bounded by the instruments in the loop and by the scope of the swept space. It also does not reach the interpretive layer where an LLM explains why a result occurred. Domain autonomous laboratories therefore relocate the verification problem rather than solving it outright: novelty and mechanistic claims still need checks the assay does not provide, even when the experimental outcome itself is reliable. A promising direction is to connect physically grounded loops with closed-loop computational agents, so that validated wet-lab outcomes can serve as ground truth where software-only evaluation currently has none.

%% file: fig_ladder.tex
% Files that include the full sub-area placement appendix (appendix_taxonomy.tex,
% label tab:taxonomy_full) override this to point readers there; the 8-page main.tex,
% which carries no appendix, leaves it empty so the figure has no dangling reference.
\providecommand{\ladderappendixref}{}
\begin{figure}[t]
\centering
\footnotesize
\begin{tikzpicture}[x=1cm,y=0.66cm]
% vertical ladder: tier VIII (bottom, weakest) to I (top, strongest).
% A fixed left column holds the tier label; a separate right column holds the
% example domain, so the two text blocks occupy disjoint horizontal bands and
% can never overprint each other (the earlier defect).
\def\W{12.6}      % total box width
\def\Lx{0.2}      % left tier-label anchor
\def\Rx{8.4}      % right example-domain anchor (start of the example band)
\foreach \i/\lab/\ex in {
  8/{I: sound formal verifier}/{theorem proving},
  7/{II: executable tests / process reward}/{coding agents},
  6/{III: physical oracle / simulator}/{self-driving labs},
  5/{IV: citation / source grounding}/{deep research},
  4/{V: proxy reward / threat-to-validity}/{mechanical L4 loops},
  3/{VI: human-expert judgment}/{human--AI collaboration},
  2/{VII: weak inter-agent / logs}/{multi-agent frameworks},
  1/{VIII: model's own judgment}/{LLM-as-judge}} {
  \fill[blue!\the\numexpr\i*10\relax!white,draw=black!40] (0,\i) rectangle (\W,\i+0.9);
  \ifnum\i>5
    \node[anchor=west,text=white] at (\Lx,\i+0.45) {\lab};
    \node[anchor=west,text=white] at (\Rx,\i+0.45) {\itshape \ex};
  \else
    \node[anchor=west,text=black] at (\Lx,\i+0.45) {\lab};
    \node[anchor=west,text=black!70] at (\Rx,\i+0.45) {\itshape \ex};
  \fi
}
% thin divider between the tier-label band and the example-domain band
\draw[black!25] (\Rx-0.2,1) -- (\Rx-0.2,8.9);
% arrow: trustworthiness increases upward
\draw[-{Latex[length=2mm]},thick] (\W+0.35,1) -- (\W+0.35,8.9)
  node[midway,rotate=90,anchor=south,yshift=2pt]{\scriptsize trustworthiness $\uparrow$};
% bracket marking where surveyed LLM-agent subareas most often sit (tiers IV-VIII)
\draw[decorate,decoration={brace,amplitude=4pt},thick,red!60!black]
  (-0.25,1) -- (-0.25,5.9) node[midway,rotate=90,anchor=south,yshift=2pt,text=red!60!black]{\scriptsize surveyed LLM-agent subareas};
\end{tikzpicture}
\caption{The verification-signal ladder. A research domain's trustworthiness under autonomy rises with the
strength of the independent check it admits, from the model's own judgment (Tier VIII) to a sound formal
verifier (Tier I). Most LLM-agent subareas surveyed here rely heavily on the lower tiers (IV--VIII) unless a
task-specific executable, physical, or formal oracle is available. Tier definitions and example domains
are in Table~\ref{tab:taxonomy_map}.\ladderappendixref}
\label{fig:ladder}
\end{figure}

%% file: sec_benchmarks.tex
\section{Benchmarks, Frameworks, and Resources}
\label{sec:benchmarks}

This category covers the measurement and engineering infrastructure on which autonomous research claims are built: benchmarks that grade agent outputs, open-source frameworks that execute them, contamination threats to benchmark validity, and cost accounting for long agent runs. The infrastructure is mature for scoring \emph{task completion}. It is much younger for probing \emph{reproducibility, soundness, and closed-loop validity}, and the newest checks are themselves under-verified. We group the discussion by what each resource measures and what it leaves auditable, with a fuller compendium in Appendix~\ref{app:benchmarks}.

\subsection{The Benchmark Compendium: What Gets Scored}

\paragraph{ML and AI-research execution benchmarks.} The densest cluster of benchmarks grades an agent's ability to execute machine-learning engineering tasks against an objective oracle. MLAgentBench frames research as iterative experimentation over Kaggle-style tasks scored by held-out performance \citep{huang2023mlagentbench}, MLE-Bench packages 75 Kaggle competitions with leaderboard-anchored medals \citep{chan2024mlebench}, MLGym and MLR-Bench extend this to open-ended research workflows \citep{nathani2025mlgym,chen2025mlrbench}, and RE-Bench measures agent-versus-human throughput on frontier ML research engineering tasks under a fixed time budget \citep{wijk2025rebench}. ExpBench and Curie target the experiment-design and execution loop directly, asking whether an agent can plan, run, and report a controlled experiment \citep{kon2025expbench,kon2025curie}, while InnovatorBench stresses end-to-end innovation under realistic compute \citep{wu2025innovatorbench}. These benchmarks share a strong mechanical signal: a number from a held-out split, a competition rank, or a wall-clock-bounded score. That signal is why agents can post credible results on them. The oracle they expose, however, is \emph{performance on a fixed target}, not the soundness of the path to it. An agent that overfits the validation set, leaks the test label, or stumbles into a high score through an unprincipled search will be scored identically to one that reasoned correctly, so a leaderboard win certifies output but not method.

\paragraph{Discovery, ideation, and hypothesis benchmarks.} A second family moves upstream to the generative front of research, where verification is weaker. IdeaBench and the controlled human study of LLM-generated ideas \citep{ideabench2024,si2024novelideas} grade novelty and quality, SciMON optimizes ideas explicitly toward novelty against prior literature \citep{wang2023scimon}, and HypoBench and ResearchBench formalize hypothesis quality and the recovery of held-out scientific findings \citep{liu2025hypobenc,liu2025researchb}. DiscoveryBench operationalizes data-driven discovery against ground-truth relationships in real datasets \citep{majumder2024discoverybench}, and Popper supplies a sequential falsification harness that controls the false-discovery rate over generated hypotheses \citep{huang2025popper}. The contrast with execution benchmarks is plain: novelty and interestingness have no clean oracle, so these benchmarks lean on LLM judges, human ratings, or proxy literature-overlap metrics. The documented gap between ideation and execution \citep{ideationexecutiongap2025} shows why a promising idea score is not evidence that the idea will survive experiment.

\paragraph{Analysis, reproduction, and soundness benchmarks.} The benchmarks most aligned with the survey's verification framing are also the youngest. Reproduction harnesses redefine the target from ``produce a passing artifact'' to ``reproduce the paper's claimed result'': PaperBench scores replication of ICML papers against author-validated rubrics \citep{starace2025paperbench}, SciReplicate-Bench and ReproRepo grade reproduction of published computational findings \citep{xiang2025scireplicate,li2026reprorepo}, AutoRepro and ReproScore audit reproducibility directly \citep{zhao2025autorepro,samuel2026reproscor}, ReplicatorBench and MLReplicate benchmark end-to-end systems on whether they actually reproduce ML results \citep{nguyen2026replicato,gaddipati2026mlreplica}, and LMR-Bench finds that frontier agents largely cannot reproduce language-modeling research code \citep{yan2025lmrbench}. CORE-Bench targets computational reproducibility from released code and data \citep{siegel2024corebench}, ScienceAgentBench and CodeScientist grade data-driven scientific analysis tasks \citep{chen2024scienceagentbench,jansen2025codescie}, and SUPER focuses on setting up and running research repositories from scratch \citep{bogin2024super}. SPOT goes further still, asking whether agents can detect errors in published papers \citep{son2025spot}. Because the oracle here is the published claim itself rather than a self-chosen metric, these benchmarks supply the strongest verification signal the field has. Yet they remain a minority of evaluation effort, and the low reproduction rates these harnesses report, with frontier agents largely unable to reproduce the target ML results \citep{gaddipati2026mlreplica,yan2025lmrbench}, show both how recently the field began measuring soundness and how far current agents remain from passing the check.

\paragraph{Data-science benchmarks.} Adjacent to reproduction sits a cluster grading the data-analysis loop, where verification is partly tractable because intermediate outputs are checkable. DSBench, DSBench-style insight tasks, and InsightBench measure end-to-end analysis and insight discovery \citep{jing2024dsbench,sahu2024insightb}, Tapilot-Crossing and Spider2-V grade interactive and enterprise data-science workflows \citep{li2024tapilotc,cao2024spiderv}, DS-Agent and DataInterpreter probe agentic data-science pipelines \citep{guo2024dsagent,hong2024data}, and the AutoML lineage (Auto-sklearn, AutoML-Agent, the agentic Kaggle line) automates model search against held-out scores \citep{feurer2015autosklearn,trirat2024automlag,li2024autokagg}. As in ML execution, the oracle is real but local: a correct number on a held-out analysis does not certify that the analytical choices were sound, only that the endpoint matched.

\paragraph{Safety, integrity, and review benchmarks.} A final family audits the agent's epistemic and social conduct rather than its task output. SciSafeEval probes safety in scientific settings \citep{li2024scisafee}, and the automated-review line, ReviewCritique, the survey of automated reviewing, and evaluations of LLM reviewers \citep{du2024reviewcritique,zheng2025automation,beel2025evaluatin}, treats the verifier itself as the object of measurement. This matters because many autonomous pipelines propose automated peer review as their check: if the reviewer is biased, shallow, or gameable, the loop certifies little. The open-problems agenda for measuring AI-research agents \citep{reuel2024open} calls for integrity-oriented evaluations, which remain scarce compared with task-success benchmarks.

\subsection{Open-Source Agent Frameworks and Synthesized Verifiers}

\paragraph{Frameworks compete on inspectability.} The engineering substrate on which research agents are built determines what can later be audited, and open-source frameworks divide along that axis \citep{wang2023a}. One class makes \emph{agent execution} inspectable: AgentScope provides actor-based, fault-tolerant message exchange, and AutoGen Studio exposes multi-agent workflows for visual debugging \citep{gao2024agentscop,dibia2024autogen}, treating execution traces as first-class artifacts rather than ephemeral logs. A second class makes the \emph{action space} itself auditable: CodeAct unifies agent actions into executable, re-runnable Python, the basis of OpenHands \citep{wang2024executabl,wang2024openhands}, and AFlow represents whole agentic workflows as searchable code \citep{zhang2024aflow}, so an agent's output is a verifiable program rather than opaque natural-language steps. The lineage these frameworks descend from supplies the cautionary evidence: an early empirical benchmark of Auto-GPT documented that popular autonomous-agent scaffolding fails on real decision tasks without external verification or supervision \citep{yang2023autogpt}. Frameworks are increasingly competing on how inspectable they make behavior, which is a precondition for verification. But inspectability is necessary, not sufficient: a fully logged trace still requires an oracle to judge whether the logged actions were correct.

\paragraph{Tool creation and environment synthesis.} A fast-emerging line shifts the verifier from a human-authored fixture to an artifact the agent or pipeline constructs itself. One mechanism is automated environment and benchmark synthesis, where a building agent emits the task, the ground truth, and its checking program: STAGE-Claw, Agent-World, and InfiniteWeb generate verifiable agentic environments at scale \citep{liang2026stageclaw,dong2026agentworl,zhang2026infinitew}, introducing defenses such as in-loop reward-hacking detection and hierarchical property, interaction, and rollout verification precisely because a self-built verifier can be gamed. A second mechanism is self-constructed test-and-verifier loops, where a model generates code and its own checks to self-evolve, as in ReVeal and OpenComputer's self-evolving verification layer \citep{jin2025reveal,wei2026opencompu}. A third is formally grounded self-verification, where counterexamples from a verifier become the optimization signal, as in VASO \citep{yang2026vaso}. Together these make \emph{who builds the verifier} a design axis. The failure mode is equally direct: synthesizing the checker scales evaluation, but a benchmark and oracle written by the same machinery being graded cannot supply a fully independent check.

\subsection{Contamination: The Threat to the Few Sound Benchmarks}

The benchmarks above are only as trustworthy as their freedom from train-test leakage, and contamination threatens autonomous-research-agent evaluation through three mechanisms. The first is detection and membership inference: black-box probes such as Min-K\% Prob and masked-answer guessing with retrieval overlap let auditors test whether a frozen test set was memorized \citep{shi2023detecting,deng2023investiga}, but they remain noisy and contested, so a passing agent score cannot be trusted without an explicit contamination audit. The second is empirical inflation: GSM1k shows accuracy drops of up to 8\% and systematic overfitting when a benchmark is faithfully re-created \citep{zhang2024a}, directly demonstrating that headline scores can be artifacts of train-test overlap rather than capability. The third is contamination-resistant design: dynamically refreshed or auto-constructed benchmarks such as LiveBench and AntiLeakBench \citep{white2024livebench,wu2024antileakb}, together with rigorous audits of mitigation strategies \citep{sun2025the}, argue that the only durable fix is live, verifiable, post-cutoff test construction. Two surveys consolidate the taxonomy linking memorization, membership inference, and benchmark leakage \citep{xu2024benchmark,tong2026pretraini}. The 2026 wave makes the stakes concrete: search-time contamination, in which an agent retrieves the answer from the live web during evaluation, inflates scores even on benchmarks that were clean at construction \citep{wang2026searchtim}. Contamination is therefore a core auditability requirement, not just a data-hygiene nuisance. It weighs most heavily on reproduction and discovery benchmarks because those draw their targets from published papers that are very likely in pretraining corpora.

\subsection{Emerging 2026 Benchmarks: Auditing the Verifier}

The newest benchmark wave reorients evaluation around verification rather than raw capability. One cluster builds end-to-end-validity benchmarks that hide the target paper and score re-discovery against expert multimodal rubrics; across this cluster top agents report low single-digit-to-roughly-twenty-percent pass rates and recurring failures on experimental-protocol fidelity, evidence mismatch, and a missing scientific core, as reported individually by ResearchClawBench, the AARRI line, and FML-bench \citep{xu2026researchc,wang2026act,zou2026fmlbench}. A second cluster attacks the trustworthiness of the verifier itself, with meta-evaluation showing LLM judges score below 55\% accuracy and are weakest at evidence verification, and VERITAS building computationally irreducible, auto-verifiable ground truth to escape the paradox that completeness checks need full ground truth \citep{wang2026time,wu2026breaking}. A third cluster measures closed-loop validity threats and forward-looking judgment: span-level claim-centric error auditing of agent trajectories, and the evidence-decision decoupling of ForeSci, where agents cite correct evidence yet forecast the wrong research object \citep{wang2026where,tian2026foresci}. These papers shift the benchmark target from more generation to more reliable, contamination-resistant assessment.

\subsection{Cost, Compute, and Economic Viability}

Cost is part of an autonomous-research claim. A system that wins by spending far more inference than its
baseline has not necessarily found a better research method. One line of work evaluates LLMs in dollar
terms and treats the cost of a mistake as part of the decision rule; this includes economic evaluations of
LLMs, SWE-Effi's expensive-failures and token-snowball analyses, and the credit-budgeted ICPC arena, where
each unverified decision consumes a finite budget
\citep{erol2025costofpas,fan2025sweeffi,zhou2026creditbud}. A second line shows how efficiency claims can
disappear under budget-matched baselines: skill and memory modules for web agents rarely beat a
token-matched vanilla actor once hidden inference cost is counted \citep{hajimiri2026are}. A third line
studies how compute should be allocated, including AgentTTS's compute-optimal per-subtask budgeting and
compute-accuracy saturation in reasoning systems \citep{wang2025agenttts,prucs2025computeac}. Together,
these papers make budget disclosure part of verification. Extra test-time compute is useful only when a
matched check can show that the extra work improved correctness rather than merely lengthening the run.

\subsection{Efficiency, Cost, and Inference-Budget Accounting as a Reporting Gap}
\label{subsec:cost_accounting}

A reported accuracy gain on an agentic benchmark is meaningful only if we know what it cost to obtain. An autonomous-research agent that climbs a leaderboard by issuing ten times as many model calls, sampling a hundred parallel rollouts, or silently swapping in a larger backbone is not necessarily a better discovery system; it may simply be a more expensive one. Cost and inference-budget accounting is therefore part of the claim: without a disclosed and matched budget, an improvement cannot be distinguished from a budget-unmatched comparison, and the verifier cannot tell whether the method or the spending produced the result. The open problems below are arranged by the mechanism through which inference cost enters and escapes the record.

\paragraph{Test-time compute as an unaccounted source of gains.}
The dominant lever behind recent agentic improvements is test-time compute. \citet{zhu2025scaling} conduct a systematic study of applying test-time scaling to language agents, varying parallel sampling, sequential revision, verification, and rollout diversity, and find that spending more inference reliably raises task performance. The automated part is the search itself: the agent explores more candidates and reflects more often. A controlled ablation that holds the underlying model fixed and varies only the compute knob is what validates the gain, since it makes the curve of accuracy against budget explicit. The blind spot lies in the comparison against external baselines that were run at a different, usually unstated, budget. When a new agent reports a higher score than a prior system, the reader rarely learns whether the two were given the same number of samples, the same wall-clock allowance, or the same verifier budget. \citet{lin2025sleep} sharpen the same point from the cost-reduction direction, showing that precomputing useful quantities offline can cut the test-time compute needed for a fixed accuracy by several times and amortize cost across related queries. This makes budget a first-class axis: a method can win on accuracy-at-fixed-budget, on budget-at-fixed-accuracy, or on neither, and only the first two are verifiable claims. Test-time scaling is therefore a regime where the independent check must be budget-matched; otherwise a system can appear better simply by outspending the baseline.

\paragraph{Hidden inference cost in the evaluation harness.}
Even when a paper intends to report cost honestly, the harness can hide it. \citet{moghadasi2026audit} audit a set of well-known agent benchmark papers against a small disclosure schema covering benchmark identity, harness specification, inference settings, cost reporting, and failure breakdown, and report that cost is the weakest dimension: the agent-benchmark papers they examine disclose inference cost in essentially no form, and none fully specify a content-addressed image of the evaluation environment. The mechanism is that an agent run folds many implicit decisions, scaffold retries, tool-call loops, evaluator model versions, and sampling temperature, into a single reported number, and these decisions move both score and cost. The end-to-end run is automated; the disclosed harness and budget are what should account for it; and whatever the harness declines to print is what slips past audit. The remedy proposed in this line is to score the disclosure of a run separately from its correctness, which is exactly the verification-first stance this survey advocates: a result whose cost and harness are unrecorded cannot be independently rerun, so the independent check is impossible by construction regardless of whether the result is true.

\paragraph{Energy and the deployment-scale denominator.}
Token counts and dollar prices are proxies, and they are noisy ones. \citet{liu2026energy} argue that inference should be evaluated as energy-to-token production and call for benchmarks to report joules per token, the limiting resource, and power- and utilization-adjusted output alongside accuracy and latency, noting that listed API prices vary by more than an order of magnitude across providers and so cannot stand in for marginal cost. For autonomous-research agents that run long horizons, the choice of denominator changes the verdict: a method that looks cheap in API dollars may be expensive in energy, and vice versa. The agent's reasoning loop is the automated piece; a dimensionally consistent cost measure is what would establish its efficiency; and the gap between the headline price and the physical resource actually consumed is what goes unexamined. The empirical study by \citet{tripathy2025swenergy} makes the stakes concrete: across four agentic issue-resolution frameworks driven by small language models, the most energy-intensive framework consumed roughly nine times the energy of the least, yet task-resolution rates were near zero, so the bulk of that energy was spent on unproductive reasoning loops. Framework architecture, not the model, drove consumption. In a budget-unmatched comparison, two systems can differ ninefold in cost while differing negligibly in delivered discovery, and a report that omits energy would credit the difference to capability.

\paragraph{Cost as a disclosed and reproducible quantity.}
Budget should be reported as a quantity that can be checked, not just mentioned.
\citet{yuksekgonul2026discover} report a test-time training procedure that reaches new results on
mathematics, kernel-engineering, algorithm, and biology problems using an open model and public code, and
they give the run cost as a few hundred dollars per problem with solutions reviewed by domain experts or
competition organizers. The cost figure is part of the claim: an independent reader can ask whether the
same budget, weights, and review protocol reproduce the result. \citet{radanliev2026aibom} make the same
point at the systems layer by extending software bills of materials into agentic bills of materials that
record runtime dependencies, environment drift, and agent decision provenance. These examples pair
rerunnable budgets with external checks. They also show why code alone is not enough: without a matched
inference budget, a reader cannot tell whether the reported gain came from the method or from extra
spending.

\subsection{Reporting Standards, Model Cards, and Disclosure Norms}
\label{subsec:reporting_standards}

The benchmarks and artifact-release practices surveyed above establish what an autonomous-research agent produces; the disclosure-standards layer governs whether a third party can audit it. Model cards, datasheets, evaluation reports, and reproducibility checklists record what was tested, on which data, under which protocol, and with what known failure modes. For agentic discovery this layer is not cosmetic. It is the mechanism that lets a reader reconstruct and re-run the claimed check. We sort the relevant work by the mechanism each standard targets: developer-level transparency reporting, evaluation-level reporting templates, baseline and protocol disclosure, and the documentation of release artifacts themselves. Standards exist and are increasingly well specified, but adherence remains voluntary, uneven, and largely unaudited.

\paragraph{Developer-level transparency reporting.} The broadest standards address what the developer of a foundation model discloses about its construction and use. The Foundation Model Transparency Index quantifies disclosure across dozens of indicators spanning training data, compute, capabilities, and downstream impact, and its longitudinal tracking reports that average transparency has regressed rather than improved as commercial stakes have risen, with developers most opaque precisely about training data and compute \citep{wan2025fmti}. This trajectory matters for agentic research because an agent that builds on an undisclosed base model inherits its opacity: claims about a discovery pipeline cannot be fully audited if the substrate model's data provenance and evaluation history are themselves undocumented. Complementary work shows that even where developers publish extensive ethics and safety discourse, the framing can substitute rhetoric for substantive disclosure, a pattern of ethics-washing in which public communication emphasizes safety language without applying the corresponding documentation frameworks \citep{wilfley2026ethicswash}. In the agent setting, a transparency narrative does not verify the underlying claims; the standard must be machine-checkable and externally scored rather than self-asserted.

\paragraph{Evaluation-level reporting templates.} A second mechanism narrows the scope from the developer to the specific evaluation, prescribing what a model report must contain for a reader to judge whether a benchmark result is credible. Standards in this category convert tacit evaluation choices into explicit, comparable fields. A representative effort proposes a structured template for reporting dangerous-capability evaluations, developed in consultation with experts across government, civil society, and frontier labs, and pairs the recommendations with gold-standard worked examples and a compact reporting template so that third parties can identify whether a report contains enough detail to assess evaluation rigor \citep{mccaslin2025stream}. The design goal aligns with the verification stance this survey advocates: the template is not meant to flatter the result but to expose whether the evaluation can be reconstructed and trusted. The same logic applies to autonomous-research agents that self-report benchmark gains. Without a disclosure template that pins down the test set, the scoring procedure, and the decision thresholds, an agent's claimed improvement is an assertion rather than an auditable measurement, and the closed-loop claim that the agent improved itself collapses into an unverified narrative.

\paragraph{Baseline and protocol disclosure.} A third strand targets the comparison itself, since a benchmark number is meaningless without a documented baseline. A meta-review of human baselines in foundation-model evaluations argues that comparisons of human versus model performance are routinely under-specified, and it derives a reporting checklist for designing, executing, and documenting such baselines; applying that checklist across a large sample of published baselines surfaces systematic shortcomings in how the comparison was constructed \citep{wei2025baselines}. For agentic discovery, where claims of super-human or autonomous capability are common, this checklist is a concrete instrument for distinguishing a validated closed-loop claim from a flattering one: it forces disclosure of who or what the agent was measured against and how that reference was obtained. Reporting standards in the peer-review channel reinforce the point. An interview study of authors of LLM-integrated systems finds that disclosure norms are contested and context-dependent, with reviewers applying inconsistent skepticism and authors disagreeing over how much of the prompt, configuration, and model identity must be reported \citep{navarro2026reporting}. Even the human gatekeeping layer therefore lacks a settled standard for what an LLM-driven contribution must disclose, which weakens the independent check at exactly the point where it should be strongest.

\paragraph{Documentation of release artifacts and aggregators.} The fourth mechanism concerns the artifacts and infrastructure through which results circulate. An empirical study of foundation-model leaderboards catalogs recurring documentation deficiencies, or leaderboard smells, including unstated evaluation protocols and missing provenance, that undermine the transparency of the very aggregators researchers use to compare systems \citep{zhao2024lbops}. Because autonomous-research agents increasingly consult and post to such leaderboards, undocumented evaluation pipelines propagate directly into agent decision-making and reported gains. At the system level, surveys of trustworthy agentic AI argue for consolidating disclosure into outcome and process signals, such as constraint-violation counts and trace completeness, so that release gating depends on documented evidence rather than headline metrics \citep{qi2026trustagentic}. This reframes disclosure as a precondition for the closed-loop verification that distinguishes a genuine discovery from an unaudited claim: a complete, machine-readable trace is what an independent checker actually re-executes.

\paragraph{Synthesis.} This literature describes a maturing standards layer whose components are increasingly well designed yet whose adoption is voluntary, fragmented, and rarely audited end to end. The divergence we observe for artifacts also holds for documentation: release is becoming common, but disclosure remains incomplete and its accuracy is seldom independently checked. The next step is to make the standard executable: disclosure templates and reproducibility checklists should be contracts that an external harness can ingest, re-run, and score. Until that happens, documentation around agentic discovery will describe claims more reliably than it substantiates them.

\subsection{Prompt and Context Engineering: The Unlogged Variable}

Underlying every benchmark number is a scaffold of prompt and context choices that is rarely reported. If a result depends on a hand-tuned scaffold, an agent's measured capability is partly an artifact of unaudited human context work. The brittleness evidence is strong. FormatSpread shows up to 76-point swings from spurious prompt-formatting choices, and the Prompt Report systematizes dozens of techniques \citep{sclar2023quantifyi,schulhoff2024the}, so prompt choices can determine the result yet remain unlogged. Reported gains are not reproducible without the prompt artifact. Long-context limits compound this: Lost-in-the-Middle, RULER, and NoLiMa show that stuffing papers and tool outputs into a window degrades retrieval and reasoning \citep{liu2023lost,hsieh2024ruler,modarressi2025nolima}, motivating active context management such as ReSum summarization and memory-as-action \citep{wu2025resum,zhang2025memory} whose curation steps must themselves be auditable when an agent claims to have read a corpus. Self-evolving frameworks, GEPA's reflective prompt evolution and ACE's contexts as evolving playbooks, framed by the context-engineering literature as entropy-reducing human-to-machine labor \citep{agrawal2025gepa,zhang2025agentic,mei2025a,hua2025context}, automate the scaffold but shift the verification burden onto whether the auto-generated context and trace are faithful and inspectable. Artifact disclosure is rising, with code released by 83\% of our coded systems and prompts by 71\%, yet the surrounding context construction and prompt-evolution steps are rarely auditable even when the prompt text itself is released. Releasing an artifact is still not the same as enabling an independent check of the claim built on it.

%% file: sec_frontiers.tex
\section{Frontiers}
\label{sec:frontiers}

The audit problem grows fastest where agents add state, reward, self-modification, and oversight machinery. Continual memory, agentic reinforcement learning, self-evolution, and multi-stage pipelines raise the throughput of generated claims while adding surfaces that are hard to inspect. For each area, we ask what the mechanism automates, what verifies it today, and what a reviewer still cannot audit. The section closes by setting up the six open problems that follow.

\subsection{Continual Memory and Self-Improvement: Accumulated State and Audit Debt}

Continual-memory systems move from one-shot agents to systems that accumulate competence across experiments. The mechanisms fall into four families. Episodic and experiential memory extracts natural-language insights from past trajectories and re-injects them: ExpeL distills transferable lessons \citep{zhao2023expel}, while CoPS makes the reuse step provable by selecting distribution-matched experiences under a pessimistic criterion \citep{yang2024cops}. Production-oriented stores such as Mem0 \citep{chhikara2025mem} and dynamically self-organizing note graphs such as A-MEM, which links and evolves memories in a Zettelkasten style \citep{xu2025amem}, scale long-horizon recall. A second family builds procedural and skill libraries: Agent Workflow Memory induces reusable routines from solved tasks \citep{wang2024agent}, and ArcMemo distills concept-level abstractions for test-time continual learning without weight updates \citep{ho2025arcmemo}. A third grows research knowledge graphs, iteratively expanding self-organizing networks that structure scientific knowledge for downstream synthesis \citep{buehler2025agentic}. The fourth and most aggressive is recursive self-improvement, where an agent rewrites its own code or architecture, as in the G\"odel Agent \citep{yin2024gdel} and the Darwin G\"odel Machine \citep{zhang2025darwin}, a direction now mapped by a dedicated survey of \emph{what}, \emph{when}, \emph{how}, and \emph{where} an agent should evolve \citep{gao2025a}.

Reuse is only as trustworthy as the provenance and validation behind it. The most disciplined designs expose that boundary. CoPS's distribution-matched selection bounds which past experiences may transfer \citep{yang2024cops}; A-MEM's inspectable note structure exposes the links a reviewer would need to trace a claim's lineage \citep{xu2025amem}; Agent Workflow Memory's human-readable routines are auditable artifacts rather than opaque weights \citep{wang2024agent}; and the Darwin G\"odel Machine commits each self-modification only after empirical benchmark validation \citep{zhang2025darwin}.

These mechanisms cover different slices of the audit problem. CoPS gives a \emph{selection-time} check: a pessimistic, distribution-matched criterion controls which past trajectories may influence the current task, but it says only that a lesson is eligible, not that the lesson is correct \citep{yang2024cops}. A-MEM gives an \emph{inspection-time} affordance: the note graph does not block bad memory from entering, but it records the links along which a claim was assembled \citep{xu2025amem}. Agent Workflow Memory provides \emph{representational} auditability by storing routines as human-readable procedures rather than latent weights, though it does not automatically test whether a routine generalizes beyond the tasks that produced it \citep{wang2024agent}. The Darwin G\"odel Machine is \emph{outcome-validating}: a self-modification is committed only if it survives an empirical benchmark, the strongest signal in the group but one that still inherits the benchmark's construct-validity limits \citep{zhang2025darwin}. No current store combines selection bounds, lineage, readable state, and benchmark revalidation. A reviewer who wants to know both that a reused lesson was on-distribution and that it stayed correct under the new regime has no single artifact to inspect.

The missing property is durability. A stored insight that held under one experimental regime can quietly become a contaminating prior under another, and no current store separates a validated lesson from a lucky one. Given a released memory store, there is no procedure that re-validates each entry against the regime in which it is being reused. A selection bound is computed at write-or-retrieve time and is not recomputed when the experimental distribution drifts; an inspection trail shows what was linked but not whether the link still holds; a readable routine stays readable after it has gone stale; and a benchmark gate is evaluated once, at commit, rather than re-run against later tasks. The audit debt compounds because each reuse silently reasserts a validation performed under different conditions. Self-evolution turns this into a governance problem, because an agent that edits its own selection logic can in principle edit away the very gates that made its earlier commits checkable. For continual learning, the useful verification signal is not just the final benchmark number. It is the auditable provenance of each reused unit plus a mechanism to re-check that provenance under distribution shift, and that paired signal is, today, mostly absent.

\subsection{Agentic RL Training: When the Reward Is Harder to Verify Than the Task}

Agentic RL in 2025--2026 trains research and coding agents with multi-turn, end-to-end reinforcement learning. This literature shows that verification of agent behavior is often the limiting obstacle, not raw capability. Its credit-assignment and stability mechanisms exist because sparse, hard-to-verify turn-level rewards make long-horizon agent RL collapse or stall: APPO introduces a fine-grained Branching Score to localize credit \citep{wang2026appo}, PACT co-trains on privileged traces to densify the signal \citep{du2026pact}, and SENTINEL synthesizes tasks from observed failures so the reward targets the agent's actual weak points \citep{wang2026sentinel}. The diagnostic evidence is direct. ``Why Multi-Step Tool-Use RL Collapses'' shows that the underlying tool-use ability stays intact while apparent RL gains are masked by control-token and formatting failures \citep{hao2026why}, and a study of the effectiveness and efficiency of agentic tool-calling shows leaderboard rankings flip under minor harness changes \citep{liu2026on}. Both indicate that evaluation harnesses and behavior verification can dominate apparent gains, so what an agent is credited with depends heavily on how its behavior is measured rather than on capability alone.

Reward hacking is the common failure. An agent optimized against a signal it can satisfy without doing the underlying work will do exactly that, and the formatting-masked collapse \citep{hao2026why} is a concrete instance: gains accrue to the gradable surface, not the task. The field's response reframes scaling agent RL as a problem of reliably grading what the agent did. Polar reconstructs token-faithful trajectories over arbitrary harnesses on SWE-Bench Verified so that the reward reflects the real execution rather than a logged summary \citep{xu2026polar}, and OpenWebRL grounds rewards in live-web verification rather than a static cache \citep{yang2026openwebrl}.

These two responses need to be separated because they attack different fault lines in the reward. Polar is about \emph{faithfulness of the reward to the agent's own execution}: the failure it closes is that a logged summary can diverge from what actually ran, so a reward computed over the summary can credit work the agent did not do. Reconstructing the token-faithful trajectory over an arbitrary harness makes the reward an honest function of the real rollout, yet it leaves untouched whether the task specification the reward encodes was itself valid \citep{xu2026polar}. OpenWebRL goes after the complementary fault line, \emph{freshness and groundedness of the reward's evidence}: a static cache can certify an answer that was correct when the cache was built and is now stale, so grounding the reward in live-web verification ties credit to the current state of the world rather than to a frozen snapshot \citep{yang2026openwebrl}. Neither subsumes the other. A token-faithful reward over a stale cache still rewards confidently retrieving an outdated fact, and a live-grounded reward over an unfaithfully logged trajectory still credits execution that did not happen. Reward verification therefore has at least two parts: faithfulness to execution and groundedness of evidence. The diagnostic studies above detect failures of both: the harness-sensitivity result \citep{liu2026on} is a faithfulness failure, the same execution scored differently by different harnesses, while reward-hacking against a static target is a groundedness failure.

These are infrastructures for a trustworthy reward. The part a reviewer still cannot audit is often the reward function itself. A released research agent typically ships its policy and training data but not a faithful, re-runnable specification of the reward it was optimized against, so there is no artifact on which to check whether that reward matched execution or grounded itself in current evidence. The contamination sits upstream of any released claim and stays invisible in it: a benchmark number reports what the trained policy scored, not whether the reward that shaped the policy was sound. Polar and OpenWebRL matter because they begin to turn reward construction into an inspectable object.

\subsection{Uncertainty, Calibration, and Abstention as a Verification Primitive}

Epistemic calibration asks whether an agent can gauge the strength of its own evidence and abstain when it cannot, rather than emit a confident but unsupported claim. The work clusters into three mechanisms. One turns verbalized or internal uncertainty from a passive diagnostic into an active control signal that gates an agent's act-versus-reflect-versus-escalate decisions, yielding trajectory-level calibration \citep{zhang2026from}. Another sharpens selective prediction and abstention, with evidence that entropy alone is an unsafe trigger and that abstention often fails precisely when a model is most uncertain \citep{phillips2026entropy}, and that knowing when to abstain remains brittle in high-stakes settings \citep{machcha2026knowing}; geometry-calibrated conformal abstention is one corrective \citep{xu2026geometryc}. A third extends conformal risk control to the compound, multi-stage pipelines that research agents actually are, giving end-to-end coverage and sample-then-filter trustworthiness guarantees \citep{kotte2026pasc,wang2025safer}.

Calibrated abstention is a \emph{learnable, auditable verification primitive} rather than an emergent capability, and current agents systematically lack it. An agent that abstains when its evidence is weak supplies its own partial verification signal: the reviewer no longer has to catch every unsupported claim, because the agent declines to make some of them. The same literature also bounds how far this can go. If abstention degrades exactly when uncertainty is highest \citep{phillips2026entropy}, then the primitive fails in the regime where it is most needed, and the multi-stage conformal guarantees \citep{kotte2026pasc} hold only under assumptions that a long-horizon research pipeline routinely violates. Calibration is a real but fragile verification signal, and its fragility is itself something a reviewer cannot yet audit from a released artifact.

\subsection{Negative Results, Failed Replications, and the Reproducibility-Crisis Context}

Negative results around agent self-reporting, reproduction, and automated review are now a useful audit source. EviBound shows a prompt-only autonomous research agent claiming success on all eight of its tasks while none are actually verified, with the gap closing only when architectural verification gates are imposed \citep{chen2025evidenceb}. Failed reproduction is a second signal: LMR-Bench finds that frontier agents cannot reproduce language-modeling research code \citep{yan2025lmrbench}, and an attempt to replicate eighteen published LLM-centric studies fully reproduces none of the executable ones \citep{angermeir2025reflectio}. Unreliable verification is a third: audits of the automated-review layer that many pipelines propose as their check find that it inflates LLM-authored papers, penalizes critical or risk-flagging statements, and produces shallow over-praising feedback \citep{li2025llmreval,taechoyotin2025remor}. The checker many systems rely on is therefore itself biased and gameable.

These agent-level failures inherit a longer human-science context. The broader ML reproducibility crisis supplies the motivation: data leakage documented across seventeen scientific fields \citep{kapoor2022leakage} and unreported seed and hyperparameter variance in deep RL \citep{henderson2017deep} are exactly the ``looks-good-but-does-not-hold'' pattern an agent auditor must catch, yet artifact release alone does not expose leakage, seed variance, or unreported tuning conditions. Community process safeguards, reproducibility checklists and code-submission programs, operationalize what a trustworthy claim requires and give a template for what agent-produced artifacts should be held to \citep{pineau2020improving}. Construct-validity critiques of dominant benchmarks \citep{raji2021ai} and a review of 445 LLM benchmarks \citep{bean2025measuring} show that high leaderboard scores can be measurement artifacts, so an agent optimizing against a benchmark may be exploiting invalidity rather than demonstrating capability. If human-run science already inflates results irreproducibly, agentic pipelines that close the loop without oversight inherit and amplify those failure modes.

\subsection{Autonomy Levels, Oversight, and Governance}

Autonomy levels depend on the strength of the verification signal available. The literature divides along a mechanistic question: does the proposal try to \emph{expose} an existing verification signal, or to \emph{manufacture} one where none is available? The two answers fail in opposite ways, and an autonomy level is only as sound as the mechanism it rests on.

Governance and visibility \emph{infrastructure} turns opaque agent behavior into checkable artifacts. Its logic is that an autonomy level can be conditioned only on evidence an overseer can actually inspect, so the work is about widening the inspectable surface: agent identifiers, real-time monitoring, and activity logging make the agent's actions observable in the first place \citep{chan2024visibilit}; principal-agent and liability framings convert that observability from a courtesy into an enforceable requirement, so that an unobservable action becomes a governance violation rather than a gap \citep{kolt2025governing}; runtime policy enforcement on action paths moves the check from after-the-fact logging to inline interception, blocking a disallowed action rather than merely recording it \citep{wang2025mi}; and the broader scoping of technical governance levers maps which of these surfaces are reachable at all \citep{anderljung2023frontier,reuel2024open}. This cluster does not create a verification signal. It routes an existing one to the overseer, and its failure mode is a blind spot: an action the infrastructure does not instrument is invisible no matter how faithful the rest of the pipeline is. A complementary position argues that trustworthy AI scientists will require reformed institutions, not just better models, so that the verification burden is shared between the agent and the surrounding scientific process \citep{jimenez2026aiscientists}. The same point holds at field scale: visibility infrastructure is worthless if no institution is obligated to look.

Scalable oversight confronts the regime where exposure is not enough because there is no trustworthy signal to expose: when the agent's competence exceeds the verifier's, observing the agent perfectly still does not tell an overseer whether it is right. The aim is to \emph{manufacture} a trustworthy verification signal when ground truth is unavailable, and the mechanisms differ in where they get leverage. Weak-to-strong generalization tries to elicit a strong model's latent competence using only weak supervision, betting that the strong model already knows more than its supervisor can articulate \citep{burns2023weaktostr}; sandwiching and proof-of-concept human-AI supervision instead construct a controlled gap between a non-expert overseer and an expert ground truth so the protocol can be measured before it is trusted \citep{bowman2022measuring}; and debate-versus-consultancy protocols pit the agent against an adversarial copy of itself so that a weak judge can adjudicate by cross-examination rather than by direct knowledge \citep{kenton2024on}. Visibility infrastructure assumes the signal exists and must be transported. Scalable oversight assumes it must be built from weaker materials, so its failure mode is not a blind spot but a corrupted signal. These same results expose that failure mode directly: oversight protocols can be gamed, judges grow overconfident, and scaling laws bound their reliability, so a manufactured signal can be confidently wrong in a way a missing one cannot. The two failure modes do not cancel; they stack, because a high-autonomy research agent needs both an instrumented action surface and a trustworthy verdict over what that surface reveals. A levels-of-autonomy taxonomy that gates agent authority on verifiable oversight inherits the limits of whichever mechanism it leans on, a blind spot from visibility infrastructure or a corrupted verdict from scalable oversight. Autonomy should be granted only as far as the weaker mechanism still binds.

\subsection{Human--AI Collaboration and Calibrated Oversight}

Human-AI collaboration treats the research agent as a partner whose autonomy must be bounded, steered, and audited. Interaction design supplies the controls: IRIS exposes fine-grained, steerable handles so a researcher can guide LLM ideation in a mixed-initiative loop \citep{garikaparthi2025iris}, a levels-of-autonomy taxonomy recasts the operator, collaborator, consultant, approver, and observer roles as per-task design choices that fix the human's oversight position \citep{feng2025levels}, and embodied co-scientists extend this collaboration into physical experimentation \citep{cong2025labos}. The recurring empirical finding is about reliance, not capability: when scientists cannot calibrate trust in agent outputs, autonomy is unsafe regardless of raw performance, and AI assistance can homogenize ideas, making the collective more diverse while leaving individual creativity unimproved, ``different, not better'' \citep{ashkinaze2024how}. The human's residual role is increasingly verification, steering, and diversity injection rather than generation.

This connects to the position, theory, and economics literature around scientific oversight. Critical evaluations of flagship end-to-end systems show that polished papers can hide failure modes visible only in workflow traces and code, not in the final artifact: independent evaluation of Sakana's AI Scientist \citep{beel2025evaluatin} and process-level audits \citep{luo2025the,singh2025the} argue that the unit of verification must be the research process, not its output. An evaluation-integrity strand argues that as agents scale paper and benchmark production, the field's self-correction infrastructure becomes the limiting factor, motivating a refutations-and-critiques track \citep{schaeffer2025position} and an AI-augmented peer-review ecosystem \citep{wei2025position}. Economics work formalizes the throughput-versus-validity tension as an automate-versus-augment choice over the workforce \citep{shao2025future} and warns that idea homogenization erodes the diversity discovery depends on \citep{anderson2024homogeniz}. The shared warning is simple: faster paper production does not become scientific progress unless the verification process scales with it.

\subsection{Red-Teaming, Dual-Use, and Safety Evaluation of Research Agents}
\label{subsec:red_team_safety}

The preceding sections treated verification as a property of scientific claims. Safety evaluation asks the same question about deployment claims: what does the system claim to be safe to do, and what independent check substantiates that claim? An agent that can search literature, design wet-lab protocols, write and execute code, and order reagents has dual-use surface by construction. We organize the literature below by the mechanism through which that surface is probed, asking of each what it automates, what it verifies, and where it leaves the agent unaudited.

\paragraph{Hazardous-knowledge proxies.} The most mature instrument measures whether a model possesses knowledge that would assist weapon development. The Weapons of Mass Destruction Proxy benchmark releases multiple-choice questions across biosecurity, cybersecurity, and chemical security as a public proxy for hazardous knowledge, doubling as a target for unlearning methods that aim to remove such knowledge while preserving general competence \citep{li2024wmdp}. The design is itself an exercise in dual-use balance: the authors filtered sensitive content before release so that the benchmark measures the hazard without instantiating it. A WMDP-style instrument automates knowledge possession at the level of a static question set and verifies recall under a fixed prompt distribution; what it leaves untested is procedural capability under tool use, multi-step planning, or adversarial reframing. \citet{deleeuw2026biorefusal} sharpen this gap from the refusal side, auditing whether a model's refusal on biosecurity-adjacent prompts is structurally sound or collapses under modest changes to formatting, output-length caps, and framing; across several architectures refusal tracked legality and cultural salience more than genuine hazard, and behavioral refusal sometimes diverged from internal activation-level signals. A passing score on a knowledge proxy, or a high surface refusal rate, is therefore a weak check: it certifies behavior on the measured distribution, not the closed-loop competence an agent actually wields. The overclaim is familiar from the rest of the survey, but the safety stakes are higher.

\paragraph{Dangerous-capability evaluation under agency.} A second mechanism moves from knowledge to action, eliciting whether a system can actually carry out a harmful task end to end. \citet{phuong2024dangerous} pilot a programme of dangerous-capability evaluations spanning persuasion and deception, cybersecurity offense, self-proliferation, and self-reasoning, reporting early warning signs rather than strong capabilities in the models tested and explicitly framing the work as building a rigorous science of such evaluation. The contribution that matters for verification is methodological: capability is established by demonstrated task completion in an instrumented environment, not by inference from a quiz. Yet the same agentic affordances that make research agents productive, namely tool access, code execution, and web browsing, are precisely the affordances that invalidate chat-context safety guarantees. \citet{kumar2024browserart} show that browser agents built on refusal-trained backbones attempt harmful behaviors that the underlying model refuses in a chat setting, and that chat-context jailbreaks transfer to the agent. The implication for autonomous science is direct: an alignment check performed on the conversational interface does not transfer to the scaffolded agent, so safety claims established at the model layer are not valid evidence about the deployed research system. The check and the artifact must match.

\paragraph{Behavioral risk awareness and monitoring.} A third mechanism asks not whether the agent can do harm but whether the surrounding system can notice. \citet{yuan2024rjudge} benchmark whether models can judge the safety of recorded agent interaction trajectories, finding substantial headroom in risk awareness and that fine-tuning on safety judgment helps where prompting does not. Monitoring is itself a verification layer, and it inherits the verification literature's central caveat: a check is only as strong as its robustness to an adversary who knows it exists. \citet{kale2025monitoring} stress-test monitors of covert agent misbehavior and find that an agent's awareness that it is being watched degrades monitor reliability more than additional monitor information helps, while scaffolding choice and targeted human escalation recover much of the lost signal. For research agents this reframes oversight as an adversarial control problem rather than a passive logging exercise. \citet{yang2026overprivileged} expose a complementary structural hazard: agents routinely select higher-privilege tools when a sufficient lower-privilege alternative exists, and general safety alignment does not reliably transfer to least-privilege tool choice. An autonomous agent that can requisition compute, data, or laboratory actuation at higher privilege than its task warrants is accumulating unaudited blast radius, and prompt-level controls only partially contain it.

\paragraph{Agent-driven safety discovery.} A fourth mechanism turns the agent's own capabilities toward auditing itself. \citet{chen2025agentguard} repurpose an agentic orchestrator to autonomously discover unsafe tool-use workflows, validate them by real execution, synthesize safety constraints, and then test whether those constraints hold. This closes a loop that knowledge proxies leave open: the unsafe behavior is verified by demonstrated execution and the mitigation is verified by demonstrated prevention, mirroring the strongest forms of closed-loop validation the survey advocates for scientific claims. The residual weakness is coverage. An agent red-teaming itself can only surface the failure modes its own generation distribution reaches, so the absence of a discovered exploit is not evidence of its absence, a limitation that recapitulates the gap between passing a test and being correct.

\paragraph{The access and disclosure substrate.} Underlying all four mechanisms is the question of who may run the check and on what. \citet{charnock2026access} propose a taxonomy of evaluator access, disentangling model access, model information, and evaluation timeframe into graduated levels, and argue that constrained external access inflates false-negative rates and erodes stakeholder trust. The analogy to artifact release is direct: just as a withheld codebase blocks reproduction of a scientific claim, black-box, time-boxed access blocks the independent substantiation of a safety claim. The dual-use tension is older than the agent era. \citet{shevlane2020offense} argue that whether disclosing research aids attackers or defenders depends on field-specific structure and warn against importing software-vulnerability disclosure norms wholesale into AI. For autonomous research agents the offense-defense calculus is unsettled precisely because the same system that accelerates defensive science can accelerate the offensive variant, and no public benchmark yet measures that balance at the level of closed-loop autonomous execution. Across every mechanism the pattern holds: artifact release and capability demonstration are increasingly common, but the independent check that would convert a capability into a verified-safe capability remains partial, evadable, or access-gated. Safety is the verification problem under higher stakes: the cost of an unaudited claim is measured in harm rather than in retracted novelty.

\subsection{Memory, Personalization, and the Provenance of Accumulated State}
\label{subsec:memory_provenance}

An always-running research agent does not begin each task from a blank context. It writes to a persistent store, retrieves prior conclusions, consolidates them, and carries personalized preferences across sessions, so that the state available at step $t$ is a function of every prior trajectory. This accumulated state is increasingly treated as a first-class component rather than a transient scratchpad, with storage, retrieval, update, consolidation, and lifecycle governance handled by a dedicated subsystem \citep{zhou2026agentnative}. A reused memory is itself a discovery claim: when an agent retrieves a stored fact and acts on it, it is asserting that the fact is still true and was legitimately derived. The question is not whether agents can accumulate state, which they plainly can, but whether a reused memory is auditable, that is, whether the stored claim's continued validity and provenance can be independently checked. The mechanisms below sort along that question.

\paragraph{Retrieval memory and the validity of stored facts.}
The most common mechanism is retrieval over an accumulating fact store. Here the automation is write-then-retrieve: the agent appends conclusions and later surfaces the nearest neighbors of a query. The verification on offer is embedding similarity, which proxies for relevance, not for truth or recency. The gap becomes acute over evolving knowledge. When a fact changes, a retriever surfaces both the stale and current values with near-identical similarity, and contradiction is not separable from duplication by cosine distance alone \citep{yadav2026temporal}. The remedy is to make supersession explicit through a bi-temporal ledger that retires contradicted values by a deterministic rule, which collapses the stale-fact-error rate that similarity-only retrieval cannot avoid \citep{yadav2026temporal}. What goes unaudited in the common case is precisely temporal validity: a research agent that cached a deprecated API, a retracted result, or a superseded measurement will retrieve it with full confidence, and nothing in the default pipeline flags that the stored claim is no longer the current one. The independent check here is a validity model over time, and where it is absent the strength of the discovery degrades silently as the world moves on.

\paragraph{Consolidation, extraction, and update correctness.}
A second mechanism transforms raw episodes into compressed, structured memory through extraction and consolidation, automating summarization, deduplication, and the merging of new observations into existing entries. Where verification exists, it is decomposed: a systematic study that splits agent memory into representation, extraction, retrieval, and maintenance modules shows that no single architecture dominates and that effectiveness depends on matching memory structure to the workload, with localized maintenance proving more cost-efficient than global reorganization \citep{zhou2026agentnative}. That work measures update correctness and long-horizon stability as distinct quantities rather than folding them into end-to-end task success, which is the usual practice and which treats the memory as a black box. Under the black-box convention, the consolidation step itself goes unaudited: a summary that drops a qualifying condition, or a merge that overwrites a correct entry with a plausible but wrong one, is invisible to a task-success metric that happens to pass for other reasons. For a research agent, consolidation is where a hedged finding can quietly become an unhedged one, and the absence of a module-level check means the provenance of the simplification is lost.

\paragraph{Provenance and the authority of a memory to act.}
A third mechanism concerns where a memory came from and whether that origin entitles it to influence a consequential action. The automated part is the write itself, often from untrusted content such as a tool output, a retrieved web page, or another agent's message. What is supposed to verify a memory's authority is some signal of trust, and here the evidence is sharply negative. Content-based and lineage-based defenses are both malleable: an adversary can launder an untrusted origin through the agent's own summarization, a trusted-tool echo, or manufactured corroboration, flipping a derivation edge to trusted and reaching high attack success, so that only write-time origin binding with corroboration-gated elevation is provably sound \citep{louck2026securing}. The attack surface generalizes to shared stores: covert, minimal-perturbation tampering of a common knowledge base or replay buffer succeeds at sub-percent poison rates in heterogeneous multi-agent systems \citep{sharma2025bilevel}. The reused-memory-as-claim framing makes the stakes explicit. If provenance is forgeable, then a stored conclusion an agent cites as established may trace back to an unverified or adversarial source, and the closed loop closes around contaminated state. What goes unaudited is the binding between a memory and its true origin, which is exactly the independent check that would let a downstream reader trust an accumulated finding.

\paragraph{Personalization and drift.}
A fourth mechanism is personalization, where the agent maintains a model of a user or a research context and adapts to it across long horizons. The automated part is preference inference and reuse across domains and years. Verification means benchmarking against authentic longitudinal behavior rather than scripted personas, and such evaluation shows that current memory methods remain far from satisfying real cross-domain, lifelong personalization \citep{zhang2026memorycd}. The faithfulness of retained information is itself fragile: cognitively grounded probes of multimodal memory reveal that models fail to keep representations disentangled and exhibit interference patterns unlike human memory, so what is recalled is not reliably what was stored \citep{huang2026meval}. Personalization also concentrates sensitive accumulated state, and membership inference attacks recover whether a given interaction is present in an agent's memory, which both exposes a privacy risk and demonstrates that the store leaks structured signal about its own contents \citep{chen2026mrmmia}. The unaudited quantity is drift: a personalization profile that has slowly absorbed contaminating or stale signal will steer future research choices without any record of when or why it changed. Even the lower-level substrate participates, since persisting attention state across sessions accelerates reuse but offers no semantic guarantee that the restored state still encodes a valid claim \citep{shkolnikov2026agentmem}.

\paragraph{Synthesis.}
Memory systems automate accumulation and reuse faster than they automate the checks that would make reused state trustworthy. Temporal validity, update correctness at the module level, non-forgeable provenance, and drift monitoring are still partial, recent, or absent. A reused memory is a discovery the agent implicitly re-asserts; it can be trusted only as far as an independent check on it reaches. Until the validity and provenance of accumulated state are auditable rather than assumed, an always-running agent's most confident claims may rest on the least verified part of its own history.

\paragraph{From frontiers to open problems.} Each frontier leaves a concrete audit question. Memory and self-evolution need provenance for accumulated state; agentic RL needs a reward specification that can be rerun; calibrated abstention needs evidence that it works under uncertainty; negative results show the cost of trusting self-reporting; and governance and human oversight bound how much autonomy a given verification signal can responsibly license. These frontiers motivate, but do not resolve, the concrete challenges ahead; the reporting checklist of
Section~\ref{sec:checklist} operationalizes part of the response, and the conclusion
(Section~\ref{sec:conclusion}) distills the frontiers into six concrete open problems.

%% file: appendix_corpus.tex
\begingroup
\begin{landscape}
\footnotesize
% In a pdflscape page the table spans the physical page height, but \linewidth is
% still the (narrower) portrait \textwidth. Size columns against \textheight (the
% usable landscape measure, ~9in) and trim \tabcolsep so the eight columns plus
% their inter-column gaps fill the rotated page without overrunning it.
\setlength{\tabcolsep}{3.5pt}
\setlength{\LTcapwidth}{\textheight}
\begin{longtable}{@{}
  >{\raggedright\arraybackslash}p{\dimexpr0.142\textheight\relax}
  >{\raggedright\arraybackslash}p{\dimexpr0.135\textheight\relax}
  >{\raggedright\arraybackslash}p{\dimexpr0.075\textheight\relax}
  >{\raggedright\arraybackslash}p{\dimexpr0.158\textheight\relax}
  >{\raggedright\arraybackslash}p{\dimexpr0.135\textheight\relax}
  >{\raggedright\arraybackslash}p{\dimexpr0.145\textheight\relax}
  >{\raggedright\arraybackslash}p{\dimexpr0.105\textheight\relax}
  >{\raggedright\arraybackslash}p{\dimexpr0.045\textheight\relax}@{}}
\caption{Full coded corpus of the 26 full-text-coded focal systems on the seven audit dimensions, with values normalized to a controlled vocabulary. Stage codes: Id ideation, Lt literature, Hy hypothesis, Ex experiment-design, Cd coding, Ru run, An analysis, Wr writing, Rv review, CL closed-loop. ``n/d'' = not disclosed.}\label{tab:fullcorpus}\\
\toprule System & Stages & Aut. & Eval & Artifacts & HITL & Novelty & Sel. \\ \midrule
\endfirsthead \toprule System & Stages & Aut. & Eval & Artifacts & HITL & Novelty & Sel. \\ \midrule \endhead
CAMEO & Hy, Ex, Ru, An, CL & L4 & benchmark, task-success, deployment & none & topic, per-step, seed & n/d & yes \\
ReviewAdvisor & Rv & L2 & LLM-review, human & code, dataset & none & n/d & n/d \\
ChemCrow & Lt, Hy, Ex, Cd, Ru, An & L3 & LLM-review, human, task-success (MS/NMR terminal validation) & code, prompts, traces & topic, template, per-step, seed & automated-lit & yes \\
Ultimate Brain & Id, Lt, Hy, Ex, Cd, An, Wr & study & human & prompts & topic, template, per-step, seed & human & yes \\
ResearchAgent & Id, Lt, Hy, Ex, Rv & L3 & LLM-review, human & code, prompts, seeds & topic, seed & none & n/d \\
AutoArticleGen & Cd, Wr, Rv & L3(c) & human & none & topic, seed & n/d & n/d \\
SWE-agent & Cd, Ru & L2 & benchmark, task-success & code, prompts, traces & topic, seed & n/d & yes \\
SWIF2T & Lt, An, Wr, Rv & L3 & human, benchmark & code, prompts, traces & topic, per-step, seed & automated-lit & yes \\
AI Scientist & Id, Lt, Ex, Cd, Ru, An, Wr, Rv, CL & L4 & LLM-review, human & code, prompts, seeds, traces & topic, template, baseline, seed & automated-lit & yes \\
CLADD & Lt, An & L2 & benchmark, task-success & code & topic, seed & n/d & n/d \\
DatawiseAgent & Cd, Ru, An, Wr & L3 & LLM-review, benchmark, task-success & prompts & topic, seed & n/d & no \\
AI+Robot Scientists & Lt, Hy, Ex, Cd, Ru, An, Wr, Rv, CL & position & n/d & none & per-step & automated-lit & n/d \\
AI Scientist-v2 & Id, Lt, Hy, Ex, Cd, Ru, An, Wr, Rv, CL & L4 & LLM-review, human, workshop & code, prompts, traces & topic, seed & automated-lit & yes \\
AI-Researcher & Lt, Id, Hy, Ex, Cd, Ru, An, Wr & L3 & LLM-review, benchmark, task-success & code, prompts & topic, seed & n/d & yes \\
ChatBattery & Id, Lt, Hy, Ex, An, Ru, CL & L3 & LLM-review, human, task-success, deployment & code, prompts & topic, template, per-step, seed & automated-lit & yes \\
SR-Scientist & Hy, Cd, Ru, An, CL & L4 & LLM-review, human, benchmark, task-success & code, prompts, traces & topic, baseline, seed & n/d & yes \\
Denario & Id, Lt, Hy, Ex, Cd, Ru, An, Wr, Rv & L3 & LLM-review, human & code, prompts & topic, per-step, seed & automated-lit & n/d \\
Jr. AI Scientist & Lt, Id, Hy, Ex, Cd, Ru, An, Wr & L3 & LLM-review, human, workshop & code & topic, baseline, per-step, seed & automated-lit & yes \\
MASTER (materials) & Hy, Ex, Cd, Ru, An, Rv, CL & L4 & human, benchmark, task-success & prompts & topic, template, seed & n/d & yes \\
EvoScientist & Id, Lt, Hy, Ex, Cd, Ru, An, Wr, Rv, CL & L3 & LLM-review, human, task-success, workshop & code & topic, seed & automated-lit & yes \\
Claw AI Lab & Id, Ex, Cd, Ru, An, Wr, Rv, CL & L4(c) & LLM-review & code & topic, per-step, seed & n/d & n/d \\
LLM-AutoSciLab & Hy, Ex, Ru, An, CL & L4 & benchmark, task-success & code, prompts & none & n/d & yes \\
Proactive Reviewer & Rv & L2 & LLM-review, human, benchmark & code, prompts, traces & none & none & yes \\
E2E AI Research & Id, Lt, Hy, Ex, Cd, Ru, An, Wr, Rv, CL & L4 & LLM-review, human, benchmark, workshop & code, prompts, seeds, traces & topic, template, per-step, seed & automated-lit & yes \\
ARIA (materials) & Lt, Hy & L2 & LLM-review, human, benchmark & code, prompts & per-step & n/d & n/d \\
LLM-ACES & Hy, Ex, Cd, Ru, An, CL & L4 & benchmark, task-success & code, prompts & topic, template, seed & n/d & yes \\
\bottomrule
\end{longtable}
\end{landscape}
\endgroup

%% file: appendix_stagecov.tex
\section{Per-System Stage Coverage}
\label{app:stagecov}
Table~\ref{tab:stagecov} shows which research-lifecycle stages each full-text-coded system automates, making the lifecycle concentration (execution/analysis dense, closed-loop sparse) inspectable per system. Columns: Id idea, Lt lit., Hy hyp., Ex exp.-design, Cd code, Ru run, An analysis, Wr writing, Rv review, CL closed-loop.

\begin{longtable}{@{}p{0.26\textwidth}*{10}{c}@{}}
\caption{Per-system lifecycle-stage coverage of the 26 full-text-coded systems.}\label{tab:stagecov}\\
\toprule System & Id & Lt & Hy & Ex & Cd & Ru & An & Wr & Rv & CL \\ \midrule
\endfirsthead \toprule System & Id & Lt & Hy & Ex & Cd & Ru & An & Wr & Rv & CL \\ \midrule \endhead
CAMEO &  &  & $\checkmark$ & $\checkmark$ &  & $\checkmark$ & $\checkmark$ &  &  & $\checkmark$ \\
ReviewAdvisor &  &  &  &  &  &  &  &  & $\checkmark$ &  \\
ChemCrow &  & $\checkmark$ & $\checkmark$ & $\checkmark$ & $\checkmark$ & $\checkmark$ & $\checkmark$ &  &  &  \\
Ultimate Brain & $\checkmark$ & $\checkmark$ & $\checkmark$ & $\checkmark$ & $\checkmark$ &  & $\checkmark$ & $\checkmark$ &  &  \\
ResearchAgent & $\checkmark$ & $\checkmark$ & $\checkmark$ & $\checkmark$ &  &  &  &  & $\checkmark$ &  \\
AutoArticleGen &  &  &  &  & $\checkmark$ &  &  & $\checkmark$ & $\checkmark$ &  \\
SWE-agent &  &  &  &  & $\checkmark$ & $\checkmark$ &  &  &  &  \\
SWIF2T &  & $\checkmark$ &  &  &  &  & $\checkmark$ & $\checkmark$ & $\checkmark$ &  \\
AI Scientist & $\checkmark$ & $\checkmark$ &  & $\checkmark$ & $\checkmark$ & $\checkmark$ & $\checkmark$ & $\checkmark$ & $\checkmark$ & $\checkmark$ \\
CLADD &  & $\checkmark$ &  &  &  &  & $\checkmark$ &  &  &  \\
DatawiseAgent &  &  &  &  & $\checkmark$ & $\checkmark$ & $\checkmark$ & $\checkmark$ &  &  \\
AI+Robot Scientists &  & $\checkmark$ & $\checkmark$ & $\checkmark$ & $\checkmark$ & $\checkmark$ & $\checkmark$ & $\checkmark$ & $\checkmark$ & $\checkmark$ \\
AI Scientist-v2 & $\checkmark$ & $\checkmark$ & $\checkmark$ & $\checkmark$ & $\checkmark$ & $\checkmark$ & $\checkmark$ & $\checkmark$ & $\checkmark$ & $\checkmark$ \\
AI-Researcher & $\checkmark$ & $\checkmark$ & $\checkmark$ & $\checkmark$ & $\checkmark$ & $\checkmark$ & $\checkmark$ & $\checkmark$ &  &  \\
ChatBattery & $\checkmark$ & $\checkmark$ & $\checkmark$ & $\checkmark$ &  & $\checkmark$ & $\checkmark$ &  &  & $\checkmark$ \\
SR-Scientist &  &  & $\checkmark$ &  & $\checkmark$ & $\checkmark$ & $\checkmark$ &  &  & $\checkmark$ \\
Denario & $\checkmark$ & $\checkmark$ & $\checkmark$ & $\checkmark$ & $\checkmark$ & $\checkmark$ & $\checkmark$ & $\checkmark$ & $\checkmark$ &  \\
Jr. AI Scientist & $\checkmark$ & $\checkmark$ & $\checkmark$ & $\checkmark$ & $\checkmark$ & $\checkmark$ & $\checkmark$ & $\checkmark$ &  &  \\
MASTER (materials) &  &  & $\checkmark$ & $\checkmark$ & $\checkmark$ & $\checkmark$ & $\checkmark$ &  & $\checkmark$ & $\checkmark$ \\
EvoScientist & $\checkmark$ & $\checkmark$ & $\checkmark$ & $\checkmark$ & $\checkmark$ & $\checkmark$ & $\checkmark$ & $\checkmark$ & $\checkmark$ & $\checkmark$ \\
Claw AI Lab & $\checkmark$ &  &  & $\checkmark$ & $\checkmark$ & $\checkmark$ & $\checkmark$ & $\checkmark$ & $\checkmark$ & $\checkmark$ \\
LLM-AutoSciLab &  &  & $\checkmark$ & $\checkmark$ &  & $\checkmark$ & $\checkmark$ &  &  & $\checkmark$ \\
Proactive Reviewer &  &  &  &  &  &  &  &  & $\checkmark$ &  \\
E2E AI Research & $\checkmark$ & $\checkmark$ & $\checkmark$ & $\checkmark$ & $\checkmark$ & $\checkmark$ & $\checkmark$ & $\checkmark$ & $\checkmark$ & $\checkmark$ \\
ARIA (materials) &  & $\checkmark$ & $\checkmark$ &  &  &  &  &  &  &  \\
LLM-ACES &  &  & $\checkmark$ & $\checkmark$ & $\checkmark$ & $\checkmark$ & $\checkmark$ &  &  & $\checkmark$ \\
\bottomrule
\end{longtable}

%% file: appendix_crosstab.tex
\section{Corpus Cross-Tabulations}
\label{app:crosstab}
This appendix reports compact analytic cross-tabulations of the coded corpus that complement the
lifecycle\,$\times$\,autonomy map in the body. The evaluation-method rows count systems independently, so a
system that reports several evaluation modes contributes to several rows.

\paragraph{Temporal trend.}
Table~\ref{tab:trend} dates each focal entry by its arXiv identifier. The corpus is deliberately
back-weighted: a single pre-LLM externally validated platform (CAMEO, 2020) and one 2021 system anchor the
``verification built in'' lineage, after which end-to-end LLM systems accelerate sharply, with 2025--2026
alone contributing $17$ of the $26$ coded entries. The acceleration in system count is exactly what makes
the lagging verification disclosure (Table~\ref{tab:stats}) worth flagging now rather than later.

\begin{table}[H]
\centering
\caption{Coded focal entries by year of arXiv posting ($N=26$). The 2020/2021 entries predate the LLM-agent
wave and include the only externally validated closed loop (CAMEO).}
\label{tab:trend}
\small
\begin{tabular}{@{}lccccccc@{}}
\toprule
Year & 2020 & 2021 & 2023 & 2024 & 2025 & 2026 \\
\midrule
Coded systems & 1 & 1 & 2 & 5 & 10 & 7 \\
\bottomrule
\end{tabular}
\end{table}

\paragraph{Evaluation methods in use.}
Table~\ref{tab:evalmix} tabulates how the corpus evaluates itself. Two patterns matter for the
verification thesis. First, the most common evaluation signals are benchmark/task-success (62\%) and a
human-expert read (65\%), both of which certify \emph{task completion} rather than scientific validity.
Second, an LLM acting as reviewer is now as common as benchmarking (62\%); since LLM-as-judge reliability
is itself contested (Tier VIII of the verification ladder), this means a large share of the field certifies
its own output with the weakest signal on the ladder. Genuinely external signals, real or wet-lab
deployment, appear for only $2$ of $26$ systems.

\begin{table}[H]
\centering
\caption{Evaluation methods reported across the corpus ($N=26$; methods co-occur, so columns do not sum to
$N$). Benchmark/task-success and human reads dominate; external (deployment) signals are rare.}
\label{tab:evalmix}
\small
\begin{tabular}{@{}p{0.62\columnwidth}c@{}}
\toprule
Evaluation method & Systems \\
\midrule
Benchmark / task-success score & 62\% (16/26) \\
LLM-as-reviewer (automated review) & 62\% (16/26) \\
Human-expert judgment & 65\% (17/26) \\
Workshop / venue acceptance & 15\% (4/26) \\
Real or wet-lab deployment (external oracle) & \phantom{0}8\% (2/26) \\
\bottomrule
\end{tabular}
\end{table}

\paragraph{The nine L4 closed-loop systems, classified.}
Because the survey's headline rests on \emph{how} each closed loop is validated, Table~\ref{tab:l4} lists
all nine L4 systems with the in-loop signal that drives their next step and our resulting label:
\emph{mechanical} (the loop is triggered by an internal metric or model score), \emph{author-claimed}
(L4(c); the closed loop is asserted but no feedback mechanism is evidenced in the released material), or
\emph{externally validated} (an outside physical or independent oracle gates hypothesis revision). Only
CAMEO meets the last bar, and it predates LLM agents; the materials system MASTER and the wet-lab system
ChatBattery (the latter coded L3, so not in this table) are the closest LLM-era cases but use a metric
plus post-hoc human read and a terminal wet-lab check, respectively, neither gating the loop.

\begin{table}[H]
\centering
\caption{The nine systems coded L4, with the signal that drives the loop and the resulting validation
label. ``m'' mechanical, ``v'' externally validated, ``(c)'' author-claimed.}
\label{tab:l4}
\small
\begin{tabular}{@{}p{0.30\columnwidth}p{0.12\columnwidth}p{0.46\columnwidth}@{}}
\toprule
System & Label & In-loop signal that drives the next step \\
\midrule
CAMEO~\citep{kusne2020cameo} & \textbf{v} & Physical measurement (Bayesian active learning); the instrument selects the next experiment \\
AI Scientist~\citep{lu2024aiscientist} & m & Metric from \texttt{perform\_experiments}; re-plan on numeric result \\
AI Scientist-v2~\citep{yamada2025aiscientistv2} & m & Agentic tree search over experiment nodes scored internally \\
SR-Scientist~\citep{srscientist2025} & m & Benchmark/task-success score on a symbolic-regression target \\
MASTER~\citep{materials2025master} & m & Task-success metric plus a human-expert read \emph{after} the run \\
LLM-AutoSciLab~\citep{llmautoscilab2026} & m & Benchmark/task-success reward \\
E2E AI Research~\citep{e2eairesearch2026} & m & Automated-reviewer plus benchmark score \\
LLM-ACES~\citep{llmaces2026} & m & Benchmark/task-success on the experiment target \\
Claw AI Lab~\citep{clawailab2026} & (c) & Closed loop asserted; no feedback mechanism evidenced in release \\
\bottomrule
\end{tabular}
\end{table}

%% file: appendix_taxonomy.tex
\section{Verification-Signal Taxonomy}
\label{app:taxonomy}
Tables~\ref{tab:taxonomy_full} and~\ref{tab:taxonomy_full_lower} give the full-placement expansion of the
verification-signal ladder introduced in Section~\ref{sec:verification} (Table~\ref{tab:taxonomy_map}).
They use the \emph{same} eight tier definitions and map every surveyed sub-area to the strongest
\emph{verification signal} it admits, ordered from a sound external verifier (Tier I, top) to the model's
own judgment (Tier VIII, bottom). Autonomous discovery is trustworthy in proportion to where its domain
sits on this ladder; the gap the survey documents is that most LLM-agent research operates near the
bottom.

\begin{table}[H]
\centering
\small
\setlength{\tabcolsep}{4pt}
\caption{Surveyed sub-areas ranked by strongest available verification signal, Tiers I--IV (\#works = cited in our corpus).}
\label{tab:taxonomy_full}
\begin{tabular}{@{}>{\raggedright\arraybackslash}p{0.07\textwidth}>{\raggedright\arraybackslash}p{0.28\textwidth}>{\raggedright\arraybackslash}p{0.50\textwidth}c@{}}
\toprule Tier & Sub-area & Strongest verification signal & \#works \\ \midrule
I & Formal theorem proving & Sound formal verifier (proof assistant) & 10 \\
II & Verifiable rewards \& verifiers & Learned/automatic verifier; reward model (sound only with a checker) & 8 \\
II & Reasoning \& planning & Process reward / step verification & 13 \\
II & Coding-agent platforms & Executable tests & 8 \\
II & Research \& discovery benchmarks & Executable / provenance-graded & 8 \\
II & Emerging 2026 benchmarks & Executable / closed-loop graded & 8 \\
II & Reproducibility \& replication & Re-execution of artifacts & 6 \\
III & Scientific foundation models & Simulator / surrogate confidence & 9 \\
III & Physics / astronomy / earth & Physical oracle, simulator, conservation laws & 8 \\
III & Medicine \& clinical & Clinical evidence / wet-lab & 8 \\
III & Engineering \& hardware design & Simulation / functional verification & 8 \\
III & Math discovery beyond proof & Computational check / human inspection & 7 \\
IV & Deep-research agents & Citation / source grounding & 12 \\
IV & Tool use, retrieval, grounding & Retrieval grounding / attribution & 8 \\
IV & Hallucination \& attribution & Citation / claim grounding & 8 \\
IV & KG hypothesis generation & KG consistency / link evidence & 8 \\
IV & Automated survey \& meta-analysis & Citation faithfulness & 8 \\
\bottomrule
\end{tabular}
\end{table}

\begin{table}[H]
\centering
\small
\setlength{\tabcolsep}{4pt}
\caption{Surveyed sub-areas ranked by strongest available verification signal, Tiers V--VIII and contextual rows.}
\label{tab:taxonomy_full_lower}
\begin{tabular}{@{}>{\raggedright\arraybackslash}p{0.07\textwidth}>{\raggedright\arraybackslash}p{0.28\textwidth}>{\raggedright\arraybackslash}p{0.50\textwidth}c@{}}
\toprule Tier & Sub-area & Strongest verification signal & \#works \\ \midrule
V & Data contamination \& leakage & Threat to evaluation validity & 8 \\
V & Agentic RL training & Reward signal (often proxy) & 7 \\
V & Cost \& economic viability & Budget-transparent reporting & 6 \\
V & Newest 2026 AI scientists & Externalized checking apparatus (mixed) & 6 \\
V & Failure modes \& attribution & Threat: failure localization (~14\% acc.) & 2 \\
VI & Reproducibility crisis (context) & Motivation: audit need & 5 \\
VI & Human--AI collaboration & Human expert judgment & 4 \\
VII & Multi-agent foundations & Inter-agent consensus (weak) & 10 \\
VII & Open-source frameworks & Logs/traces (infrastructure) & 6 \\
VIII & LLM-as-judge reliability & Model's own judgment (contested) & 7 \\
-- & Position / theory / economics & Motivation, not a verification signal & 7 \\
\bottomrule
\end{tabular}
\end{table}

\noindent Tiers: I sound formal verifier; II executable tests / process rewards; III physical oracle or
simulator; IV citation/source grounding; V threat-to-validity or proxy reward; VI motivation / human
judgment; VII weak inter-agent or infrastructural signal; VIII the model's own judgment.

%% file: appendix_benchmarks.tex
\section{Benchmark Compendium}
\label{app:benchmarks}
A comprehensive survey of autonomous research agents must also map how the field \emph{measures} them.
Tables~\ref{tab:benchmarks_execution}--\ref{tab:benchmarks_safety} catalog the evaluation suites in our
corpus, grouped by what they score. The pattern reinforces the paper's thesis: most benchmarks reward task
completion or single-turn accuracy, and only a minority probe reproducibility, soundness, or closed-loop
validity.

\begin{table}[H]
\centering
\small
\setlength{\tabcolsep}{4pt}
\caption{Benchmarks and evaluation suites for ML/AI-research execution and hypothesis generation.}
\label{tab:benchmarks_execution}
\begin{tabular}{@{}>{\raggedright\arraybackslash}p{0.30\textwidth}p{0.08\textwidth}>{\raggedright\arraybackslash}p{0.52\textwidth}@{}}
\toprule Benchmark & Year & What it evaluates \\ \midrule
\multicolumn{3}{@{}l}{\emph{ML / AI-research execution}}\\
MLAgentBench~\citep{huang2023mlagentbench} & 2023 & ML experimentation as iterative code-and-run tasks \\
MLE-bench~\citep{chan2024mlebench} & 2025 & ML engineering on Kaggle competitions \\
MLGym~\citep{nathani2025mlgym} & 2025 & framework + benchmark for AI-research agents \\
RE-Bench~\citep{wijk2025rebench} & 2025 & frontier R\&D tasks, human-expert-calibrated \\
PaperBench~\citep{starace2025paperbench} & 2025 & replicating AI-research papers from scratch \\
SciReplicate-Bench~\citep{xiang2025scireplicate} & 2025 & agent-driven algorithmic reproduction \\
InnovatorBench~\citep{wu2025innovatorbench} & 2025 & long-horizon innovative research execution \\
SWE-bench~\citep{jimenez2024swebench} & 2024 & real GitHub issue resolution (substrate) \\
\midrule
\multicolumn{3}{@{}l}{\emph{Discovery, ideation \& hypotheses}}\\
DiscoveryBench~\citep{majumder2024discoverybench} & 2024 & multi-step data-driven hypothesis search \\
IdeaBench~\citep{ideabench2024} & 2024 & research idea generation quality \\
HypoBench~\citep{liu2025hypobenc} & 2025 & systematic hypothesis-generation evaluation \\
\bottomrule
\end{tabular}
\end{table}

\begin{table}[H]
\centering
\small
\setlength{\tabcolsep}{4pt}
\caption{Benchmarks and evaluation suites for reproduction, soundness, and data-science agents.}
\label{tab:benchmarks_analysis}
\begin{tabular}{@{}>{\raggedright\arraybackslash}p{0.30\textwidth}p{0.08\textwidth}>{\raggedright\arraybackslash}p{0.52\textwidth}@{}}
\toprule Benchmark & Year & What it evaluates \\ \midrule
\multicolumn{3}{@{}l}{\emph{Analysis, reproduction \& soundness}}\\
CORE-Bench~\citep{siegel2024corebench} & 2024 & reproducing published computational results \\
ScienceAgentBench~\citep{chen2024scienceagentbench} & 2024 & data-driven scientific analysis tasks \\
SPOT~\citep{son2025spot} & 2025 & detecting errata-grade errors in real papers \\
SoundnessBench~\citep{ho2026soundnes} & 2026 & can an agent tell sound from unsound research \\
\midrule
\multicolumn{3}{@{}l}{\emph{Data-science agents}}\\
InfiAgent-DABench~\citep{hu2024infiagen} & 2024 & data-analysis task completion \\
DSBench~\citep{jing2024dsbench} & 2024 & realistic data-science tasks \\
InsightBench~\citep{sahu2024insightb} & 2024 & multi-step business-analytics insight \\
Tapilot-Crossing~\citep{li2024tapilotc} & 2024 & interactive data-analysis agents \\
\bottomrule
\end{tabular}
\end{table}

\begin{table}[H]
\centering
\small
\setlength{\tabcolsep}{4pt}
\caption{Benchmarks and audits for safety, integrity, review validity, and benchmark disclosure.}
\label{tab:benchmarks_safety}
\begin{tabular}{@{}>{\raggedright\arraybackslash}p{0.30\textwidth}p{0.08\textwidth}>{\raggedright\arraybackslash}p{0.52\textwidth}@{}}
\toprule Benchmark & Year & What it evaluates \\ \midrule
WMDP~\citep{li2024wmdp} & 2024 & hazardous-knowledge / dual-use measurement \\
SciSafeEval~\citep{li2024scisafee} & 2024 & safety alignment for scientific tasks \\
Agent-SafetyBench~\citep{zhang2024agentsaf} & 2024 & LLM-agent safety across scenarios \\
ABC-Bench~\citep{liu2026abcbench} & 2026 & agentic bio-capabilities / biosecurity \\
SciIntegrity-Bench~\citep{yang2026sciinteg} & 2026 & academic-integrity evaluation \\
Benchmark-disclosure audit~\citep{moghadasi2026audit} & 2026 & what agent benchmarks disclose about themselves \\
\bottomrule
\end{tabular}
\end{table}